\documentclass[11pt,preprint]{aastex}

\newcommand{\cs}{$^{13}$CS}
\newcommand{\occs}{O$^{13}$CS}
\newcommand{\hhs}{H$_2$S}
\newcommand{\hh}{H$_2$}
\newcommand{\coo}{C$^{18}$O}

\shorttitle{Observations of sulfur-bearing molecules}
 \shortauthors{Li et al.}

\begin{document}


\title{SULFUR-BEARING MOLECULES IN MASSIVE STAR-FORMING REGIONS: OBSERVATIONS OF OCS, CS, \hhs\ AND SO}

\author{Juan Li\altaffilmark{1}, Junzhi Wang\altaffilmark{1}, Qingfeng Zhu\altaffilmark{2}, Jiangshui Zhang\altaffilmark{3}, Di Li\altaffilmark{4, 5}}

\altaffiltext{1}{Shanghai Astronomical observatory, 80 Nandan RD, Shanghai 20030, China; lijuan@shao.ac.cn}

\altaffiltext{2}{Astronomy Department, University of Science and Technology, Chinese Academy of Sciences, Hefei 210008, China}

\altaffiltext{3}{Center for astrophysics, Guangzhou university, Guangzhou 510006, China}

\altaffiltext{4}{National Astronomical Observatories, Chinese Academy of Sciences A20 Datun Road, Chaoyang District, Beijing 100012, China}

\altaffiltext{5}{Key Laboratory of Radio Astronomy, Chinese Academy of Sciences, Nanjing, 210008, China}

\begin{abstract}

We studied sulfur chemistry of massive star-forming regions through single-dish submillimeter spectroscopy. OCS, \occs, \cs, \hhs\ and SO transitions were observed toward a sample of massive star-forming regions with embedded UCH II or CH II regions. These sources could be divided into H II-hot core and H II-only sources based on their CH$_3$CN emission. Our results show that the OCS line of thirteen sources is optically thick, with optical depth ranging from 5 to 16. Column densities of these molecules were computed under LTE conditions. CS column densities were also derived using its optically thin isotopologue \cs. \hhs\ is likely to be the most abundant gas-phase sulfuretted molecules in hot assive cores. Both the column density and abundance of sulfur-bearing molecules decrease significantly from H II-hot core to H II-only sources. Ages derived from hot core models appear to be consistent with star-formation theories, suggesting that abundance ratios of [CS]/[SO], [SO]/[OCS] and [OCS]/[CS] could be used as chemical clocks in massive star-forming regions.

\end{abstract}

\keywords{ISM: clouds - ISM: molecules - ISM:
abundances - molecular processes - radio lines: ISM}

\section{INTRODUCTION}

Sulfur-bearing molecules are not only excellent tracers of early protostellar evolution (van der Tak et al. 2003), but also of specific interest because of its rapid evolution in warm gas (Charnley 1997; Hatchell, 1998a, b). Unlike nitrogen chemistry or oxygen chemistry, most of the molecules involved are easy to detect at mm and submm wavelengths and column densities can be calculated easily, thus sulfur is a potential good clock on time scales relevant to the embedded phase of star formation.

Chemical evolution of sulphur-bearing molecules have been well studied in low-mass star-forming regions, e.g., Buckel \& Fuller et al. (2003) studied chemical evolution of \hhs, SO and SO$_2$ in low-mass star-forming regions. They demonstrated that the chemical evolution of sulphur-bearing species was a potentially valuable probe of chemical timescales in low-mass star-forming regions. However, this issue has been less developed in the context of high-mass star-forming regions. Hatchell et al. (1998a, 1998b) studied the sulphur chemistry of eight sources associated with UCH II regions. The observed abundance ratios vary little from source to source.
Van der Tak et al. (2003) studied the sulfur chemistry in nine high-mass protostars and found that high-energy transitions probing the inner parts of the protostars where the temperature exceeds 100 K were mandatory for using S-bearing molecules as chemical clocks. They proposed that OCS was the main carrier of sulfur on grains, based on the high excitation temperature of the molecule and its high abundance in the protostars.

Since hydrogenation is expected to be the most efficient process on grains, \hhs\ could be formed on interstellar grains (van Dishoeck \& Blake 1998). It is now generally accepted that during the cold collapse phase, sulphur atoms freeze out onto grains and remain there in the form of \hhs\ until core heating begins. At this point, \hhs\ is evaporated from the grains, and rapidly undergoes reactions which drive the production of SO and SO$_2$ (e.g., Charnley 1997; Wakelam et al. 2004, 2011). The initial destruction of \hhs\ by H$_3$O$^+$ is more efficient in the high-mass objects than in low-mass protostars, as water is more abundant there (van der Tak et al. 2006). Viti et al. (2004) extended chemical models of high-mass star-forming cores by including the experimental results on desorption from grains in the evaporation of icy mantles formed in star-forming regions and developed a time-dependent gas-grain model. They found that distinct chemical events occur at specific grain temperatures. They showed that, provided there is a monotonic increase in the temperature of the gas and dust surrounding the protostar, the changes in the chemical evolution of each species due to differential desorption are important. They proposed that the abundance ratio of sulphur-bearing species, like [CS]/[\hhs], [OCS]/[\hhs], [CS]/[SO], [SO]/[OCS], can be used to investigate the evolution of the early phases of massive star-formation, and that sulphur-bearing species are good 'chemical clocks'. Esplugues et al. (2014) modeled the sulphur chemistry evolution of the hot core and the plateau in Orion KL using the chemical models in Viti et al. (2004), with two different phases in calculations. Phase I starts with the collapsing cloud and the depletion of atoms and molecules onto grain surfaces, while Phase II starts when a central protostar is formed and the evaporation from grains takes place. They reproduced the observations of sulphur-bearing molecules (SO, SO$_2$, CS, OCS, \hhs, and H$_2$CS) in the hot core and plateau.

Herpin et al. (2009) observed sulfur-bearing species of two mid-infrared quiet and two brighter massive cores, with one bright massive core associated with hypercompact H II (HCH II) regions. They found that SO and SO$_2$ relative abundances increased with time, while CS and OCS decreased. This trend seems to be in contradiction with chemical model of Viti et al. (2004), and more observations are needed to investigate whether this is biased by the small observing sample.

Considering that the systematic understanding of the chemistry of sulfur-bearing molecules in massive star-forming regions is far from complete, we carried out observations of OCS, \occs, \cs, \hhs\ and SO toward a sample of 36 massive star-forming regions with embedded H II regions and studied the evolution of molecular column densities and abundance ratios. These results are compared with chemical models of hot core. In this paper we first introduce the observations and data reductions in \S\ \ref{observation}. In \S\ \ref{result}, we present the observing results. The analysis and discussion of the results are presented in \S\ \ref{discussion}, followed by a summary in \S\ \ref{summary}.

\section{OBSERVATIONS AND DATA REDUCTIONS}
\label{observation}

The observations were conducted in July 2013 at the 10.4 m telescope of the Caltech Submillimeter Observatory (CSO) using
the 230 receiver, which is double-side band (DSB) receiver. The data were obtained in position switching mode with an offset of 30\arcmin\ in azimuth.
The Fast Fourier Transform Spectrometer (FFTS2) provides a 4-GHz window of DSB coverage with a velocity resolution of 0.35 km s$^{-1}$ at 230 GHz.
The total integration times ranged from 3 to 20 minutes, while system temperatures ranged from 250 to 400 K. OCS 19-18 (231.060991 GHz), \occs\ 19-18 (230.317500 GHz), \cs\ 5-4 (231.220768 GHz), \hhs\ 2$_{2,0}$-2$_{1,1}$ (216.710437 GHz) and SO $5_6-4_5$ (219.949433 GHz) were observed simultaneously within the receiving band, with OCS, \occs\ and \cs\ transitions observed in USB, while \hhs\ and SO transitions observed in LSB. \coo\ 2-1 transition was obtained simultaneously, which was used to derive column density of \hh\ and molecular abundances. Table 1 gives details about the observed transitions.

Part of our sample is selected from Shirley et al. (2003) of massive star-forming cores associated with water masers. The observing center was the water maser position from the catalog of Cesaroni et al. (1988). Other sources are chosen from Zhu et al. (2008), Wood \& Churchwell (1989), Kurtz et al. (1994) and Beck et al. (1998). Table 1 lists detailed information of these 36 sources. Classification of associated H II region and references are also listed. Most of these sources are associated with UCH II regions, while W75OH is associated with HCH II region (Li et al. 2012). G10.6-0.4, W33IRS1, W33IRS2, W33IRS3 and W51M are associated with CH II regions (Wood \& Churchwell 1989; Beck et al. 1998). S76E is associated with H II regions (Li et al. 2012).

Data Reduction was carried out using the GILDAS/CLASS package\footnote{\tt
http://www.iram.fr/IRAMFR/GILDAS.} developed at IRAM. Linear baselines were subtracted and calibrated onto T$_{MB}$ using a beam efficiency of 0.69. The final spectra have rms noise levels (in T$_{MB}$) ranging from 0.03 to 0.14 K at the velocity resolution of 0.35 km s$^{-1}$.

\section{RESULTS}
\label{result}

The whole 8-GHz spectra data indicate that some sources show many high-excitation lines from complex molecules like CH$_3$CN, whereas others show mainly low-excitation lines like $^{13}$CO and C$^{18}$O (Li et al., in prep.). High-excitation energy lines require high temperatures to excite, so the no-detections mean that there is no evidence for hot cores in line-poor sources. Since massive star-formation takes place on rather short timescales and in clustered environments, the transitions from one into the next stage are smooth and not always clearly distinct. There might be overlaps among the hot massive core (HMC) and UCH II region stages. Due to the large beam, regions with different evolutionary stages in their proximity can also be observed as one objects showing the characteristics of different stages. This situation was noted by Hatchell et al. (1998a), and they divided their UCH II sample into line-rich and line-poor sources. Thus we also divided our sources into two groups. Sources with detectable CH$_3$CN emission (220 GHz) above 2$\sigma$ ($\sim$0.1 K) level are labeled as H II-hot core (H II-HC), while sources without detectable CH$_3$CN emission above 2$\sigma$ ($\sim$0.1 K) level are labeled as H II-only. The brightness of CH$_3$CN emission does not correlate with distance, implying intrinsic differences in the source properties, rather than simply beam-dilution effects (Hatchell et al. 1998a). For the H II-HC sources, though they are already surrounded by an ionized medium, they still show rich hot core chemistry. On the contrary, the H II-only sources without signature of hot core could better represent chemistry in UCH II stage. Thus the H II-HC sources are regarded to be younger than the H II-only sources because they contain younger components (Gerner et al. 2014). There are 25 H II-HC sources and eleven H II-only sources in our sample. The classification information is also listed in Table 2.

Spectra from each of the molecules toward all the sources are given in Figure 1 to 4 (on-line only). SO emission is stronger than other lines in all the sources except for W31, G10.6-0.4 and W42, in which it is weaker than \hhs\ emission. OCS emission was detected in 26 sources, most of which are H II-HC sources. \occs\ emission was detected in 15 sources, all of which have been detected in OCS except for G35.20-1.74. \cs\ was detected in 33 sources, while \hhs\ was detected in 30 source, including all the H II-HC sources and some H II-only sources. SO was detected in all the sources, with integrated intensities ranging from 0.82 (G78.44+2.66) to 61.97 $K~km~s^{-1}$ (W51M). We fitted the detected lines with Gaussian profiles. Most lines could be satisfactorily fitted with a single Gaussian except for SO, which is consistent with van der Tak et al. (2003). The Gaussian fitting results, including peak temperature (T$_{mb}$), integrated line intensity, LSR velocity and full-width half-maximum (FWHM) linewidth of OCS, \occs, \cs, \hhs\ and \coo\ are presented in Table 3-5. For SO which could not be satisfactorily fitted with a single Gaussian, we present the integrated intensity, LSR velocity and linewidth in Table 4.

\section{ANALYSIS AND DISCUSSIONS}
\label{discussion}

\subsection{OCS Optical Depth}

The OCS/\occs\ intensity ratio is significantly lower than the expected abundance ratio of about 45, suggesting that OCS is optically thick. This is consistent with SMA observations of G10.6-0.4 (Liu et al. 2010) and PdBI observations of the intermediate mass protostar NGC 7129 FIRS 2 (Fuente et al. 2014). Fuente et al. (2014) estimated an opacify of the OCS 19-18 line of about 12 in NGC7129 FIRS 2, while Schilke et al. (1997) calculated an optical depth of about 3.5 for Orion KL. We estimated the OCS optical depth by assuming that the excitation temperature is the same for both isotopologues. Tereco et al. (2010) found an average value of $^{12}$C/$^{13}$C = 45 in study of sulfur carbon molecules of Orion KL, thus we adopted $^{12}$C/$^{13}$C of 45 for all these sources. With these assumptions we can write the following relation (Myers, Linke \& Benson 1983),
\begin{equation}
\frac{T_{O^{13}CS}}{T_{OCS}}= \frac{1-exp(-\tau_{OCS}/45)}{1-exp(-\tau_{OCS})},
\end{equation}


The computed $\tau_{OCS}$ is tabulated in Table 3. OCS 19-18 emission are optically thick in all the sources detected in \occs, with optical depth range from 5 to 16. It appears that OCS lines are saturated in hot cores. We found that \occs\ is stronger than OCS in DR21S and G35.20-1.74, which is possibly caused by large temperature gradient. Observations with higher angular resolution are needed to investigate this issue.

\subsection{Column Density and Abundance}

Molecular column densities have been estimated for all observed species. We assumed that the emission is extended and uniformly distributed within the beam. Therefore, the beam filling factors is $\sim$1. For all species, we assume local thermodynamic equilibrium (LTE) conditions and optically thin emission. We used the following formula to estimate molecular column densities (Vasyunina et al. 2014):
\begin{equation}
N_{tot}=\frac{8\pi}{\lambda^3 A} \frac{1}{J_{\mu}(T_{ex})-J_{\mu}(T_{bg})} \frac{1}{1-exp(-h\nu/kT_{ex})}\frac{Q_{rot}}{g_u exp(-E_l/kT_{ex})}\int T_{mb}dv,
\end{equation}

where $\lambda$ is the rest wavelength of the transition, $A$ is the Einstein coefficient, $g_u$ is the upper state degeneracy, $J_{\mu}(T_{ex})$ and $J_{\mu}(T_{bg})$ are values of the Plank function at excitation and background temperatures, respectively, $Q_{rot}$ is the partition function, and $E_l$ is the energy of the lower level (Lee et al. 2009). For $g_u$, $A$ and $E_l$, we used values from The Cologne Database for Molecular Spectroscopy (CDMS) (M$\ddot{u}$ller et al. 2001, 2005) for most of the species. We assumed an uniform excitation temperature $T_{ex}$ of 50 K (van der Tak et al. 2003) and calculated the partition function $Q_{rot}$ for every source by interpolating data from the CDMS for assumed $T_{ex}$.

We did not detect emission from $^{33}$SO (217.831762 GHz) above 2$\sigma$ ($\sim$0.1 K) in these sources, suggesting that $^{32}$SO emission is almost
optically thin. Previous observations also indicate that $^{32}$SO was at most moderately optically thick (e.g., Buckle \& Fuller 2003), thus we derived
SO column densities with the assumption of optically thin. Crockett et al. (2014) observed multi-J transitions of \hhs\ and its isotopologues at frequencies ranging from 480 to 1907 GHz.
They found that \hhs\ is optically thick, with optical depth ranging from 2 to 40. However, given that we did not observe isotoplogues of \hhs, we assume that it is optically thin for simplicity. The column density and abundance of \hhs\ might be underestimated because of this, but the
comparison between H II-HC and H II-only groups will not be affected too much. \coo\ column densities were also derived with assumption of optically thin. Using the abundance of N(\hh)/N(\coo)=7$\times$ $10^6$ (Castets \& Langer 1995), the N(\hh) based on \coo\ data are derived. To derive the CS column density, we used the optically thin emission of \cs\ by assuming $^{12}$C/$^{13}$C=45 (Tercero et al. 2010). For sources detected in \occs\ emission, the OCS column density was also derived from its less abundant isotopologue \occs, again assuming a $^{12}$C/$^{13}$C ratio of 45.
For sources detected in OCS but not detected in \occs, the OCS column densities were derived from OCS emission by assuming optical depth of 5. 
Three of our sources, including G5.89-0.39, G29.96-0.02 and G34.26+0.15, were also observed by Hatchell et al. (1998a, 1998b). It is noted that the column densities derived here are in agreement with their results. Column density for \hh, CS, OCS, \hhs\ and SO molecules, as well as abundance for \hh, CS, OCS, \hhs\ and SO molecules are listed in Table 6. Median values of the column density and abundance for different evolutionary sequence are also shown.

The number distribution of the CS, OCS, \hhs\ and SO column densities for H II-HC and H II-only sources are shown in Figure 5, while the number distribution of CS, OCS, \hhs\ and SO abundances are shown in Figure 6. Since we took into account sources undetected while calculating the median values, some median values donot fall in the middle of the sources shown. For example, for H II-only sources, OCS was detected in less than half sources, only upper limit for the median value could be obtained. We could see a trend of decreasing column density and abundance with the evolution of the clumps from these figures, which is confirmed by the median values. The median value of \hh\ column densities are 2.2$\times 10^{23}$ cm$^{-2}$ and 9.9$\times 10^{22}$ cm$^{-2}$ for H II-HC sources and H II-only sources, respectively. The decrease of \hh\ column densities might be caused by the expansion of H II regions. The decrease of CS, OCS, \hhs\ and SO column density and abundance strongly suggest that these molecules are probably destroyed by the UV-radiation from the forming stars. For H II-HC sources, median values of CS, OCS, \hhs\ and SO are 6.4$\times 10^{14}$, 1.2$\times 10^{15}$, 7.9$\times 10^{14}$ and 2.3$\times 10^{14}$ cm$^{-2}$, respectively. Since \hhs\ is assumed to be optically thin in the calculation, the column density and abundance of \hhs\ might be underestimated, and column density and abundance of \hhs\ is likely to be higher than OCS. All together these results suggest that \hhs\ is probably the main sulphur carrier molecules in massive hot cores, while column density and abundance of OCS are generally higher than CS and SO. This trend is not consistent with hot core model of Viti et al. (2014), in which CS abundance is always higher than OCS. OCS was found to be the most abundant sulfur-bearing species in starburst galaxy NGC 253 (Martin et al. 2005). Maity \& Kaiser (2013) investigated the formation of sulfur species in interstellar ices during the irradiation of CS$_2$ and O$_2$ ices with energetic electrons at 12 K, and the sulfuretted molecules produced during the irradiation were SO$_2$, SO$_3$ and OCS. The large overabundance of OCS compared with the time-dependent sulfur chemistry model seems to support the idea that OCS is injected into the gas phase from grain mantles by low-velocity shocks (Martin et al. 2005). It would be helpful to investigate the relationship between OCS abundance and low-velocity shock tracers such as HNCO (Rodr\'{\i}guez-Fern\'{a}ndez et al. 2010; Li et al. 2013).

Esplugues et al. (2014) modeled the sulphur chemistry evolution of the hot core and the plateau in Orion KL using the chemical models in Viti et al. (2004), with two different phases in calculations. Phase I starts with the collapsing cloud and the depletion of atoms and molecules onto grain surfaces. In this phase, the material collapses and atoms and molecules are depleted onto grain surfaces. The density increases with time according to the so-called modified collapse. In Phase II, the density is constant, a central protostar is formed, and the sublimation from grains takes place due to the warming up of the region. A time-dependent evaporation model was employed, as instantaneous sublimation is a more appropriate approximation only if the mass of the central stars is very high. We found that their hot core model of Orion KL could reproduce the general trend (\hhs\ $>$ OCS $>$  CS $>$  SO) observed in this paper. According to their predictions, such trend are reproduced at about 7.5$\times10^4$ years after the central star switches on (see Figure 12 of Esplugues et al. 2014). The models that reproduce the the median column densities and abundances of sulphur-bearing species are those with an initial sulphur abundances of 0.1 times the sulphur solar abundance (0.1S$_{\odot}$) and a density of \textbf{about $10^7$} cm$^{-3}$. The column density of SO is higher than CS by an order of magnitude (See Table 2 in Esplugues et al. (2014)) in the Orion hot core, whereas we found CS $>$ SO, suggesting that our sources are on average older than the Orion hot core.

\subsection{Abundance Ratios}

Abundance ratios of chemically related molecules and with similar excitations conditions are thought to be more accurate than absolute molecular abundances (Ginard et al. 2012). Previous chemical models and observations all suggest that abundance ratios of sulphur-bearing molecules can be used to investigate the evolution of the early phases of massive star-formation, and that sulphur-bearing species are good 'chemical clocks' (e.g., Charnley 1997; Viti et al. 2004; Wakelam et al. 2004, 2011; Herpin et al. 2009).

According to hot core model of Orion KL (Esplunges et al. 2014), for a 10-M$_{\odot}$ star with an initial sulphur abundances of 0.1S$_{\odot}$ and a density of at least \textbf{$10^7$ cm$^{-3}$}, (i) [SO]/[OCS] $>$1 up to 70 000 yr and $<$1 after that; (ii) [CS]/[SO]$<$1 up to 70 000 yr and $>$1 after that; (iii) [OCS]/[CS] $>$1 up to 85 000 yr and $<$1 after that. Both [OCS]/[\hhs] and [CS]/[\hhs] are not chemical clock candidate, since they are always smaller than 1. Herpin et al. (2009) investigated possible correlations of the molecular abundance ratios with the order of evolution of four objects, and also proposed that molecular ratios like [OCS]/[\hhs], [CS]/[\hhs], [SO]/[OCS], [CS]/[SO] might be good indicator of evolution. However, their trend is different from that of Esplunges et al. (2014). We calculated abundance ratios and present them in Table 7. Individual abundance ratios for H II-HC sources are discussed below, in which the chemistry is dominated by hot cores.

[OCS]/[\hhs]. The median value for H II-HC source is 1.8, which is higher than prediction in Esplunges et al. (2014). This inconsistency is likely to be caused by the assumption of optical thin of \hhs.

[CS]/[\hhs]. The median values for H II-HC is 0.8, which is not in contradiction with model prediction.

[SO]/[OCS]. The median values for H II-HC is 0.2, suggesting to be elder than 70 000 yr. [SO]/[OCS] is smaller than 1 for all the sources except for M8E, which belongs to H II-only group.

[CS]/[SO]. The median values for H II-HC and H II-only regions are 2.9 and 1.3, respectively. Both of them are larger than 1, suggesting to be elder than 70 000 yr (Esplunges et al. 2014).

[OCS]/[CS]. The median values for H II-HC is 2.0, suggesting to be younger than 85 000 yr (Esplunges et al. 2014).

According to analysis above, most of these H II-HC sources should be older than 70 000 yr, and younger than 85 000 yr. Both McKee \& Tan (2003) and Mac Low et al. (2007) derived typical timescales for the formation of a massive star of about 10$^5$ yr. Gerner et al. (2014) also found that the total chemical timescale for the high-mass star formation process is on the order of 10$^5$ yr. It appears that our result is in agreement with theoretical estimates and chemical evolution model, suggesting that [SO]/[OCS], [CS]/[SO] and [OCS]/[CS] could be used as `chemical clock` in massive star-forming regions.

For H II-only sources, abundance ratios were obtained for only a few sources. Comparing with H II-HC sources, the abundance ratios have not changed significantly, suggesting that the sulfur species that are left are still from relatively young components.  Hatchell et al. (1998a) proposed that the non-detection of hot gas may be caused by low column density of hot gas or small source size of these sources. However, the obvious decline of abundance imply that models that taking into account effects of UV-radiation from the forming stars are needed to explain results of H II-only sources.

As is mentioned above, massive star-formation always take place on rather short timescales and in clustered environments. Due to the large beam of our observations, some regions with embedded H II regions also show the characteristics of hot core, thus observations with higher resolution could better constrain the age of these sources.

\section{SUMMARY}
\label{summary}

We have conducted a systematic study of sulphur-bearing molecules toward 36 massive star-forming regions with embedded H II regions using the CSO telescope. We divided these sources into H II-HC and H II-only to characterize their evolutionary stages. OCS emission is optically thick for thirteen sources, with optical depth ranging from 5 to 16. We calculated column density and abundance of these molecules under LTE conditions, and \hhs\ seems to be the most abundant sulphuretted molecules. The hot core model of Orion KL could reproduce the general trend (\hhs\ $>$ OCS $>$  CS $>$  SO) observed in this paper. The models that reproduce the the median column densities and abundances of sulphur-bearing species are those with an initial sulphur abundances of 0.1 times the sulphur solar abundance (0.1S$_{\odot}$) and a density of at least $5\times 10^6$ cm$^{-3}$. We found that the column density and abundance of CS, OCS, \hhs\ and SO decrease significantly from H II-HC to H II-only sources. We derived abundance ratios of [OCS]/[\hhs], [CS]/[\hhs], [SO]/[OCS], [CS]/[SO] and [OCS]/[CS], and made comparison between ages predicted by hot core models with massive star-formation models. We conclude that these abundance ratios are potential good chemical clocks in massive star-forming regions.

This work is partly supported by China Ministry of Science and Technology under State Key Development Program for Basic Research (2012CB821800), and partly supported by the Natural Science Foundation of China under grants of 11103006.

\begin{table*}
    \begin{center}
\caption{Observed transitions.}
  \begin{tabular}{l c c cccccc}
    \hline
    \hline
Transition & Frequency (GHz) & E$_u$ (K) &  Eins. A. (s$^{-1}$) \\
\hline
 H$_2$S $2_{20}-2_{11}$ & 216.710437   &  84.0   &  4.87$\times 10^{-5}$    \\
 SO $5_6-4_5$           & 219.949433   &  35.0   & 1.36$\times 10^{-4}$      \\
 OCS 19-18              & 231.060991   &  110.2  &  3.58$\times 10^{-5}$     \\
 O$^{13}$CS 19-18       & 230.317500   &  109.9  &  3.54$\times 10^{-5}$    \\
 $^{13}$CS 5-4          & 231.220768   &  33.1   &  2.51$\times 10^{-4}$     \\
 \coo\ 2-1              & 219.560354   &  15.8   &  6.01$\times 10^{-3}$     \\
\hline
  \end{tabular}
  \end{center}
  \label{tab:freq}
\end{table*}

\begin{table*}
    \begin{center}
\caption{Source sample.}
  \begin{tabular}{lccccc}
    \hline
    \hline
Source name     & RA (J2000)  & DEC            &  association    &  Ref.   &  chemically    \\
                & (h m s)     & ($^{\deg}$)    &                 &         &       \\
                \hline
W3(OH)       & 02:27:04.69    &  61:52:25.5     &    UCH II   &  Li2012    &    H II-HC       \\
RCW142       & 17:50:15.13    &  -28:54:31.5     &    UCH II    & Shirley2003 & H II-HC             \\
G5.89-0.39   & 18:00:30.39   & -24:04:00.1      &   UCH II     &  Zhu2008    &  H II-HC          \\
M8E          & 18:04:53.25    & -24:26:42.4     &     UCH II   & Shirley2003 &  H II-only            \\
G8.67-0.36   & 18:06:19.03   & -21:37:32.3      &    UCH II   & Li2012       &  H II-HC       \\
W31          & 18:08:38.32   & -19:51:49.7     &      UCH II   & Shirley2003 &  H II-HC            \\
G10.6-0.4    & 18:10:28.70   & -19:55:48.6     &     CH II   &  WC1989   &    H II-HC        \\
G11.94-0.62  & 18:14:01.10   & -18:53:24.2     &  UCH II  &  Zhu2008     &  H II-HC        \\
W33cont      & 18:14:13.67    &  -17:55:25.2    &    UCH II      &  Shirley2003  &  H II-HC           \\
W33irs1      &  18:14:15.00    & -17:55:54.9     &    CH II   &   Beck1998       &  H II-HC      \\
W33irs2      & 18:14:14.49    & -17:55:14.9      &   CH II    &  Beck1998        &  H II-HC     \\
W33irs3      &  18:14:13.30   &  -17:55:40.0      &   CH II    &  Beck1998       &  H II-HC      \\
G14.33-0.64  & 18:18:54.71    & -16:47:49.7      &   UCH II     &  Shirley2003   & H II-HC           \\
W42          & 18:38:12.46    &  -07:12:10.1     &   UCH II     &  Li2012        & H II-HC      \\
G29.96-0.02  & 18:46:03.90    & -02:39:22.0     &   UCH II    &  Zhu2008         & H II-HC     \\
G30.54+0.02  & 18:46:59.5     & -02:07:26     &    UCH II      &  Zhu2008      &  H II-only      \\
G33.92+0.11  & 18:52:50.00    &  00:55:28.9     &   UCH II     &   Zhu2008  &  H II-only           \\
G34.26+0.15  & 18:53:18.53    &  01:14:57.9     &    UCH II   & WC1989      &  H II-HC       \\
S76E         & 18:56:10.43    &  07:53:14.1     &     H II     & Li2012     &  H II-HC         \\
G35.20-0.74  & 18:58:12.73   &  01:40:36.5     &    UCH II     &  Shirley2003 & H II-HC              \\
G35.20-1.74  & 19:01:47.00   &  01:13:07.9     &   UCH II    &  Kurtz1994    &  H II-only          \\
G37.543-0.11 &  19:00:15.70  & 04:03:12.9    &   UCH II    &  WC1989    &  H II-only     \\
G43.89-0.78  & 19:14:26.2    &  09:22:34     &   UCH II    & Zhu2008     &  H II-only     \\
G45.45+0.06  & 19:14:21.30   &  11:09:12.9     &  UCH II     &  Zhu2008      &  H II-only        \\
W51M         & 19:23:43.86   &  14:30:29.4      &   CH II     &  Li2012      &  H II-HC         \\
W51D         & 19:23:39.90   & 14:31:06.0      &   UCH II    & Zhu2008       &  H II-HC       \\
G61.48+0.09  & 19:46:47.33   &  25:12:45.6      &  UCH II    &  Zhu2008      &  H II-only        \\
K3-50A       &  20:01:45.59    &  33:32:42.0      &  UCH II    &  Zhu2008    &  H II-only          \\
G78.44+2.66  &  20:19:29.21  & 40:56:36.6    &     UCH II  &  Kurtz1994     & H II-only      \\
W75N         & 20:38:36.93     &  42:37:37.4     &    UCH II    &  Li2012    &  H II-HC          \\
DR21S        &  20:39:00.80  & 42:19:29.7       &    UCH II      &  Li2012   &  H II-HC           \\
W75(OH)      & 20:39:01.00    & 42:22:49.8      &   HCH II      &  Li2012    & H II-HC           \\
G81.7+0.5    &  20:39:00.90    &  42:19:44.0     &   UCH II    &  Kurtz1994  & H II-only         \\
CepA         & 22:56:18.13    & 62:01:46.3      &    UCH II      & Li2012    & H II-HC           \\
NGC7538      & 23:13:44.85    & 61:26:50.6      &   UCH II     &  Shirley2003  &   H II-HC           \\
NGC7538A     & 23:13:45.59    & 61:28:18.0       &  UCH II    &   Zhu2008     &  H II-HC         \\
\hline
  \end{tabular}
  \end{center}
  Ref.: Beck1998: Beck, S. C., Kelly, D. M. \& Lacy, J. H., 1998, ApJ, 115, 2504; \\
  Kurtz1994: Kurtz, S., Churchwell, E. \& Wood, D.O.S. 1994, ApJ, 91, 659; \\
  Li2012: Li, J., Wang, J.Z., Gu, Q.S. et al. 2012, ApJ, 745, 47; \\
  Shirley2003: Shirley, Y. L., Evans II, N. J., Young, K. E. et al. 2003, ApJS, 149, 375; \\
  WC1989: Wood, D.O.S. \& Churchwell, E. 1989, ApJS, 69, 831; \\
  Zhu2008: Zhu, Q.-F., Lacy, J. H., Jaffe, D.T., 2008, ApJS, 177, 584
  \label{tab:source}
\end{table*}

\begin{deluxetable}{lrrrrrrrrrrrr}
\tabletypesize{\scriptsize} \rotate
\tablecaption{\label{tab:dlhresult} Results of OCS 19-18, \occs\ 19-18 Transitions and Opacity of OCS} \tablewidth{0pt}
\tablehead{
\multicolumn{1}{l}{SOURCE} &\multicolumn{4}{c}{OCS 19-18}&\multicolumn{4}{c}
{\occs\ 19-18}&\multicolumn{1}{c}{Opacity }\\
\colhead{}             & \colhead{T$_{mb}$}  &
\colhead{$\int$T$_{mb}$d$\nu$}  & \colhead{V$_{LSR}$}    &
\colhead{FWHM}         & \colhead{T$_{mb}$}  &
\colhead{$\int$T$_{mb}$d$\nu$}  & \colhead{V$_{LSR}$}    &
\colhead{FWHM}         &
\colhead{}        \\
\colhead{}             & \colhead{(K)}          &
\colhead{($K~km~s^{-1}$)}  & \colhead{($km~s^{-1}$)}   &
\colhead{($km~s^{-1}$)}   & \colhead{($K$)}          &
\colhead{($K~km~s^{-1}$)}  & \colhead{($km~s^{-1}$)}   &
\colhead{($km~s^{-1}$)}    &
\colhead{} } \startdata
W3(OH)      & 0.33(.04) &1.78(0.10) & -46.47(.14)  &  5.09(.33)   & 0.03(.04) & 0.26(.13)  & -45.43(1.97) &  8.00(5.86)          &   7.2             \\
RCW142      & 0.35(.04) &2.33(0.09) & 17.02(.12)   &  6.34(.30)   & 0.06(.03) & 0.41(0.09) & 17.79(.64) &  6.35(1.34)            &   8.7               \\
G5.89-0.39  & 0.42(.06) &2.14(0.10) &  9.23(.11) &  4.76(.26)     & 0.04(.01) & 0.29(0.04)    &  11.47(.42)   &  5.49(1.45)      &   6.6             \\
M8E         & 0.06(.03) &0.14(0.06) & 9.94(.50) & 3.36(.87)       &  $<$0.04 &  $<$0.09      & -   &    -                        &  -             \\
G8.67-0.36  & 0.19(.06) &0.78(0.10) & 35.52(.24) &  3.79(.53)     & $<$0.06    &  $<$0.25      & -   &  -                        &   -            \\
W31         & 0.67(.14) &6.68(0.42) & 66.33(.28) &  9.49(.71)     & 0.23(.06) & 1.68(0.14)  & 68.32(.29)  & 6.75(.61)            &   13.1             \\
G10.6-0.4   & 0.46(.09) &3.48(0.19) & -2.70(.19) & 6.96(.45)      & 0.09(.03) & 0.39(0.06)    &  -1.77(.36)   &  4.46(.79)       &   5.3            \\
G11.94-0.62 & 0.03(.03) &0.10(0.04) & 38.64(.56)    & 2.98(1.12)  & $<$0.03  &   $<$0.10       & -   &  -                        &   -             \\
W33cont     & 0.22(.04) &1.23(0.07) & 35.02(.16) &  5.32(.38)     & $<$0.04  &  $<$0.22       & -   &  -                         &   -             \\
W33irs1     & 0.13(.06) &0.91(0.17) & 36.61(.62) & 6.36(1.32)     & $<$0.09 &   $<$0.63      & -   &  -                          &   -             \\
W33irs2     & 0.23(.04) &1.03(0.07) & 35.20(.14) &  4.33(.39)     & $<$0.03  &  $<$0.13       & -   &  -                         &    -             \\
W33irs3     & 0.20(.04) &1.09(0.10) & 35.70(.21)  & 4.98(.53)     & $<$0.06  &  $<$0.33       & -   &  -                         &    -            \\
G14.33-0.64 & 0.22(.04) &0.88(0.07) & 22.52(.15)    & 3.78(.37)   & 0.10(.04) & 0.12(0.07)    &  23.13(.40)   & 2.60(.99)        &   6.6            \\
W42         & 0.35(.07) &2.49(0.17) & 110.9(.23) & 6.78(.56)      & 0.07(.03) & 0.36(0.07)    & 11.3(.46) & 4.80(1.03)           &   7                \\
G29.96-0.02 & 0.23(.03) &2.10(0.09) & 97.76(.15) &  8.28(.45)     & 0.04(.03) & 0.23(0.06)    & 98.17(.64) & 4.81(1.42)          &   5.2                \\
G30.54+0.02  &   -      &   -       &    -       &   -            &    -      &   -           &     -      &   -                 &   -      \\
G33.92+0.11 & 0.09(.07) &0.48(0.13) & 110.6(.84) &  5.57(1.25)    & $<$0.07  &  $<$0.37      & -   &  -                          &   -                 \\
G34.26+0.15 & 0.90(.07) &7.03(0.75) & 58.49(.39) & 7.35(.94)      & 0.16(.38)  & 0.90(0.17)     & 58.53(.56) & 5.50(1.06)        &   6.1              \\
S76E        & 0.14(.03) &0.57(0.06) & 32.01(.16)  &  3.67(.42)    & $<$0.03  &  $<$0.43         & -   &  -                       &    -              \\
G35.20-0.74 & 0.09(.03) &0.48(0.06) & 32.55(.28) &  5.33(.77)     & $<$0.03  & $<$0.16   & -   &  -                              &    -               \\
G35.20-1.74 &  -       &   -       & -            &  -             & 0.07(.06)  & 0.46(0.13)     & 45.49(.84) &  5.61(.80)        &    -          \\
G37.543-0.11 & -          &  -     & -           &   -             &   -      &   -        &   -       &   -          &      -          \\
G43.89-0.78  & -          &  -    & -           &    -            &   -      &    -       &   -       &   -         &   -            \\
G45.45+0.06 &  -        & -      & -            &  -              &   -     -  &   -       & -         &  -                      &    -              \\
W51M        & 1.32(.06)&11.94(0.68) & 56.54(.23) & 8.49(.58)      & 0.26(.07) & 2.22(0.17)      & 58.14(.32) & 7.91(.68)         &    9.3                \\
W51D        & 0.35(.09) &2.90(0.26) & 60.07(.35) &  7.71(.75)     & $<$0.16(0.07) & $<$0.79(.19)  & 60.19(0.54) & 4.60(1.26)     &    -            \\
G61.48+0.09 & -   &  -     & -   &  -                             & -        &   -       & -   &  -                              &    -           \\
K3-50A      & -     &  -     & -   &  -                           & -        &   -       & -   &  -                              &    -           \\
G78.44+2.66  &       &          &             &                 &          &            &       &                                &           \\
W75N        & 0.35(.04) &2.19(0.10) & 9.11(.14)  & 5.94(.32)      & 0.07(.06) & 0.48(0.13)      & 10.98(.89)   & 5.97(2.08)      &   11.1             \\
DR21S       & 0.09(.04) &0.17(0.06) &  -1.78(.29)   &  1.79(.58)  & 0.09(.04) & 0.19(0.06)      &  -1.99(.37)  &  2.10(.87)      &   -          \\
W75(OH)     & 0.32(.04) &1.86(0.10) & -2.82(.13)  &  5.43(.33)    & 0.10(.04) & 0.57(0.14)      & -1.61(.68)  & 5.39(1.83)       &  16.5           \\
G81.7+0.5   & -    &  -    & -   &  -                             & -        &   -       & -   &  -                              &    -           \\
CepA        & -    &  -    & -   &  -                             & -         &   -       & -   &  -                             &    -             \\
NGC7538     & 0.19(.04) &0.96(0.09) &  -55.88(.24)   & 4.63(.49)  & $<$0.04  &  $<$0.20      & -   &  -                          &    -              \\
NGC7538A    & 0.14(.04) &0.83(0.07) & -57.31(.25)   & 5.22(.52)   & 0.04(.03) & 0.22(0.09)  &  -56.08(.92)  &  4.76(1.91)        &  13.9             \\
      \enddata \\
\end{deluxetable}

\begin{deluxetable}{lrrrrrrrrrrrr}
\tabletypesize{\scriptsize} \rotate
\tablecaption{\label{tab:dlhresult} Results of \cs\
5-4, \hhs\ 2$_{2,0}$-2$_{1,1}$ and SO 5$_6$-4$_5$ Transitions} \tablewidth{0pt}
\tablehead{
\multicolumn{1}{l}{SOURCE} &\multicolumn{4}{c}{\cs\
5-4}&\multicolumn{4}{c}
{\hhs\ 2$_{2,0}$-2$_{1,1}$}&\multicolumn{3}{c}{SO 5$_6$-4$_5$ }\\
\colhead{}             & \colhead{T$_{mb}$}  &
\colhead{$\int$T$_{mb}$d$\nu$}  & \colhead{V$_{LSR}$}    &
\colhead{FWHM}         & \colhead{T$_{mb}$}  &
\colhead{$\int$T$_{mb}$d$\nu$}  & \colhead{V$_{LSR}$}    &
\colhead{FWHM}         &
\colhead{$\int$T$_{mb}$d$\nu$}  & \colhead{V$_{LSR}$}    &
\colhead{FWHM}         \\
\colhead{}             & \colhead{($K$)}          &
\colhead{($K~km~s^{-1}$)}  & \colhead{($km~s^{-1}$)}   &
\colhead{($km~s^{-1}$)}   & \colhead{($K$)}          &
\colhead{($K~km~s^{-1}$)}  & \colhead{($km~s^{-1}$)}   &
\colhead{($km~s^{-1}$)}    &
\colhead{($K~km~s^{-1}$)}  & \colhead{($km~s^{-1}$)}   &
\colhead{($km~s^{-1}$)} } \startdata
W3(OH)      & 0.64(.06 )  & 3.00(.13 )  & -46.92(.09)  & 4.49(2.12)     &0.54(.04) &2.94(0.10) & -47.45(.09) & 5.18(.20)    &  20.08      &     -46.93     &      6.06      \\
RCW142      & 1.16(.04 )  & 6.41(.07 )  &  17.29(.03)   &  5.18(.08)    &1.09(.06) &7.09(0.13) & 17.49(.05) & 6.12(.51)     &  13.14     &      17.08     &      9.31        \\
G5.89-0.39  & 0.97(.06 )  & 5.84(.28 )  &   9.40(.12)  & 5.62(0.34)     &0.94(.03) &6.62(0.09) &  9.43(.04)  & 6.61(.11)    &  47.63      &      10.21     &      19.44     \\
M8E         & 0.70(.03 )  & 1.70(.04 )  &  10.65(.03)  & 2.27(.08)      &0.14(.04) &0.36(0.06) & 10.77(.18) & 2.36(.34)     &  7.89     &      10.67     &      2.92          \\
G8.67-0.36  & 0.54(.03 )  & 2.45(.07 )  &  34.96(.07) & 4.33(0.17)      &0.65(.04) &2.67(0.14) & 35.14(.10)  & 3.87(.24)    &  7.18      &      35.11     &      6.89      \\
W31         & 0.54(.06 )  & 6.88(1.00 ) &  66.35(.76) & 11.964(2.41)    &1.59(.06) &13.67(0.61) & 67.13(.19) & 8.03(.44)    &  10.46     &      67.35     &      10.38      \\
G10.6-0.4   & 2.20(.04 ) & 15.01(.12 )  &  -3.09(.02)  & 6.39(.06)      &3.80(.04) &28.94(0.13)  & -2.77(.02) &  7.16(.04)  &  24.32     &     -2.77     &      7.74       \\
G11.94-0.62 & 0.30(.03 )  & 1.26(.06 )  &  38.03(.08)  & 3.96(.21)      &0.13(.03) &0.68(0.06) & 38.17(.19)  & 4.66(.41)    &  2.83     &      38.23     &      5.52       \\
W33cont     & 1.75(.04 )  & 8.70(.09 )  &  35.02(.02)  & 4.66(.05)      &2.20(.06) &11.51(0.12) & 35.15(.02) & 4.89(.06)    &  17.02     &      35.42     &      6.02      \\
W33irs1     & 0.99(.07 )  & 5.77(.16 )  &  35.78(.07)  & 5.48(0.19)     &0.81(.09) &4.55(0.19) & 35.92(.11) & 5.27(.25)     &  11.02     &      35.82     &      5.43       \\
W33irs2     & 1.70(.04 ) &  7.09(.09 )  &  34.86(.02) & 3.93(.06)       &1.87(.06) &8.16(0.10) & 34.91(.03) & 4.09(.06)     &  14.97     &      35.22     &      5.20      \\
W33irs3     & 1.62(.06 )  & 8.84(.12 )  &  35.46(.03) & 5.11(.08)       &2.06(.07) &12.10(0.14) & 35.56(.03) & 5.54(.08)    &  18.39     &      35.59     &      5.93      \\
G14.33-0.64 & 0.39(.04 )  & 1.83(.22 )  &  22.70(.37) & 4.40(2.77)      &0.33(.04) &1.26(0.09) &  22.42(.13) &  3.61(.30)   &  11.95     &      21.46     &      7.60      \\
W42         & 0.71(.04 )  & 3.91(.51 )  &  110.9(.31) & 5.12(.84)       &1.14(.06) &6.04(0.12) & 110.0(.04) & 4.94(.12)     &  7.48     &      110.53     &      8.52      \\
G29.96-0.02 & 0.97(.03 )  & 4.26(.61 )  &  97.92(.28)  & 4.14(.71)      &1.06(.06) &4.97(0.12) & 97.92(.05) &  4.41(.12)    &  8.64     &      97.74     &      8.90       \\
G30.54+0.02  &    -       &   -         &    -         &  -             &     -    &  -        &  -        &      -        &  0.49     &   48.93     &  4.17   \\
G33.92+0.11 & 0.46(.07 )  & 1.41(.12 )  &  107.7(.11)  & 2.84(.28)      &0.26(.06) &0.75(0.14) & 107.4(.29) & 2.73(.54)     &  3.03     &      107.93     &      3.28      \\
G34.26+0.15 & 1.90(.09 )  &10.41(.19 )  &  58.20(.04)  & 5.14(.12)      &1.90(.09) &12.35(0.22) & 58.64(.05) & 6.10(.13)    &  27.65     &      58.67     &      7.40      \\
S76E        & 0.38(.03 )  & 1.23(.06 )  &  32.66(.07)  & 3.06(.18)      &0.10(.03) &0.22(0.06) & 32.68(.22) & 2.00(.56)     &  10.07     &      33.26     &      6.68     \\
G35.20-0.74 & 0.13(.03 )  & 0.78(.06 )  &  33.45(.22) & 5.36(.48)       &0.10(.04) &0.43(0.07) &  33.74(.36) & 3.93(.71)    &  4.34     &      33.55     &      6.34      \\
G35.20-1.74 & 0.26(.06 )  & 0.54(.09 )  &  43.69(.14) & 2.00(.42)       &0.10(.04) &0.49(0.19) &  44.22(.74) & 4.61(2.63)   &  3.58     &      43.64     &      4.85      \\
G37.543-0.11 &0.11(.05)   &0.45(.11)    & 52.62(.52)  & 3.86(.84)       &    -     &   -       &   -         &   -          &  0.86     &  53.20 &  3.57   \\
G43.89-0.78  &  -         &  -          &     -       &  -              &    -     &    -       &    -       &  -           &  1.93    &  54.15  &  6.15   \\
G45.45+0.06 & 0.22(.04 )  & 1.07(.10 )  &  58.29(.21) & 4.59(.45)       & -        &  -         &  -         &  -           &  2.76     &      58.12     &      5.57     \\
W51M        & 1.99(.07 ) &17.83 (3.16 ) &  56.71(.71) & 8.42(1.87)      &2.84(.07) &23.30(0.20) & 56.70(.03) & 7.70(.08)    &  61.97     &      57.54     &      11.24    \\
W51D        & 0.80(.10 )  & 5.94(.25 )  &  60.74(.15) & 7.01(.34)       &0.81(.12) &6.26(0.29) & 60.96(.16) & 7.20(.38)     &  28.38     &      60.61     &      11.62  \\
G61.48+0.09 & 0.15(.06 )  & 0.28(.07 )  &  22.06(.23) & 1.17(.39)       &-         &-          & -          & -             &  2.11     &      22.10     &      2.72     \\
K3-50A      & 0.09(.04 )  & 0.65(.12 )  & -25.25(.62)  & 7.53(1.58)     &0.09(.04) &0.81(0.13) & -24.46(.77) & 8.77(1.56)   &  6.89     &     -24.33     &      5.71        \\
G78.44+2.66  &-       &   -         &  -          &     -           &   -      & -          &      -     &  -           &   0.82    &   1.37          &   1.63  \\
W75N        & 0.61(.04 )  & 2.72(.09 )  &   9.43(.06)    & 4.17(.15)    &0.55(.07) &2.35(0.12) & 9.23(.10) & 4.05(.23)      &  32.22     &      9.76     &      7.20        \\
DR21S       & 0.65(.07 )  & 2.35(.12 )  &  -2.56(.08)  & 3.36(.21)      &0.16(.06) &1.29(0.16) & -2.42(.44)  &  7.35(1.18)  &  27.36     &     -1.25     &      15.18      \\
W75OH       & 0.77(.04 )  & 4.20(.09 )  &  -2.70(0.06) & 5.12(.13)      &0.54(.06) &3.16(0.13) & -2.93(.11) & 5.54(.26)     &  30.56     &     -2.65     &      6.72       \\
G81.7+0.5   & 0.72(.12 )  & 2.46(.22 )  &  -2.91(.14)  & 3.22(.37)      &0.29(.17) &1.25(0.30) & -3.09(.50) & 4.02(1.10)    &  25.26     &     -1.66     &      14.83      \\
CepA        & 0.20(.04 )  & 0.67(.07 )  & -10.36(.17)  & 3.04(.41)      &0.26(.06) &1.12(0.14) & -10.20(.28) &  4.10(.67)   &  14.89     &     -9.90     &      7.11       \\
NGC7538     & 0.41(.06 )  & 1.83(.12 )  & -55.81(.13)  & 4.25(.28)      &0.17(.06) &0.87(0.14) & -56.12(.39) &  4.75(.99)   &  16.38     &     -55.98     &      6.12      \\
NGC7538A    & 0.55(.04 )  & 1.80(.07 )  & -57.27(.06)  & 3.04(.15)      &0.30(.06) &1.39(0.12) & -57.31(.16) & 4.23(.41)    &  14.67     &     -57.47     &      5.13      \\
      \enddata \\
\end{deluxetable}

\begin{table*}
    \begin{center}
\caption{Observing Results of \coo.}
  \begin{tabular}{lcccccccc}
    \hline
    \hline
Source name     & T$_{mb}$ & $\int$T$_{mb}$d$\nu$  &  V$_{LSR}$  &  FWHM     \\
                & ($K$)       & ($K~km~s^{-1}$)      & ($km~s^{-1}$)  &  ($km~s^{-1}$)    \\
                \hline
W3(OH)     &  4.48(.05)   &  19.70(.09)   &  -46.77(.01)    &  4.13(.02)     \\
RCW142     & 9.63(.06)    &  49.74(.13)  &  17.28(.01)    &  4.85(.02)      \\
G5.89-0.39  & 9.39(.19)   &  43.91(.37)   &  9.23(.02)    &  4.39(.04)     \\
M8E     &  7.30(.04)   & 17.76(.06)  & 10.71(.01)     & 2.29(.01)      \\
G8.67-0.36  & 8.04(.06)   & 45.65(.13)  &  35.34(.01)    &   5.34(.02)     \\
W31         & 5.35(.05)   & 34.73(.10)  &  66.96(.01)    &  6.09(.02)       \\
G10.6-0.4   & 13.80(.08)    & 88.45(.17)  & -3.06(.01)     &  6.01(.01)     \\
G11.94-0.62 & 5.26(.04)    & 25.49(.08)  & 38.27(.01)     & 4.55(.02)      \\
W33cont &  16.0(.05)   & 92.27(.11)   & 35.16(.01)     &  5.40(.01)      \\
W33irs1 & 13.20(.19)     & 83.39(.41)   & 35.51(.01)     &  5.93(.03)     \\
W33irs2  & 18.80(.32)    & 87.85(.62)   &  34.96(.01)    &  4.40(.04)      \\
W33irs3 & 15.60(.12)    & 100.6(.26)   & 35.59(.01)     &  6.06(.02)      \\
G14.33-0.64 & 5.16(.04)    & 19.55(.08)  &  22.29(.01)    &  3.56(.02)      \\
W42         & 6.63(.08)    & 31.01  &  110.1(.01)    &  4.40(.03)      \\
G29.96-0.02 & 7.94(.16)    & 37.21(.32)  & 97.75(.02)     &  4.40(.05)     \\
G30.54+0.02 & 2.02(.09)    & 7.37(.15)  & 47.81(.03)     & 3.43(.08)       \\
G33.92+0.11 & 8.29(.10)    & 28.18(.17)  &  107.6(.01)    &  3.19(.02)       \\
G34.26+0.15 & 14.50(.09)   & 78.75(.20)  &  57.58(.01)    &  5.11(.02)      \\
S76E     & 3.40(.06)    & 11.82(.10)  & 32.92(.01)     &  3.27(.03)     \\
G35.20-0.74  &  2.97(.10)   & 10.64(.01)  & 33.81(.03)     & 3.37(.05)      \\
G35.20-1.74  & 3.34(.05)   & 18.71(.11)  & 43.10(.02)     &  5.27(.04)     \\
G37.543-0.11 &  2.01(.08)    & 7.86(.15)  & 52.74(.03)     & 3.68(.08)       \\
G43.89-0.78 &  4.41(.20)    & 17.40(.35)  & 54.02(.04)     & 3.70(.09)       \\
G45.45+0.06  & 3.40(.06)   & 15.50(.12)  &  58.63(.02)    &  4.28(.04)     \\
W51M         & 9.20(.14)   & 90.07(.39)  &  57.53(.02)    &  9.20(.05)     \\
W51D         & 8.73(1.07)   & 63.87(2.73)  & 61.47(.13)     & 6.87(.38)       \\
G61.48+0.09  & 4.07(.07)   & 12.48(.12)  & 21.80(.01)     &  2.88(.02)       \\
K3-50A  &  2.96(.04)    & 21.29(.03)  & -23.54(.01)     &  6.77(.02)       \\
G78.44+2.66 &  3.53(.13) & 7.91(.17)  & 0.96(.02)     & 2.11(.05)       \\
W75N    &  9.34(.11)    & 38.59(.21)  & 9.67(.01)     &  3.88(.03)       \\
DR21S    & 9.02(0.16)   & 36.67(0.30) &  -2.49(.02)    &  3.82(.04)      \\
W75(OH)  & 10.7(.05)    & 48.31(.10)  & -2.83(.01)     &  4.25(.01)     \\
G81.7+0.5  & 9.95(.21)     & 36.45(.36) & -2.51(.02)     & 3.44(.04)       \\
CepA  & 8.24(.09)    & 32.18(.15)  &  -10.50(.01)    &  3.67(.02)    \\
NGC7538  & 3.64(.10)    &  21.53(.21)  & -56.42(.03)   &  5.56(.06)      \\
NGC7538A & 6.06(.12)    & 26.96(.24)  &  -56.96(.02)    & 4.18(.05)         \\
\hline
  \end{tabular}
  \end{center}
  \label{tab:c18o}
\end{table*}

\begin{deluxetable}{lrrrrrrrrrrrr}
\tabletypesize{\scriptsize} \rotate
\tablecaption{\label{tab:dlhresult} Column Densities and Abundances.} \tablewidth{0pt}
\tablehead{
\multicolumn{1}{l}{SOURCE} &\multicolumn{5}{c}{Column Density ($cm^{-2}$)}&\multicolumn{4}{c}
{Abundance}\\
\colhead{}     &\colhead{\hh}     & \colhead{CS}  &
\colhead{OCS}   &
\colhead{\hhs}         & \colhead{SO}  &
\colhead{CS}  & \colhead{OCS}    &
\colhead{\hhs}  & \colhead{SO}       } \startdata
W3(OH)      &    1.12$\times 10^{23}$ &   4.61$\times 10^{14}$ & 1.40$\times 10^{15}$      & 5.09$\times 10^{14}$ & 3.05$\times 10^{14}$   & 4.10$\times 10^{ -9}$   &  1.21$\times 10^{-8}$  &  4.53$\times 10^{ -9 }$  &     2.71$\times 10^{-9 }$   \\
RCW142      &    2.84$\times 10^{23}$ &   9.84$\times 10^{14}$ & 2.16$\times 10^{15}$     & 1.23$\times 10^{15}$ & 2.00$\times 10^{14}$    & 3.47$\times 10^{ -9}$   &  7.61$\times 10^{-9}$  &  4.33$\times 10^{ -9 }$  &     7.04$\times 10^{-10}$   \\
G5.89-0.39  &    2.50$\times 10^{23}$ &    8.97$\times 10^{14}$ & 1.53$\times 10^{15}$    & 1.15$\times 10^{15}$ & 7.24$\times 10^{14}$    & 3.58$\times 10^{ -9}$   &  6.10$\times 10^{-9}$  &  4.58$\times 10^{ -9 }$  &     2.89$\times 10^{-9 }$  \\
M8E         &    1.01$\times 10^{23}$ &   2.61$\times 10^{14}$ & $^a$8.20$\times 10^{13}$  & 6.24$\times 10^{13}$ & 1.20$\times 10^{14}$   & 2.57$\times 10^{ -9}$   &  8.20$\times 10^{-10}$  &  6.15$\times 10^{ -0 }$  &     1.18$\times 10^{-9 }$  \\
G8.67-0.36  &    2.60$\times 10^{23}$ &   3.76$\times 10^{14}$ & $^a$4.57$\times 10^{14}$    & 4.62$\times 10^{14}$ & 1.09$\times 10^{14}$ & 1.44$\times 10^{ -9}$   &  1.76$\times 10^{-9}$  &  1.77$\times 10^{ -9 }$  &     4.19$\times 10^{-10}$    \\
W31         &    1.98$\times 10^{23}$ &    1.06$\times 10^{15}$ & 8.86$\times 10^{15}$    & 2.37$\times 10^{15}$ & 1.59$\times 10^{14}$    & 5.33$\times 10^{ -9}$   &  4.47$\times 10^{-8}$  &  1.19$\times 10^{ -8 }$  &     8.02$\times 10^{-10}$     \\
G10.6-0.4   &    5.04$\times 10^{23}$ &    2.30$\times 10^{15}$ & 2.06$\times 10^{15}$    & 5.01$\times 10^{15}$ & 3.70$\times 10^{14}$    & 4.56$\times 10^{ -9}$   &  4.07$\times 10^{-9}$  &  9.94$\times 10^{ -9 }$  &     7.33$\times 10^{-10}$     \\
G11.94-0.62 &    1.45$\times 10^{23}$ &   1.93$\times 10^{14}$ & $^a$5.85$\times 10^{13}$  & 1.18$\times 10^{14}$ & 4.30$\times 10^{13}$   & 1.33$\times 10^{ -9}$   &  4.03$\times 10^{-10}$  &  8.10$\times 10^{ -10}$   &    2.95$\times 10^{-10}$       \\
W33cont     &    5.26$\times 10^{23}$ &   1.34$\times 10^{15}$ & $^a$7.20$\times 10^{14}$  & 1.99$\times 10^{15}$ & 2.59$\times 10^{14}$   & 2.53$\times 10^{ -9}$   &  1.37$\times 10^{-9}$  &  3.79$\times 10^{ -9 }$  &     4.91$\times 10^{-10}$      \\
W33irs1     &    4.75$\times 10^{23}$ &   8.86$\times 10^{14}$ & $^a$5.35$\times 10^{14}$  & 7.88$\times 10^{14}$ & 1.67$\times 10^{14}$   & 1.86$\times 10^{ -9}$   &  1.13$\times 10^{-9}$  &  1.65$\times 10^{ -9 }$  &     3.52$\times 10^{-10}$      \\
W33irs2     &    5.01$\times 10^{23}$ &   1.09$\times 10^{15}$ & $^a$6.05$\times 10^{14}$  & 1.41$\times 10^{15}$ & 2.27$\times 10^{14}$   & 2.17$\times 10^{ -9}$   &  1.20$\times 10^{-9}$  &  2.82$\times 10^{ -9 }$  &     4.54$\times 10^{-10}$     \\
W33irs3     &    5.74$\times 10^{23}$ &   1.36$\times 10^{15}$ & $^a$6.40$\times 10^{14}$  & 2.10$\times 10^{15}$ & 2.80$\times 10^{14}$   & 2.36$\times 10^{ -9}$   &  1.12$\times 10^{-9}$  &  3.65$\times 10^{ -9 }$  &     4.87$\times 10^{-10}$     \\
G14.33-0.64 &    1.11$\times 10^{23}$ &   2.81$\times 10^{14}$ & 6.3$\times 10^{14}$      & 2.18$\times 10^{14}$ & 1.82$\times 10^{14}$    & 2.52$\times 10^{ -9}$   &  5.67$\times 10^{-9}$  &  1.95$\times 10^{ -9 }$  &     1.63$\times 10^{-9 }$       \\
W42         &    1.77$\times 10^{23}$ &    6.00$\times 10^{14}$ & 1.89$\times 10^{15}$    & 1.05$\times 10^{15}$ & 1.14$\times 10^{14}$    & 3.39$\times 10^{ -9}$   &  1.07$\times 10^{-8}$  &  5.91$\times 10^{ -9 }$  &     6.43$\times 10^{-10}$         \\
G29.96-0.02 &    2.12$\times 10^{23}$ &   6.54$\times 10^{14}$ & 1.21$\times 10^{15}$     & 8.61$\times 10^{14}$ & 1.31$\times 10^{14}$    & 3.08$\times 10^{ -9}$   &  5.71$\times 10^{-9}$  &  4.05$\times 10^{ -9 }$  &     6.19$\times 10^{-10}$        \\
G30.54+0.02 &    4.20$\times 10^{22}$ &                         &                          &                      & 7.44$\times 10^{12}$   &   -                     &  -                     &   -                     &     1.77$\times 10^{-10}$      \\
G33.92+0.11 &    1.61$\times 10^{23}$ &   2.17$\times 10^{14}$ & $^a$2.82$\times 10^{14}$  & 1.30$\times 10^{14}$ & 4.61$\times 10^{13}$   & 1.34$\times 10^{ -9}$   &  1.80$\times 10^{-9}$  &  8.08$\times 10^{ -10}$   &    2.86$\times 10^{-10}$        \\
G34.26+0.15 &    4.49$\times 10^{23}$ &   1.60$\times 10^{15}$ & 4.74$\times 10^{15}$     & 2.14$\times 10^{15}$ & 4.20$\times 10^{14}$    & 3.55$\times 10^{ -9}$   &  1.05$\times 10^{-8}$  &  4.76$\times 10^{ -9 }$  &     9.36$\times 10^{-10}$       \\
S76E        &    6.74$\times 10^{22}$ &    1.89$\times 10^{14}$ & $^a$3.34$\times 10^{14}$ & 3.81$\times 10^{13}$ & 1.53$\times 10^{14}$   & 2.80$\times 10^{ -9}$   &  5.00$\times 10^{-9}$  &  5.65$\times 10^{ -10}$   &    2.27$\times 10^{-9 }$      \\
G35.20-0.74 &    6.07$\times 10^{22}$ &   1.20$\times 10^{14}$ & $^a$2.82$\times 10^{14}$  & 7.45$\times 10^{13}$ & 6.60$\times 10^{13}$   & 1.97$\times 10^{ -9}$   &  4.65$\times 10^{-9}$  &  1.22$\times 10^{ -9 }$  &     1.08$\times 10^{-9 }$   \\
G35.20-1.74 &    1.07$\times 10^{23}$ &   8.29$\times 10^{13}$ & 2.43$\times 10^{15}$     & 8.49$\times 10^{13}$ & 5.44$\times 10^{13}$    & 7.77$\times 10^{ -10}$  &  2.27$\times 10^{-8}$  &  7.95$\times 10^{ -10}$   &    5.09$\times 10^{-10}$     \\
G37.543-0.11&    4.48$\times 10^{22}$ &    6.91$\times 10^{13}$  &                         &                      & 1.30$\times 10^{13}$   & 1.54$\times 10^{ -9}$   &   -                    &   -                    &     2.91$\times 10^{-10}$      \\
G43.89-0.78&     9.91$\times 10^{22}$ &                          &                         &                      & 2.93$\times 10^{13}$   &   -                     &   -                    &   -                    &     2.95$\times 10^{-10}$      \\
G45.45+0.06 &    8.84$\times 10^{22}$ &   1.64$\times 10^{14}$ &  -                       & -                    & 4.19$\times 10^{13}$    & 1.85$\times 10^{ -9}$   &   -                    &  4.70$\times 10^{ -10}$   &    4.74$\times 10^{-10}$        \\
W51M        &    5.13$\times 10^{23}$ &   2.73$\times 10^{15}$ & 1.17$\times 10^{16}$     & 4.04$\times 10^{15}$ & 9.42$\times 10^{14}$    & 5.32$\times 10^{ -9}$   &  2.27$\times 10^{-8}$  &  7.85$\times 10^{ -9 }$  &     1.83$\times 10^{-9 }$     \\
W51D        &    3.64$\times 10^{23}$ &   9.12$\times 10^{14}$ & $^a$1.70$\times 10^{15}$  & 1.08$\times 10^{15}$ & 4.31$\times 10^{14}$   & 2.50$\times 10^{ -9}$   &  4.67$\times 10^{-9}$  &  2.97$\times 10^{ -9 }$  &     1.18$\times 10^{-9 }$     \\
G61.48+0.09 &    7.11$\times 10^{22}$ &  4.28$\times 10^{13}$ &  -                         & -                   & 3.20$\times 10^{13}$  & 6.03$\times 10^{ -10}$   &   -                    &  4.18$\times 10^{ -8 }$  &     4.50$\times 10^{-10}$        \\
K3-50A      &    1.20$\times 10^{23}$ &    9.98$\times 10^{13}$ &  -                      & 1.40$\times 10^{14}$ & 1.05$\times 10^{14}$    & 8.22$\times 10^{ -10}$  &   -                    &  1.15$\times 10^{ -9 }$  &     8.62$\times 10^{-10}$        \\
G78.44+2.66&     4.50$\times 10^{22}$ &                         &                         &                      & 1.25$\times 10^{13}$    &   -                     &   -                    &   -                    &     2.76$\times 10^{-10}$      \\
W75N        &    2.20$\times 10^{23}$ &    4.18$\times 10^{14}$ & 2.53$\times 10^{15}$    & 4.07$\times 10^{14}$ & 4.90$\times 10^{14}$    & 1.89$\times 10^{ -9}$   &  1.14$\times 10^{-8}$  &  1.85$\times 10^{ -9 }$  &     2.22$\times 10^{-9 }$       \\
DR21S       &    2.09$\times 10^{23}$ &    3.61$\times 10^{14}$ & 1.0$\times 10^{15}$     & 2.23$\times 10^{14}$ & 4.16$\times 10^{14}$    & 1.72$\times 10^{ -9}$   &  4.78$\times 10^{-9}$  &  1.06$\times 10^{ -9 }$  &     1.98$\times 10^{-9 }$      \\
W75(OH)     &    2.75$\times 10^{23}$ &    6.45$\times 10^{14}$ & 3.00$\times 10^{15}$    & 5.47$\times 10^{14}$ & 4.65$\times 10^{14}$    & 2.34$\times 10^{ -9}$   &  1.09$\times 10^{-8}$  &  1.98$\times 10^{ -9 }$  &     1.68$\times 10^{-9 }$        \\
G81.7+0.5   &    2.08$\times 10^{23}$ &   3.78$\times 10^{14}$ &  -                       & 2.17$\times 10^{14}$ & 3.84$\times 10^{14}$    & 1.81$\times 10^{ -9}$   &   -                    &  1.04$\times 10^{ -9 }$  &     1.84$\times 10^{-9 }$       \\
CepA        &    1.83$\times 10^{23}$ &    1.03$\times 10^{14}$ &  -                      & 1.94$\times 10^{14}$ & 2.26$\times 10^{14}$    & 5.60$\times 10^{ -10}$  &   -                    &  1.05$\times 10^{ -9 }$  &     1.23$\times 10^{-9 }$     \\
NGC7538     &    1.23$\times 10^{23}$ &    2.81$\times 10^{14}$ & $^a$5.65$\times 10^{14}$   & 1.51$\times 10^{14}$ & 2.49$\times 10^{14}$ & 2.28$\times 10^{ -9}$   &  4.60$\times 10^{-9}$  &  1.22$\times 10^{ -9 }$  &     2.02$\times 10^{-9 }$    \\
NGC7538A    &    1.54$\times 10^{23}$ &    2.76$\times 10^{14}$ & 1.16$\times 10^{15}$      & 2.41$\times 10^{14}$ & 2.23$\times 10^{14}$   &1.79$\times 10^{ -9}$    & 7.54$\times 10^{-9}$   & 1.56$\times 10^{ -9 }$   &    1.45$\times 10^{-9 }$       \\
median$_{H II-HC}$& 2.2$\times 10^{23}$ & 6.4$\times 10^{14}$  & 1.2$\times 10^{15}$ & 7.9$\times 10^{14}$&  2.3$\times 10^{14}$  &   2.5$\times 10^{ -9}$ & 5.0$\times 10^{ -9}$  & 2.8$\times 10^{ -9}$  & 1.1$\times 10^{ -9}$    \\
median$_{H II-only}$&9.9$\times 10^{22}$& 8.3$\times 10^{13}$  & $<8.2\times 10^{13}$   & 4.2$\times 10^{13}$   &  4.2$\times 10^{13}$  & 8.2$\times 10^{ -10}$ & $<8.1 \times 10^{-10}$  & 4.7$\times 10^{-10}$ & 4.5$\times 10^{-10}$        \\
      \enddata \\
note: The column density of CS is derived using \cs. The column density of OCS is derived using \occs\ except for those labeled with $^a$, which is derived using OCS by assuming optical depth of 5.
\end{deluxetable}

\begin{table*}
    \begin{center}
\caption{Abundance Ratio.}
  \begin{tabular}{lcccccccc}
    \hline
    \hline
 Source       & OCS/\hhs   & CS/\hhs       &   SO/OCS       &  CS/SO     & OCS/CS  \\
 \hline
W3(OH)      &     2.69   &    0.90         &    0.22        &      1.51  &  2.97    \\
RCW142      &     1.76   &    0.80         &       0.09     &      4.93  &  2.19  \\
G5.89-0.39  &     1.33   &    0.78         &       0.47     &      1.24  &  1.70   \\
M8E         &     1.32   &    4.19         &     1.46       &      2.18  &  0.31  \\
G8.67-0.36  &     0.99   &    0.81         &    0.24        &      3.44  &  1.21     \\
W31         &     3.74   &    0.45         &        0.02    &      6.65  &  8.38       \\
G10.6-0.4   &     0.41   &    0.46         &        0.18    &      6.23  &  0.89       \\
G11.94-0.62 &     0.50   &    1.64         &   0.73         &      4.50  &  0.30         \\
W33cont     &     0.36   &    0.67         &    0.36        &      5.16  &  0.54        \\
W33irs1     &    0.68    &    1.12         &   0.31         &      5.29  &  0.60      \\
W33irs2     &    0.43    &    0.77         &    0.38        &      4.79  &  0.56       \\
W33irs3     &    0.31    &    0.65         &     0.44       &      4.86  &  0.47       \\
G14.33-0.64 &   2.90     &    1.29         &       0.29     &      1.55  &  2.25        \\
W42         &   1.81     &    0.57         &       0.06     &      5.28  &  3.16         \\
G29.96-0.02 &   1.41     &    0.76         &        0.11    &      4.98  &  1.85          \\
G30.54+0.02 &     -      &     -           &        -       &       -    &   -         \\
G33.92+0.11 &   2.17     &    1.67         &    0.16        &      4.70  &  1.30     \\
G34.26+0.15 &   2.22     &    0.75         &        0.09    &      3.80  &  2.96         \\
S76E        &   8.76     &    4.96         &     0.45       &      1.23  &  1.77        \\
G35.20-0.74 &    3.78    &    1.61         &    0.24        &      1.82  &   2.35       \\
G35.20-1.74 &  28.55     &    0.98         &  0.02          &      1.52  &   29.23     \\
G37.543-0.11&     -      &     -           &        -       &       -    &   -         \\
G43.89-0.78 &     -      &     -           &        -       &       -    &   -         \\
G45.45+0.06 &     -      &    -            &            -   &      3.92  &   -         \\
W51M        &    2.90    &    0.68         &        0.08    &      2.90  &  4.28        \\
W51D        &    1.57    &    0.84         &    0.25        &      2.11  &  1.86       \\
G61.48+0.09 &    -       &    0.17         &          -     &      16.05 &  -        \\
K3-50A      &    -       &    0.71         &          -     &      0.95  &  -        \\
G78.44+2.66 &     -      &     -           &        -       &       -    &  -          \\
W75N        &   6.21     &    1.03         &       0.19     &      0.85  &  6.05        \\
DR21S       &   4.48     &    1.61         &      0.42      &      0.87  &  2.77       \\
W75(OH)     &   5.49     &    1.18         &       0.15     &      1.39  &  4.66         \\
G81.7+0.5   &   -        &    1.74         &         -      &      0.98  &   -       \\
CepA        &   -        &    0.53         &         -      &      0.45  &   -       \\
NGC7538     &   3.75     &    1.86         &    0.44        &      1.13  &   2.01      \\
NGC7538A    &   4.81     &    1.15         &       0.19     &      1.24  &   4.19       \\
median$_{H II-HC}$  & 1.8 &     0.8        &     0.2        &    2.9     &   2.0    \\
median$_{H II-only}$ &  - &     -         &        -        &   1.3      &   -     \\
\hline
  \end{tabular}
  \end{center}
  \label{tab:c18o}
\end{table*}

\begin{figure}[htbp]
   \centering
\includegraphics[scale=0.3]{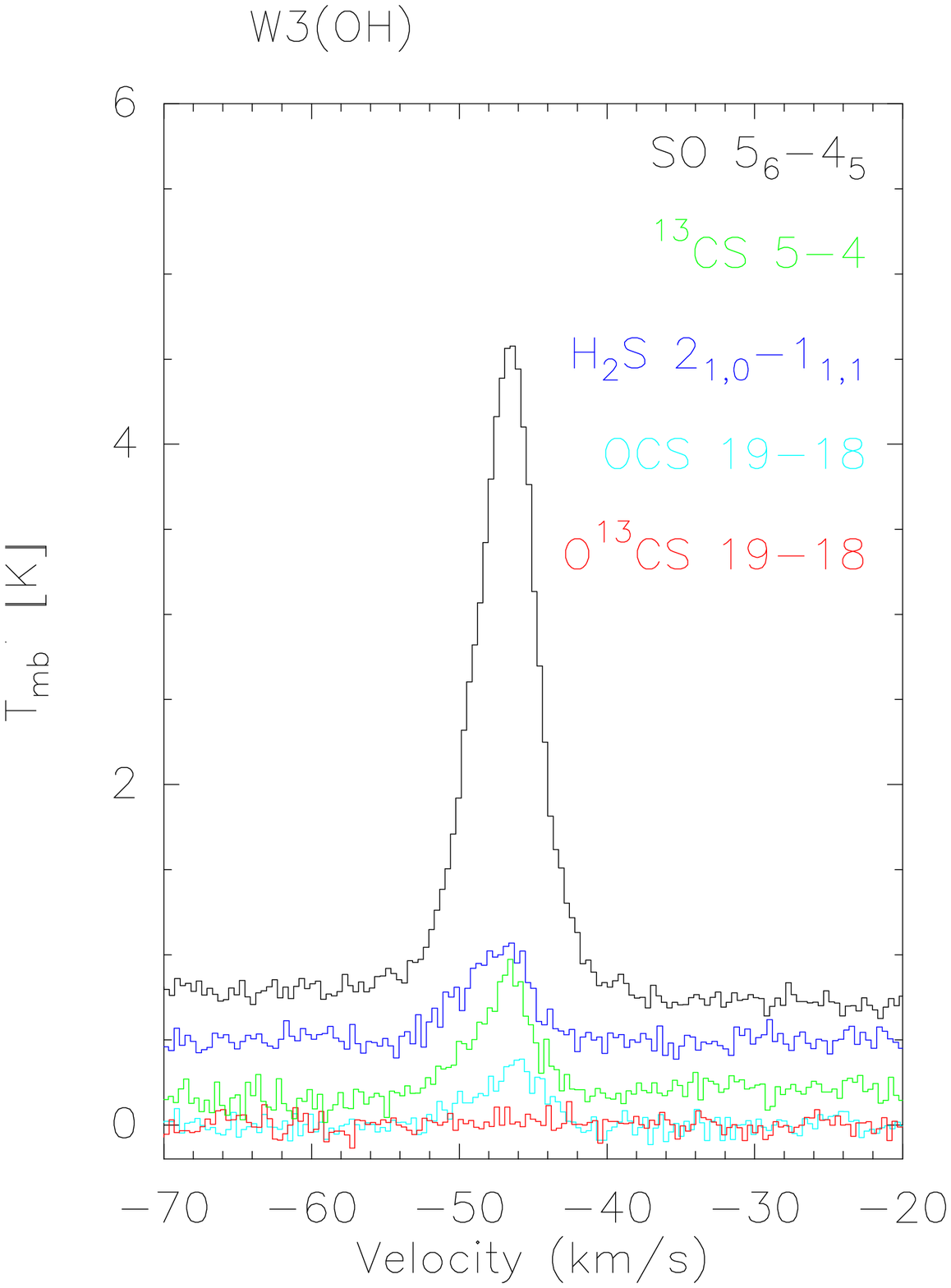}
\includegraphics[scale=0.3]{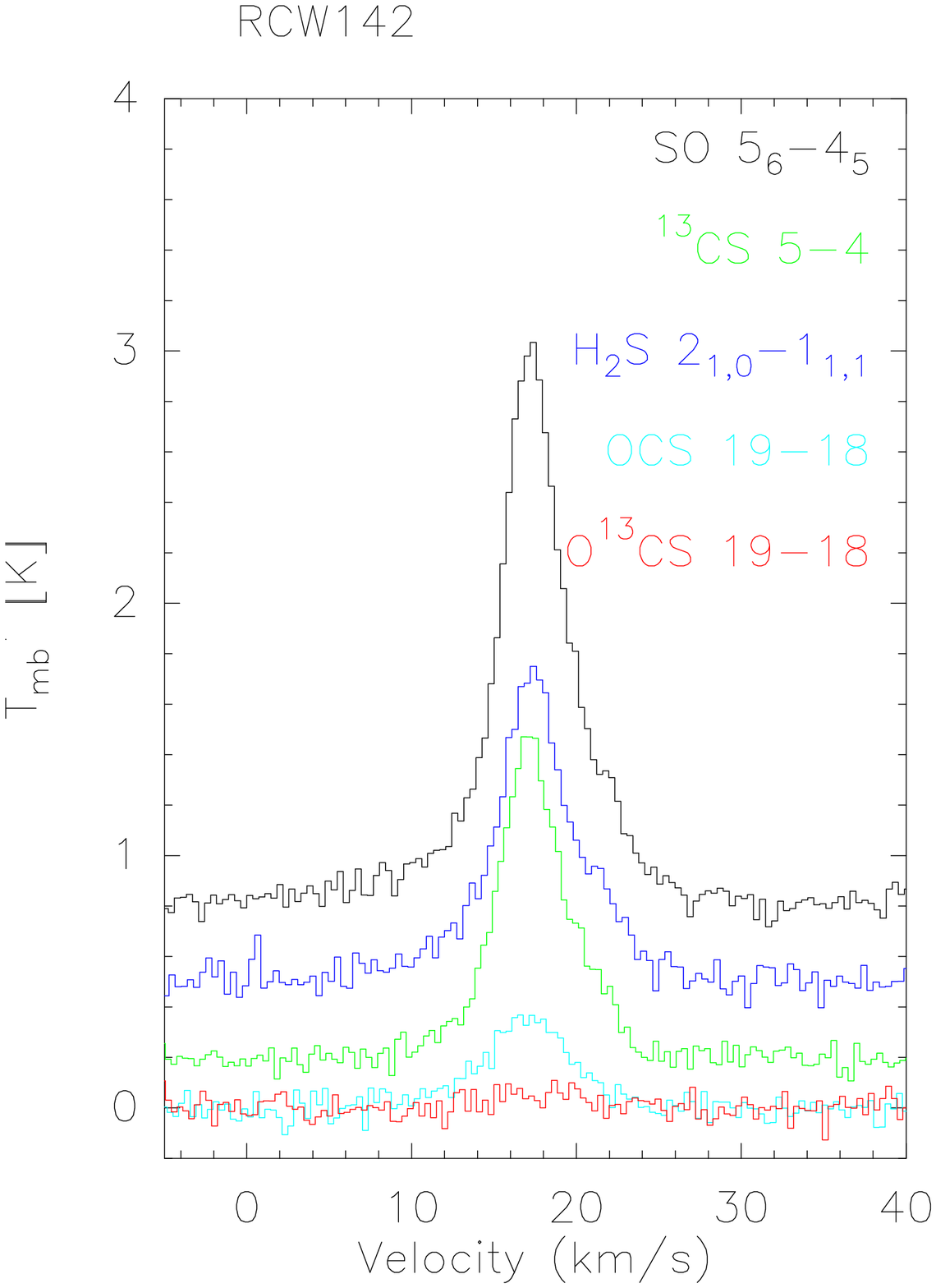}
\includegraphics[scale=0.3]{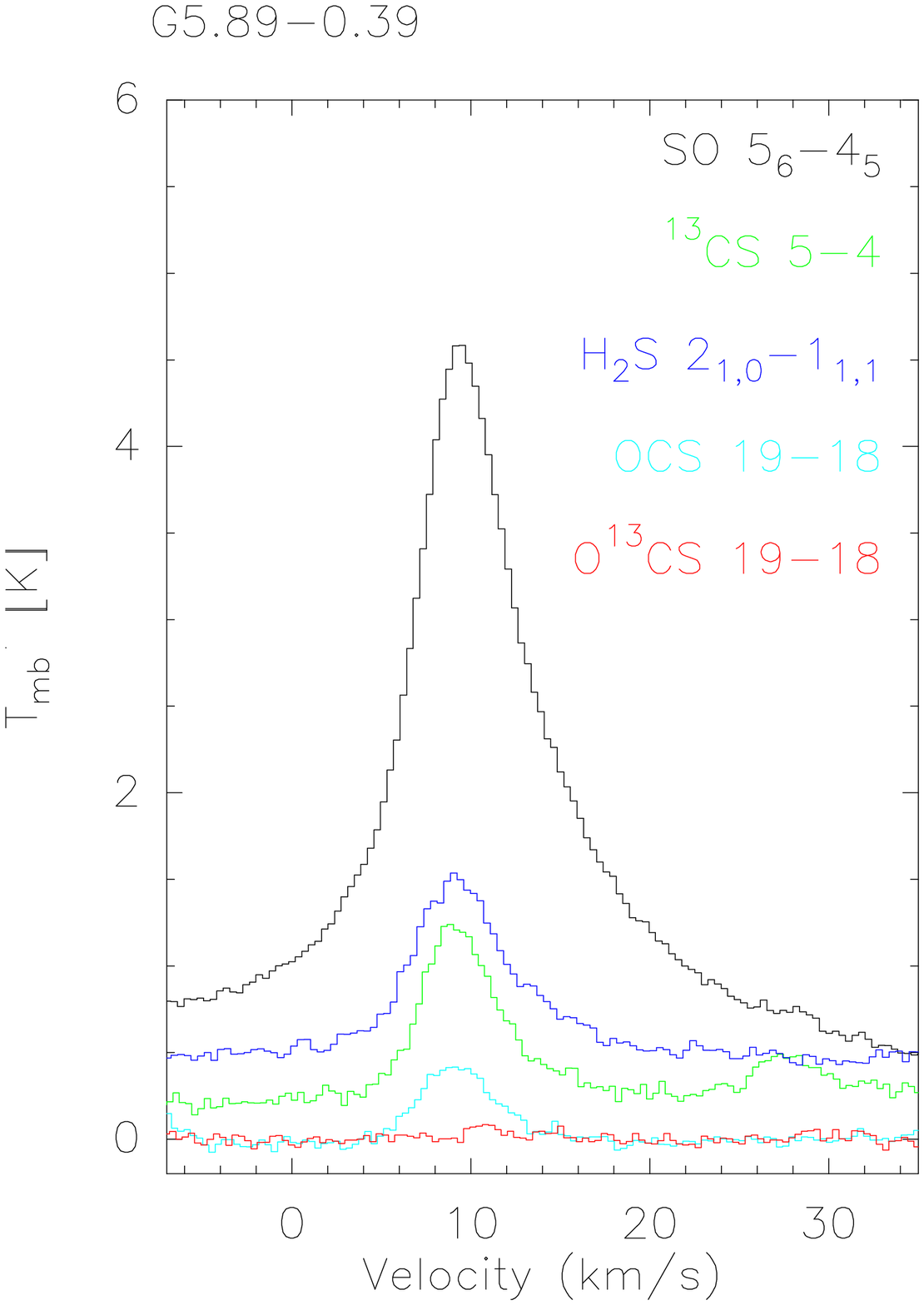}
\includegraphics[scale=0.3]{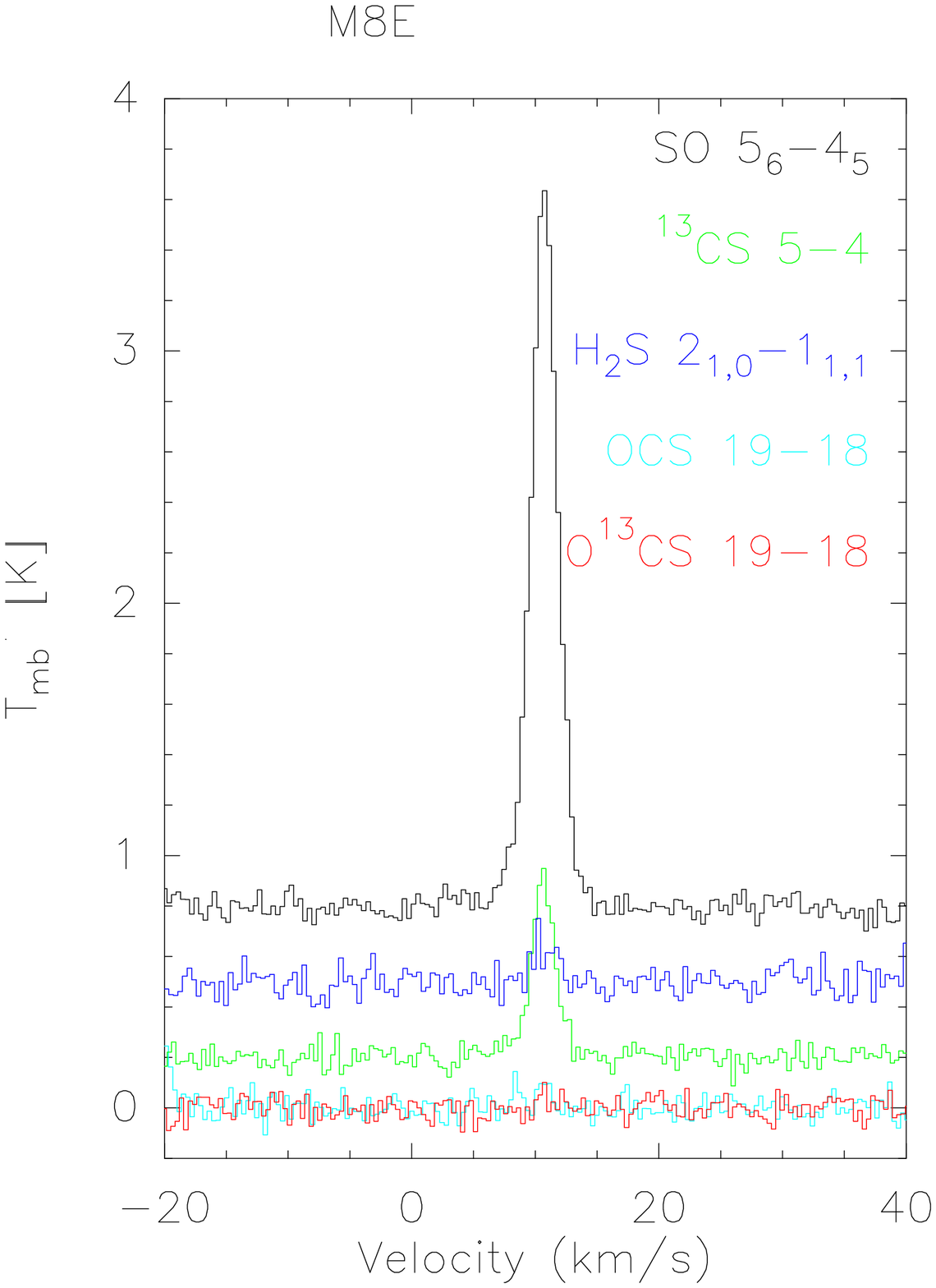}
\includegraphics[scale=0.3]{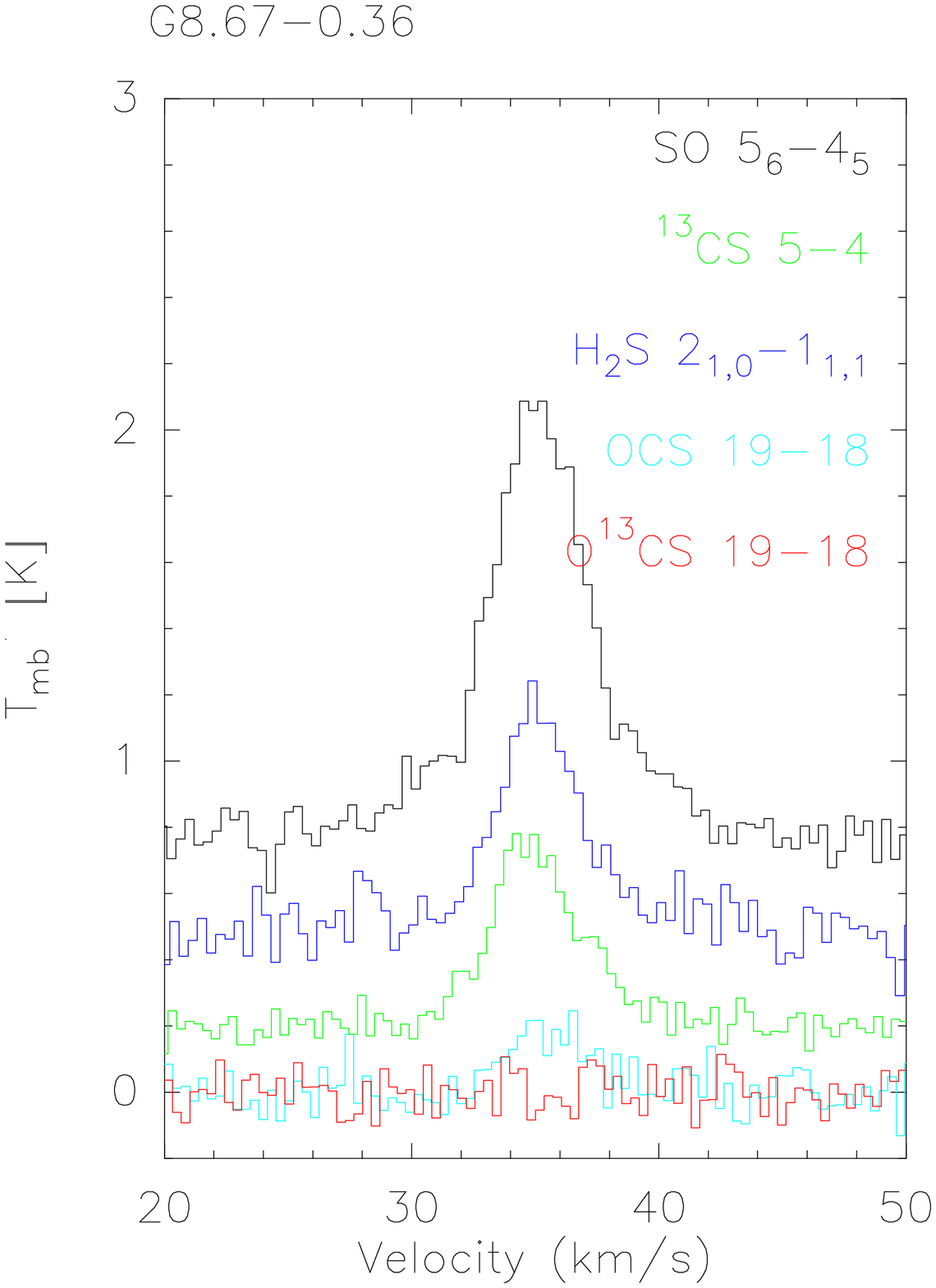}
\includegraphics[scale=0.3]{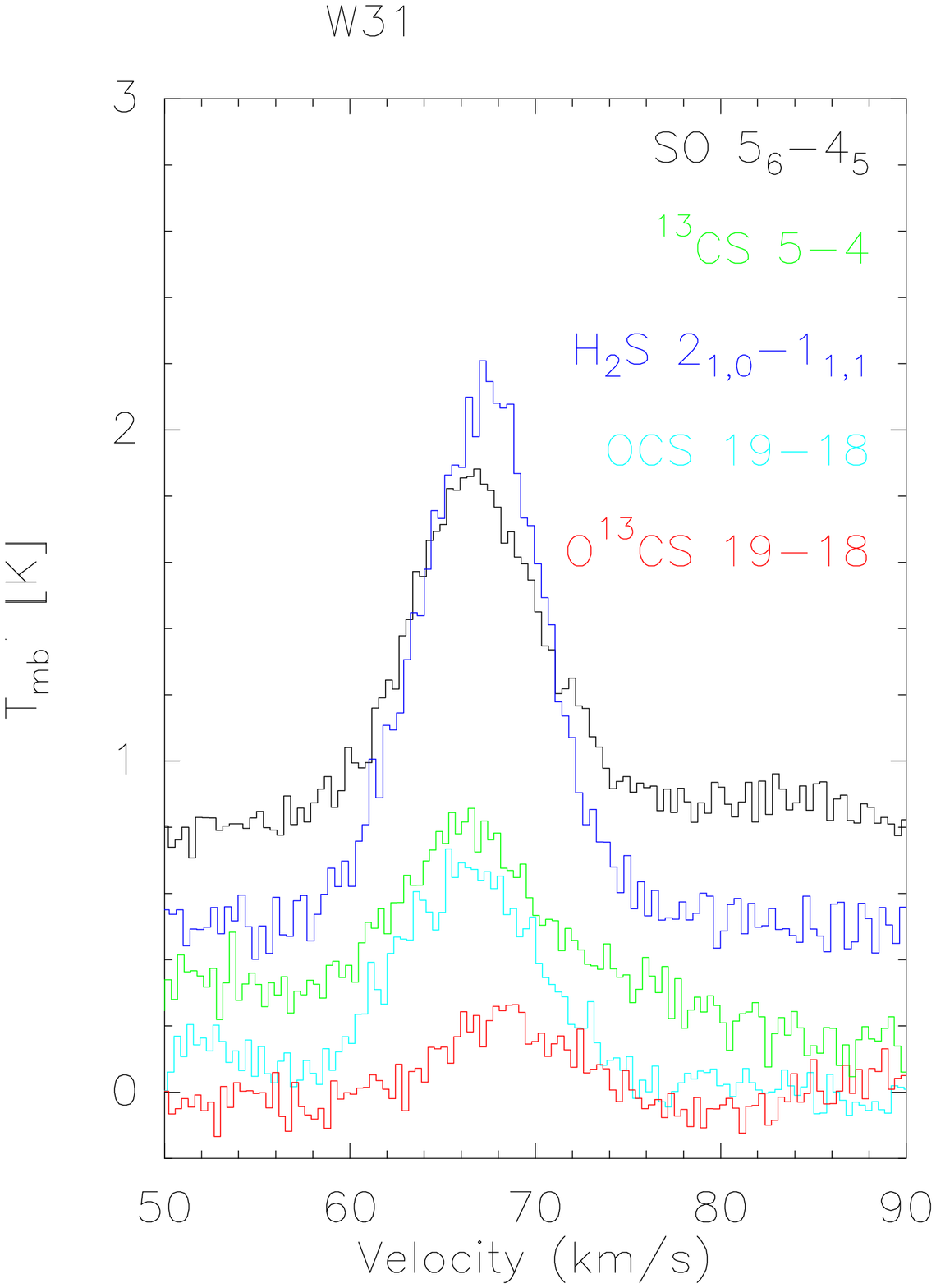}
\includegraphics[scale=0.3]{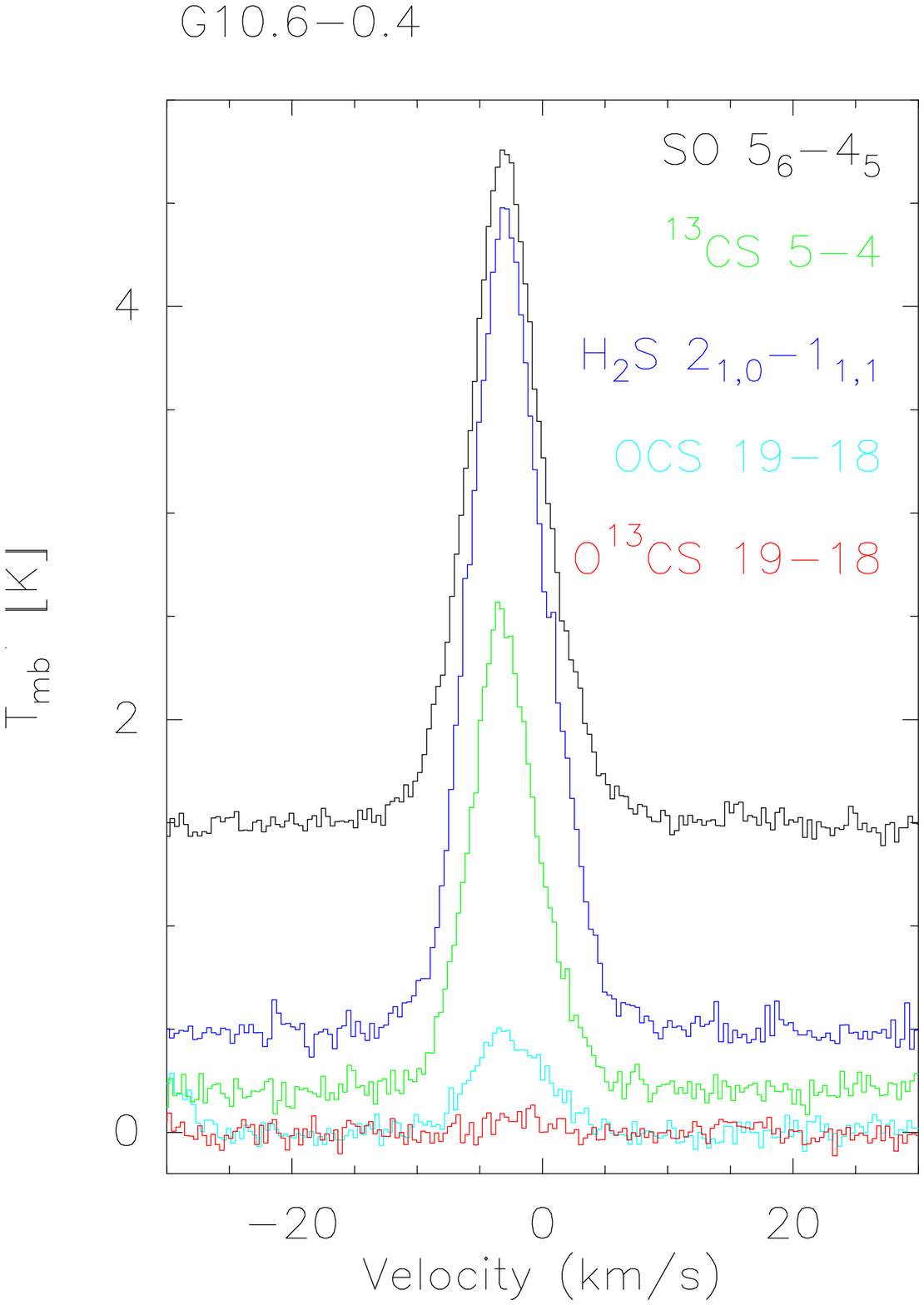}
\includegraphics[scale=0.3]{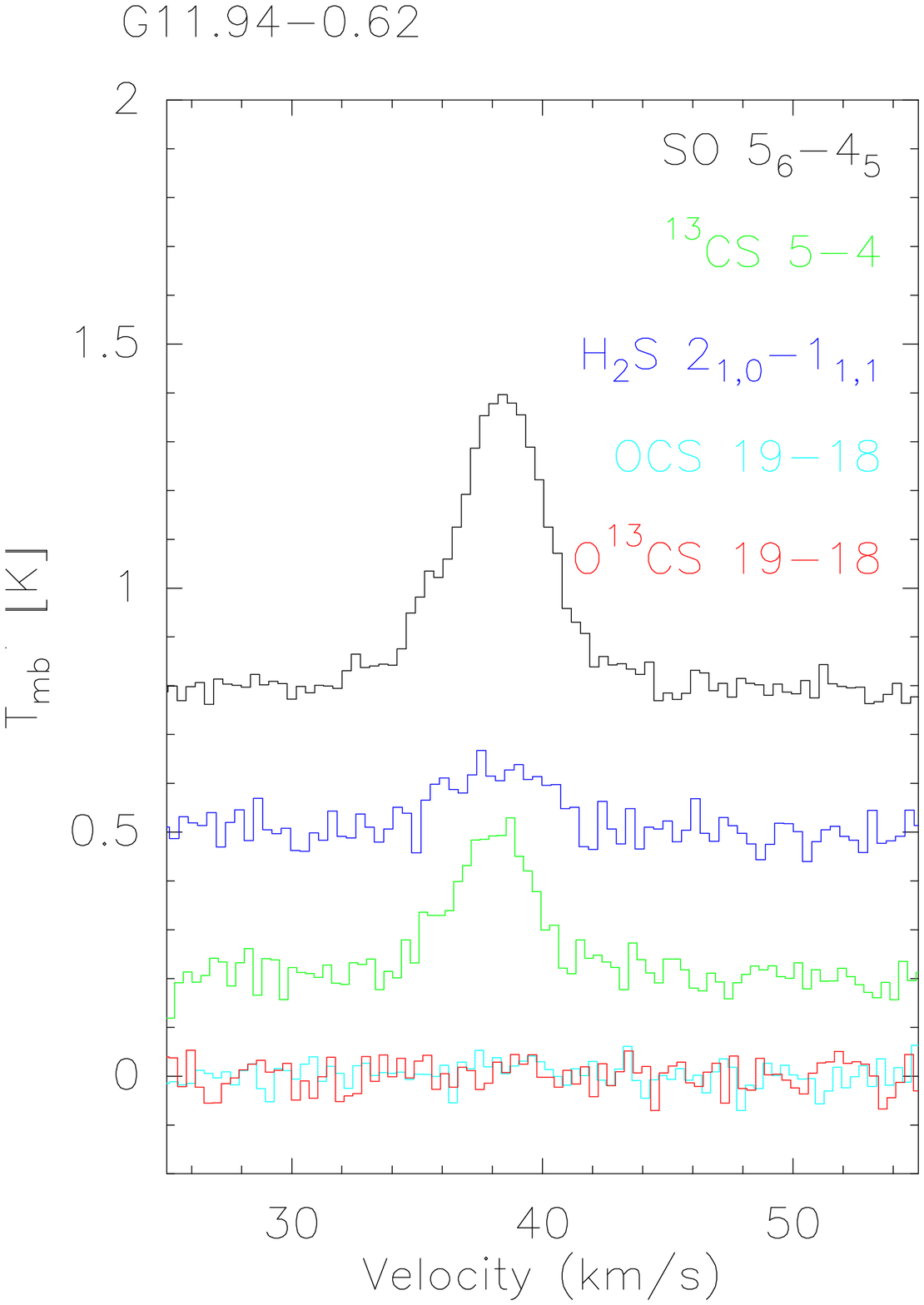}
\includegraphics[scale=0.3]{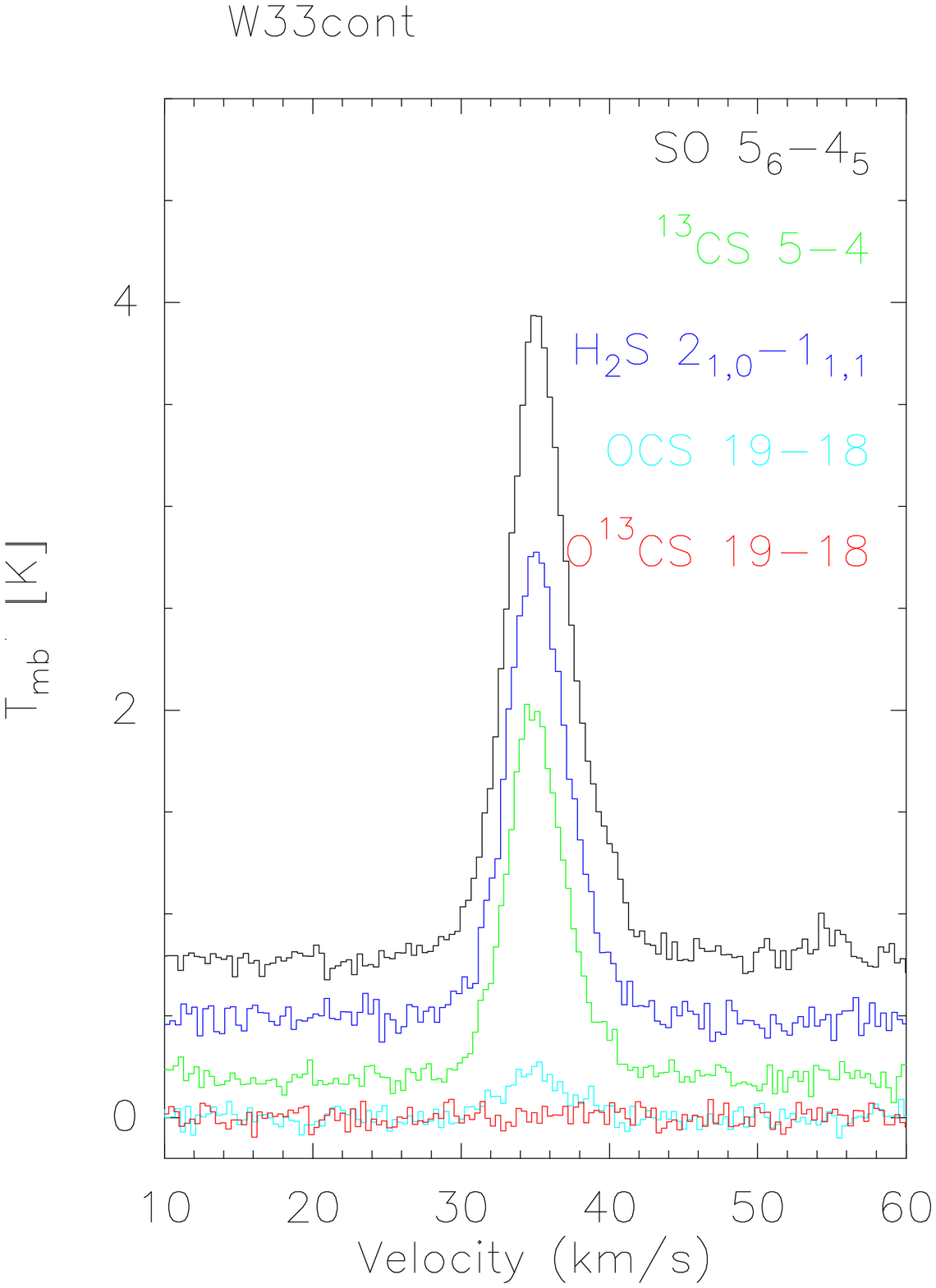}
\caption{Spectra of \cs, OCS, \occs, \hhs\ and SO for all the observed sources. The line identifications are labeled in the upper right of each figure.}
\end{figure}

\begin{figure}[htbp]
   \centering
\includegraphics[scale=0.3]{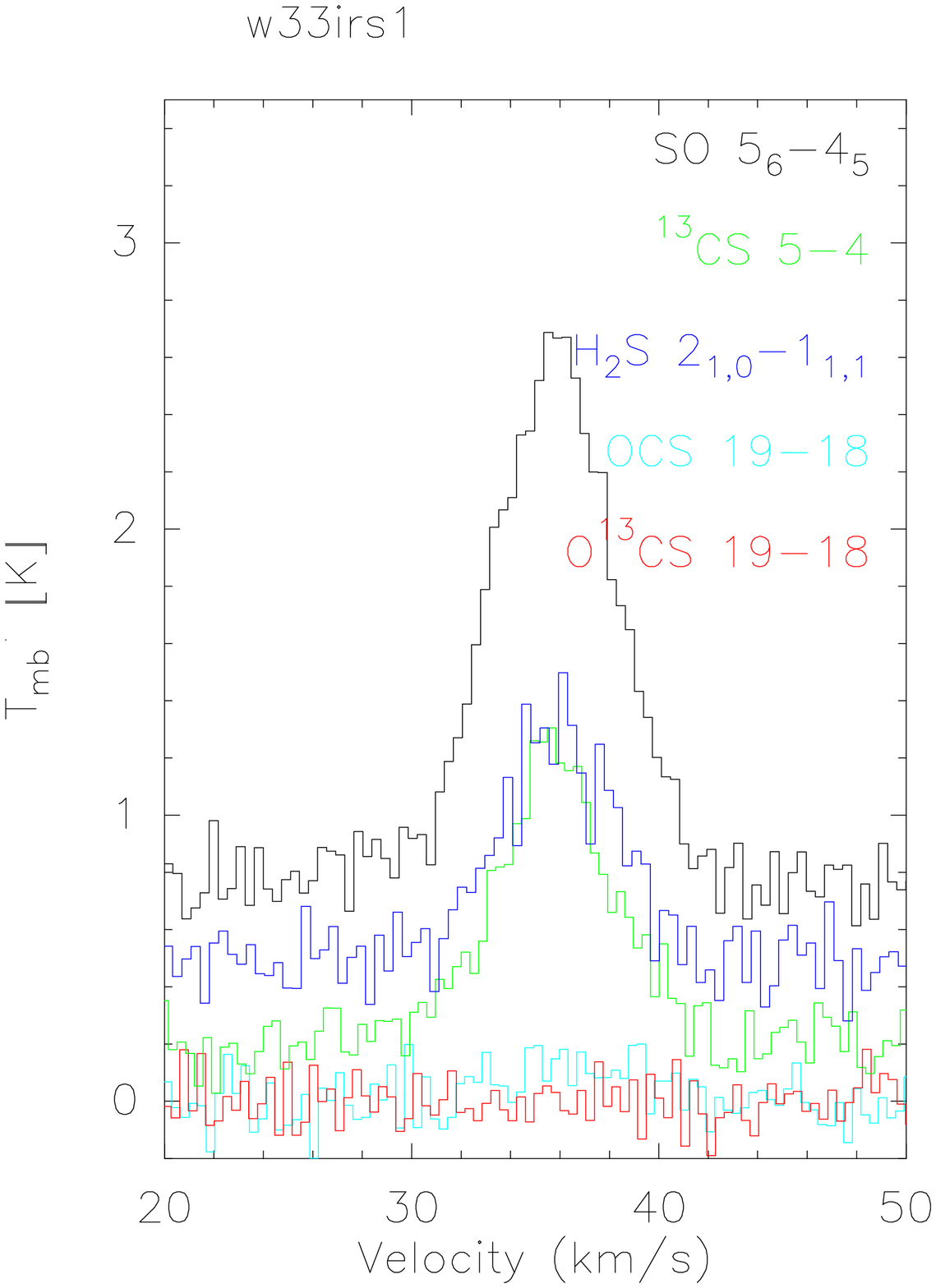}
\includegraphics[scale=0.3]{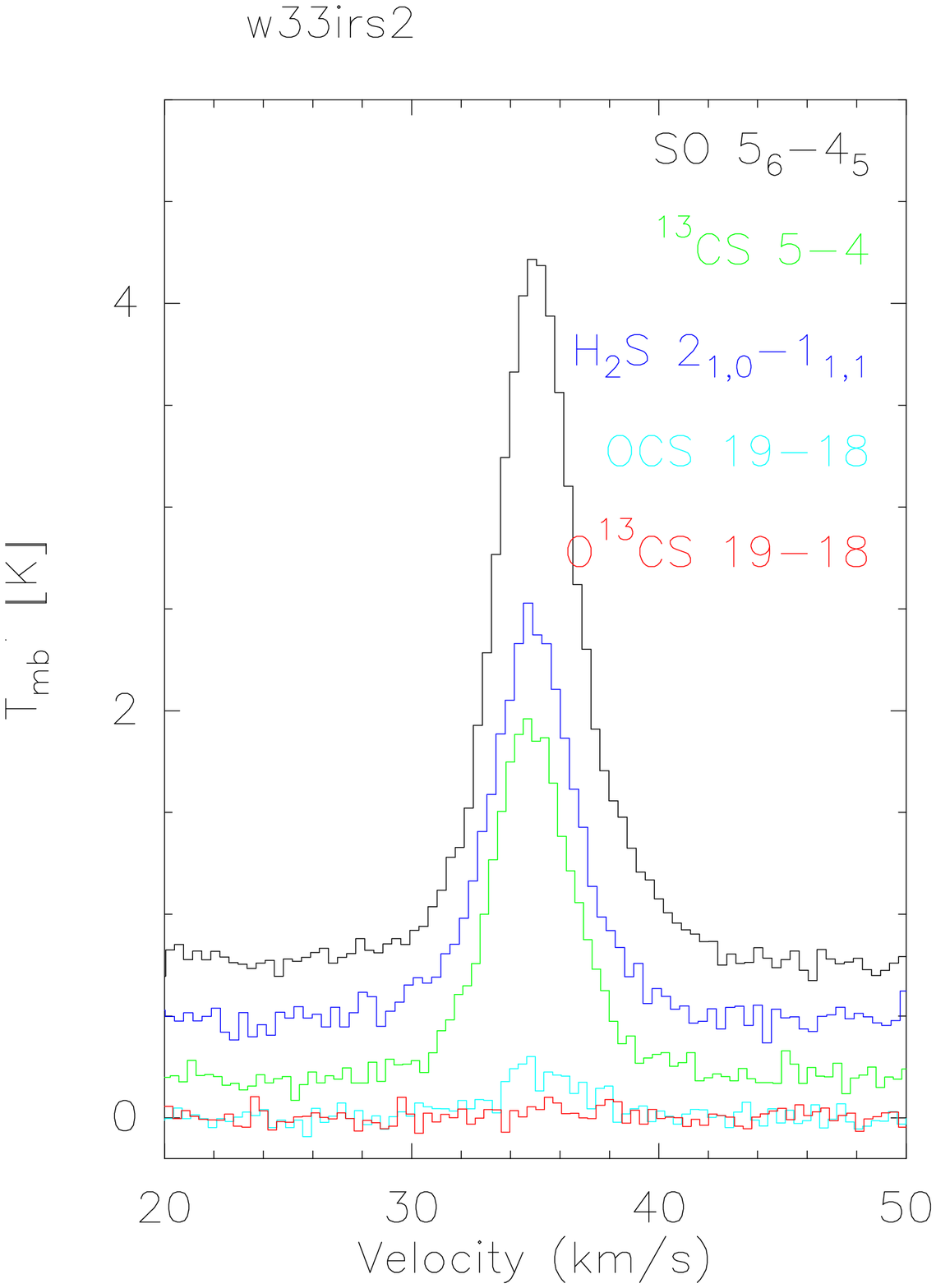}
\includegraphics[scale=0.3]{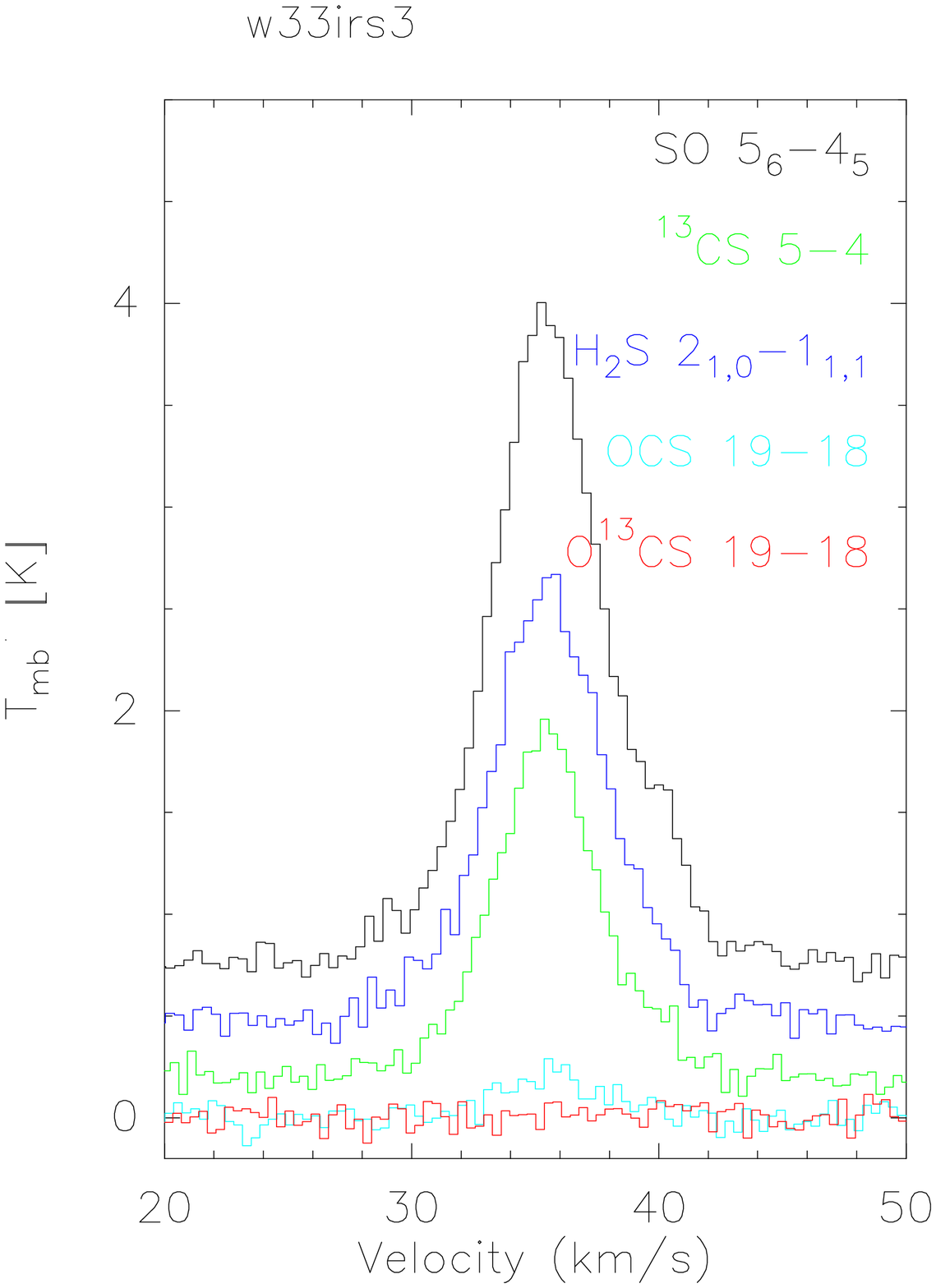}
\includegraphics[scale=0.3]{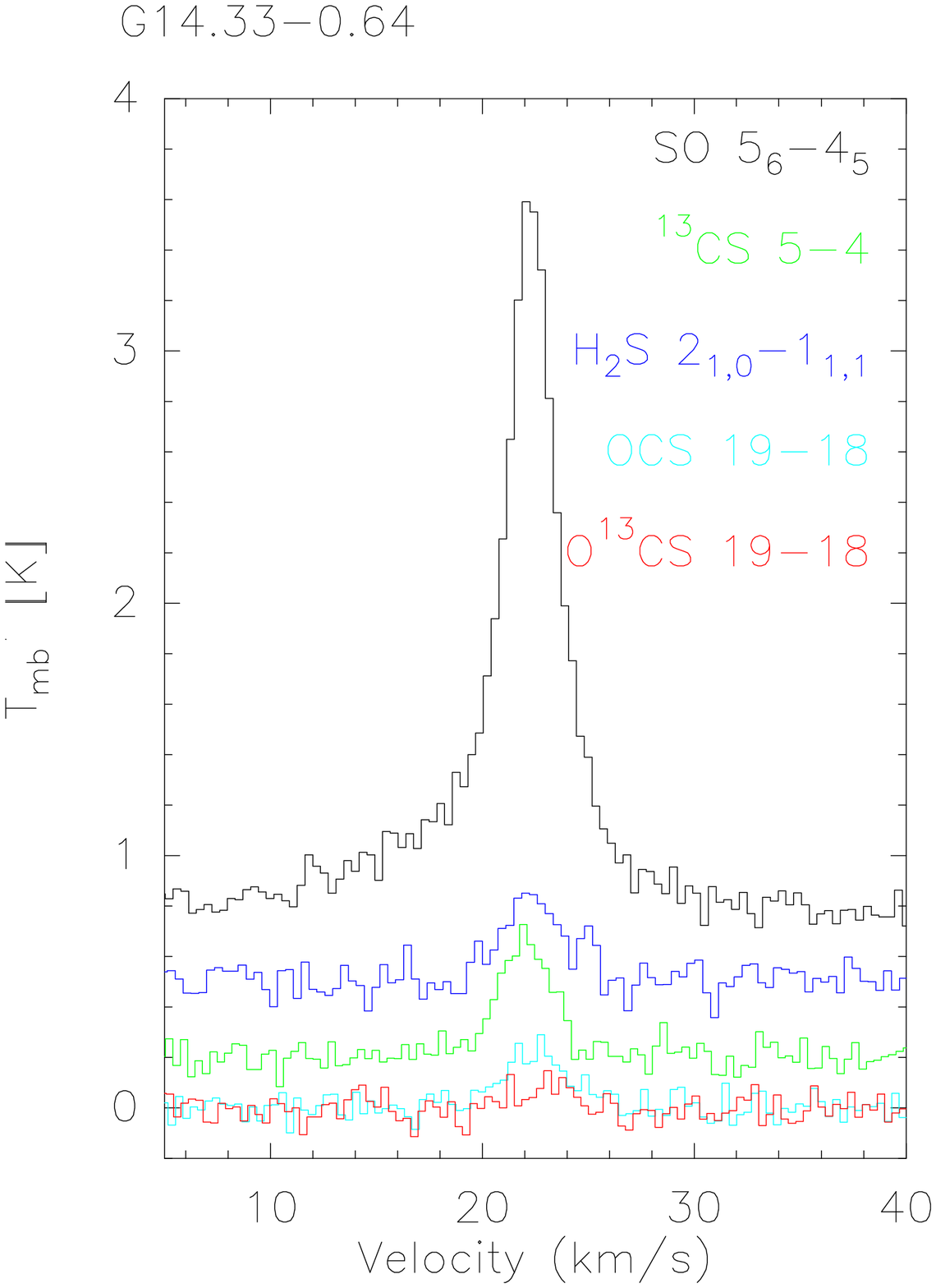}
\includegraphics[scale=0.3]{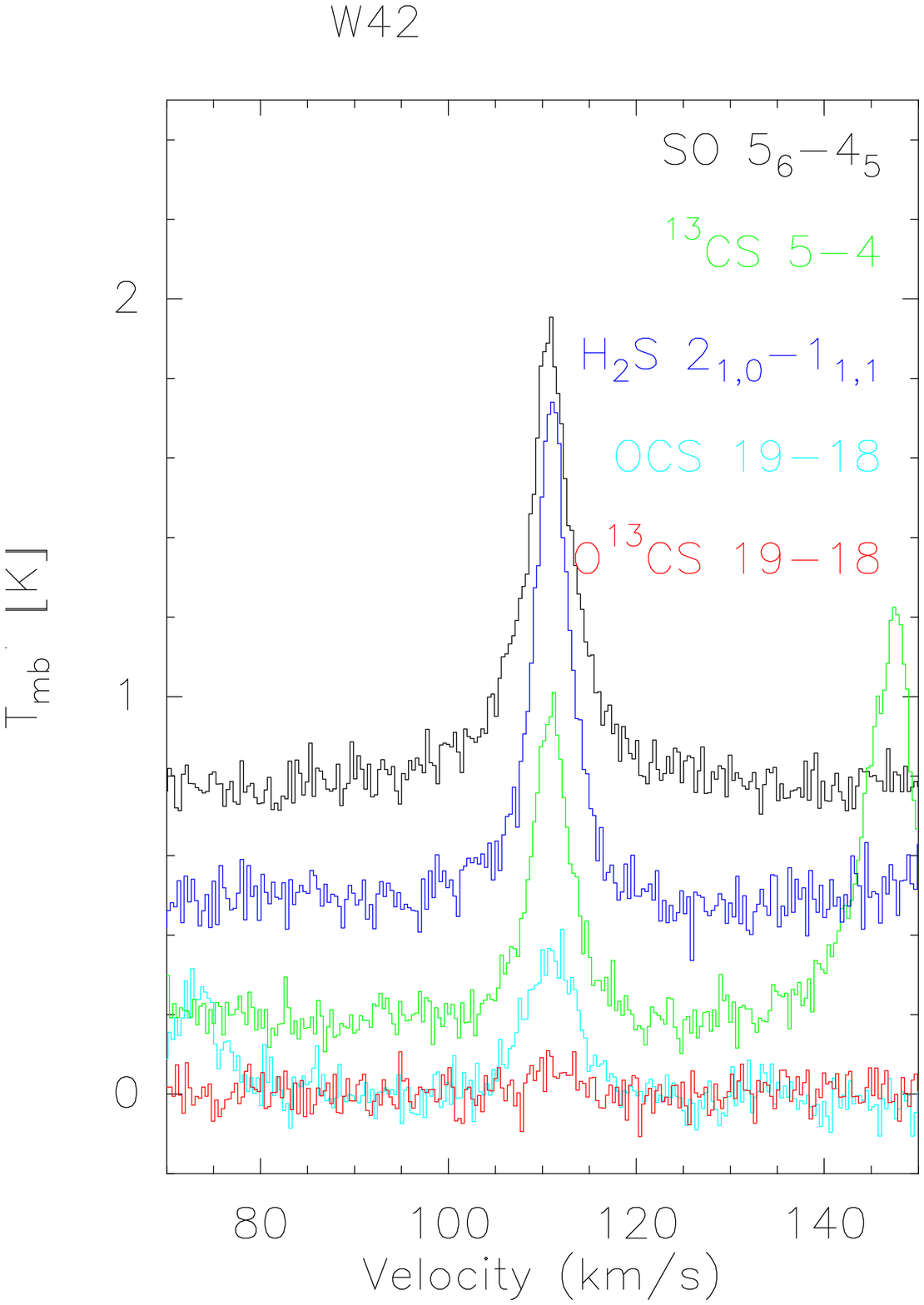}
\includegraphics[scale=0.3]{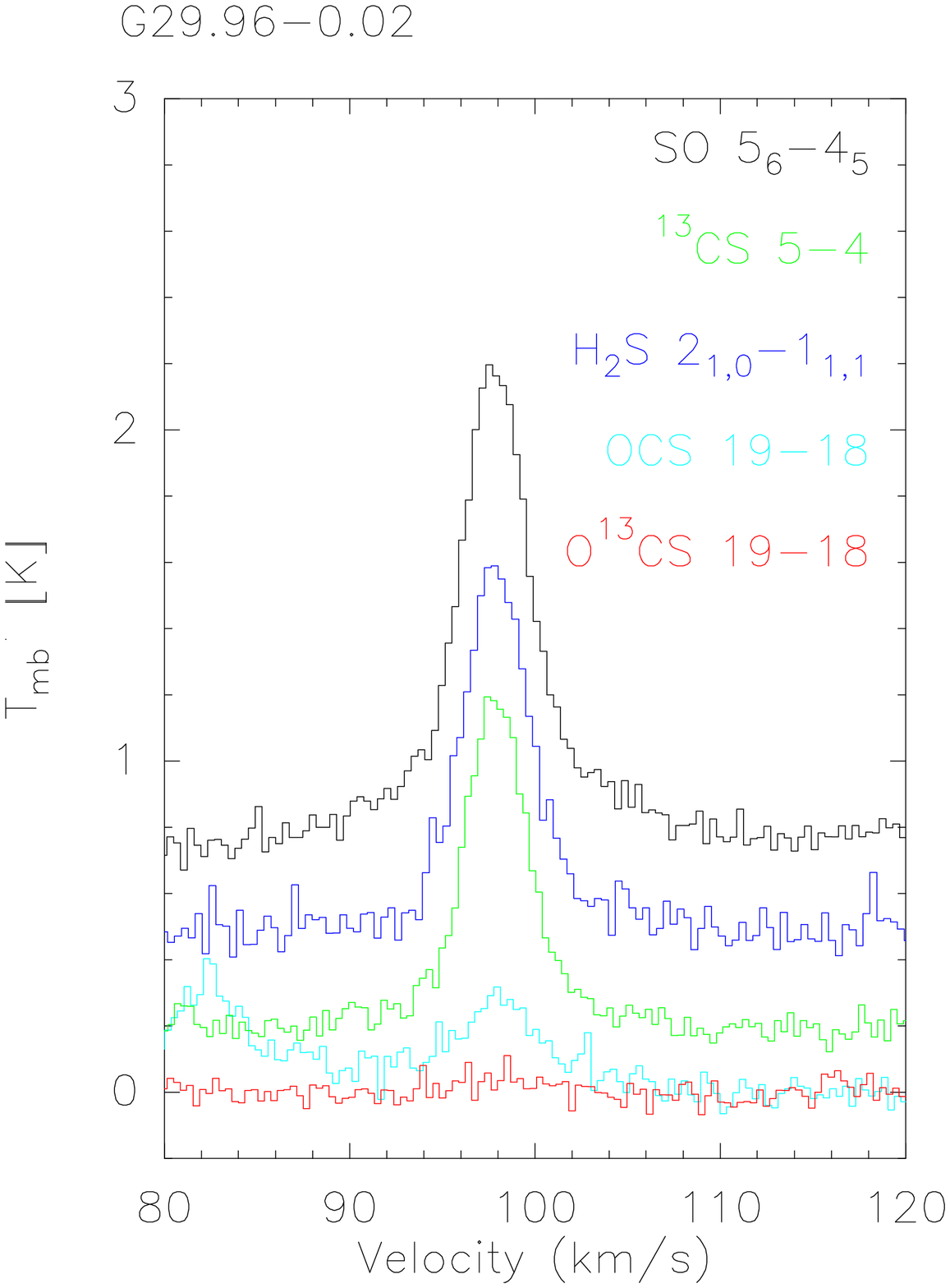}
\includegraphics[scale=0.3]{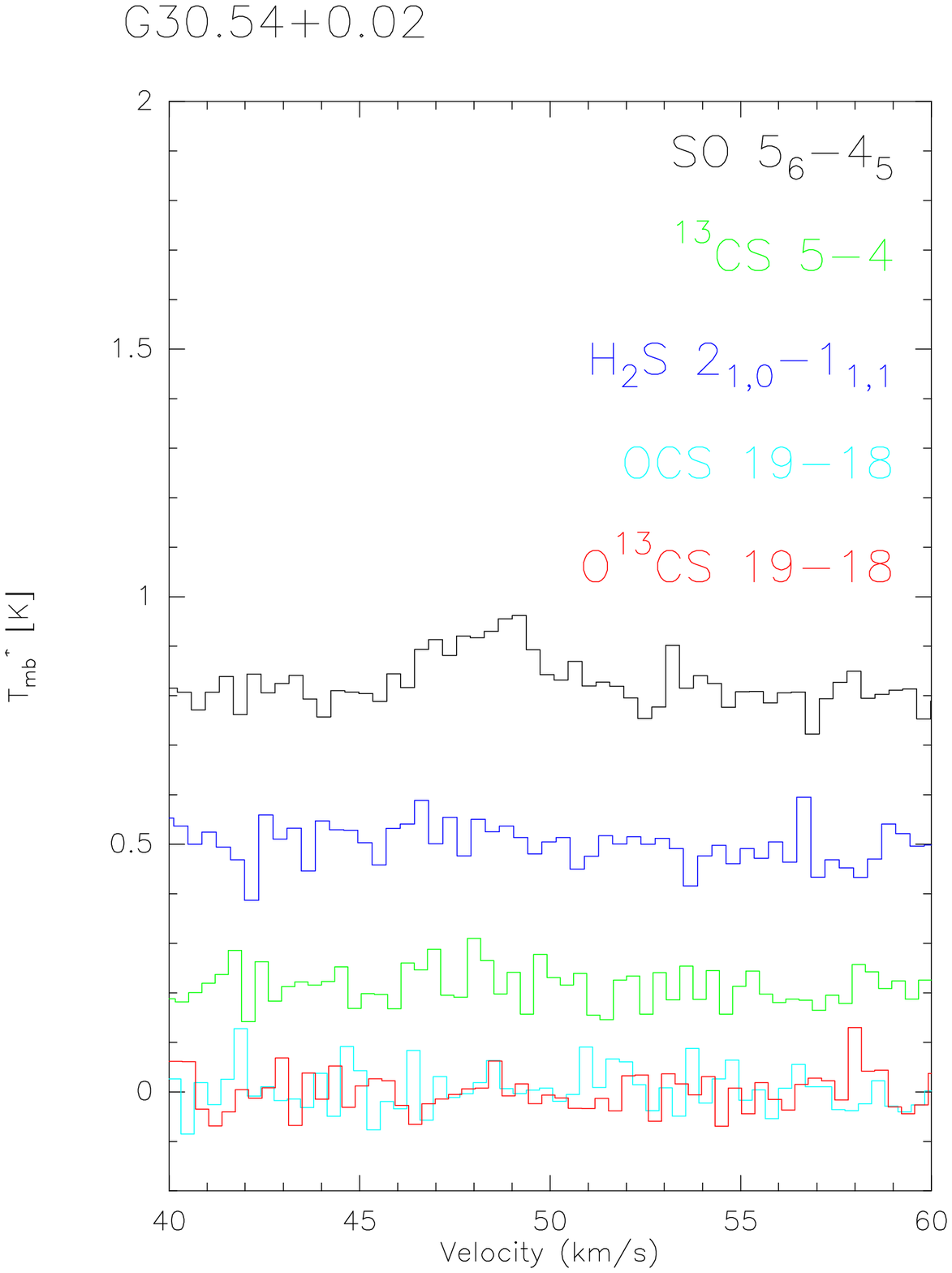}
\includegraphics[scale=0.3]{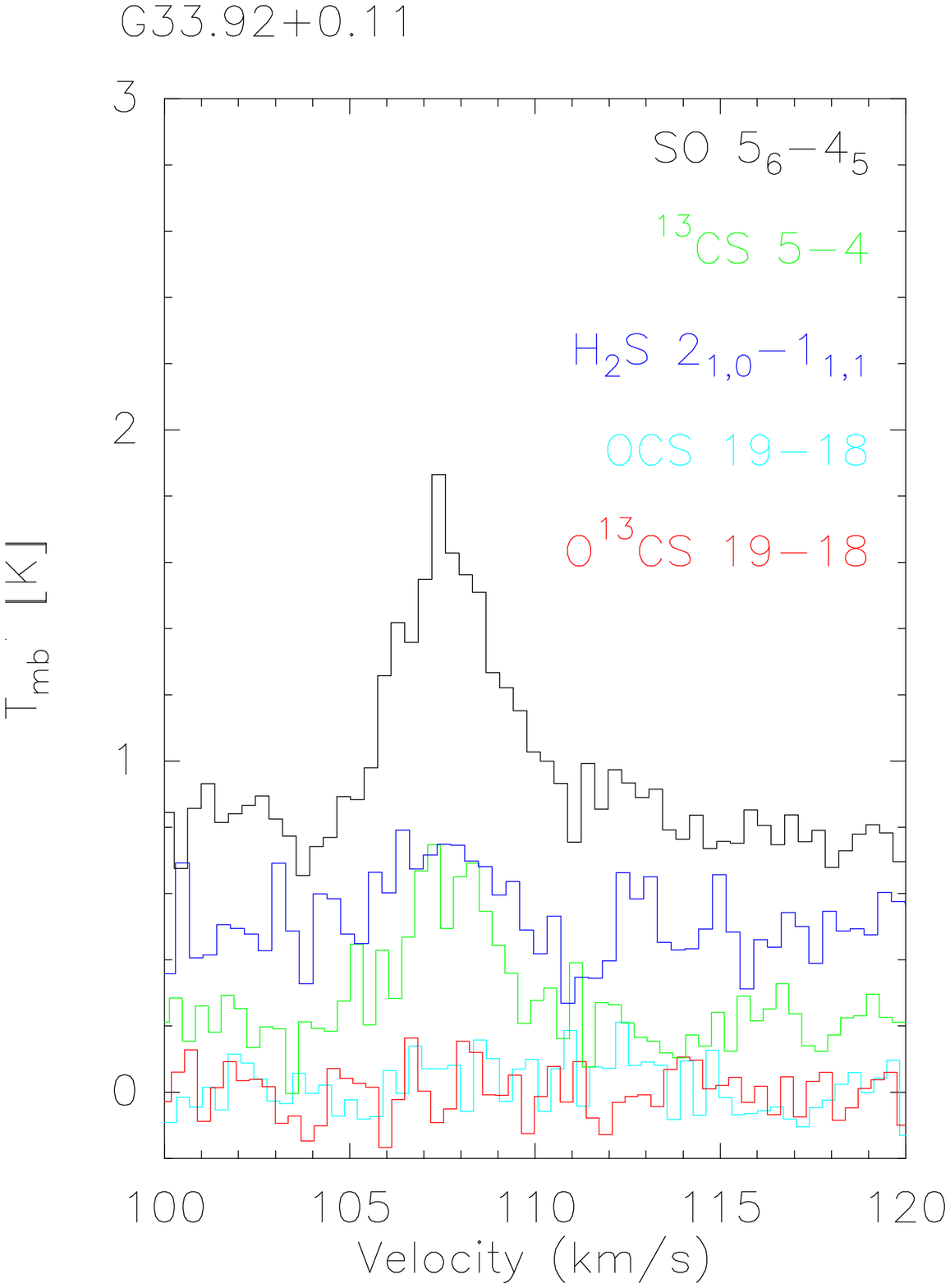}
\includegraphics[scale=0.3]{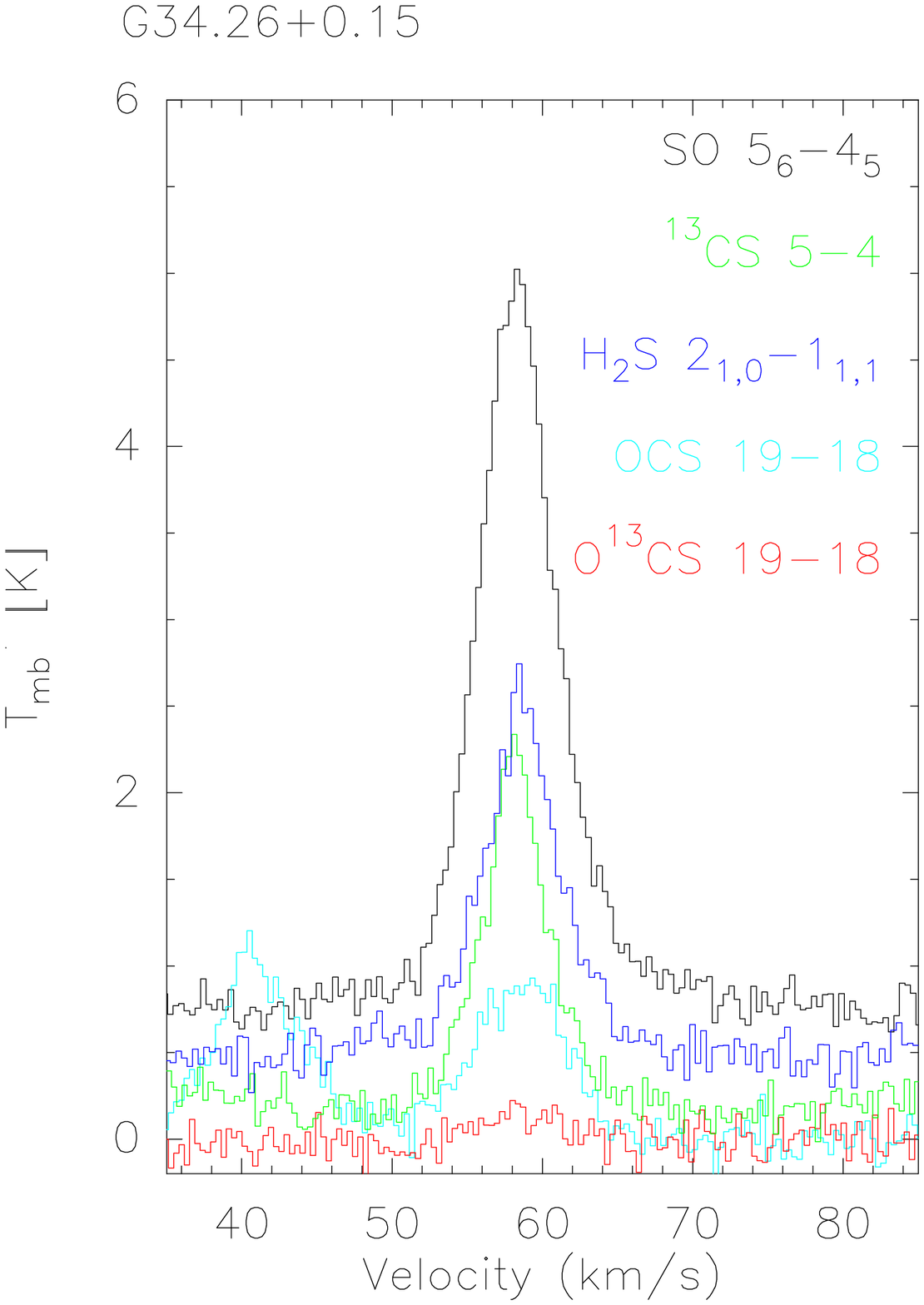}
\caption{Figure 1 continued.}
\end{figure}

\begin{figure}[htbp]
   \centering
\includegraphics[scale=0.3]{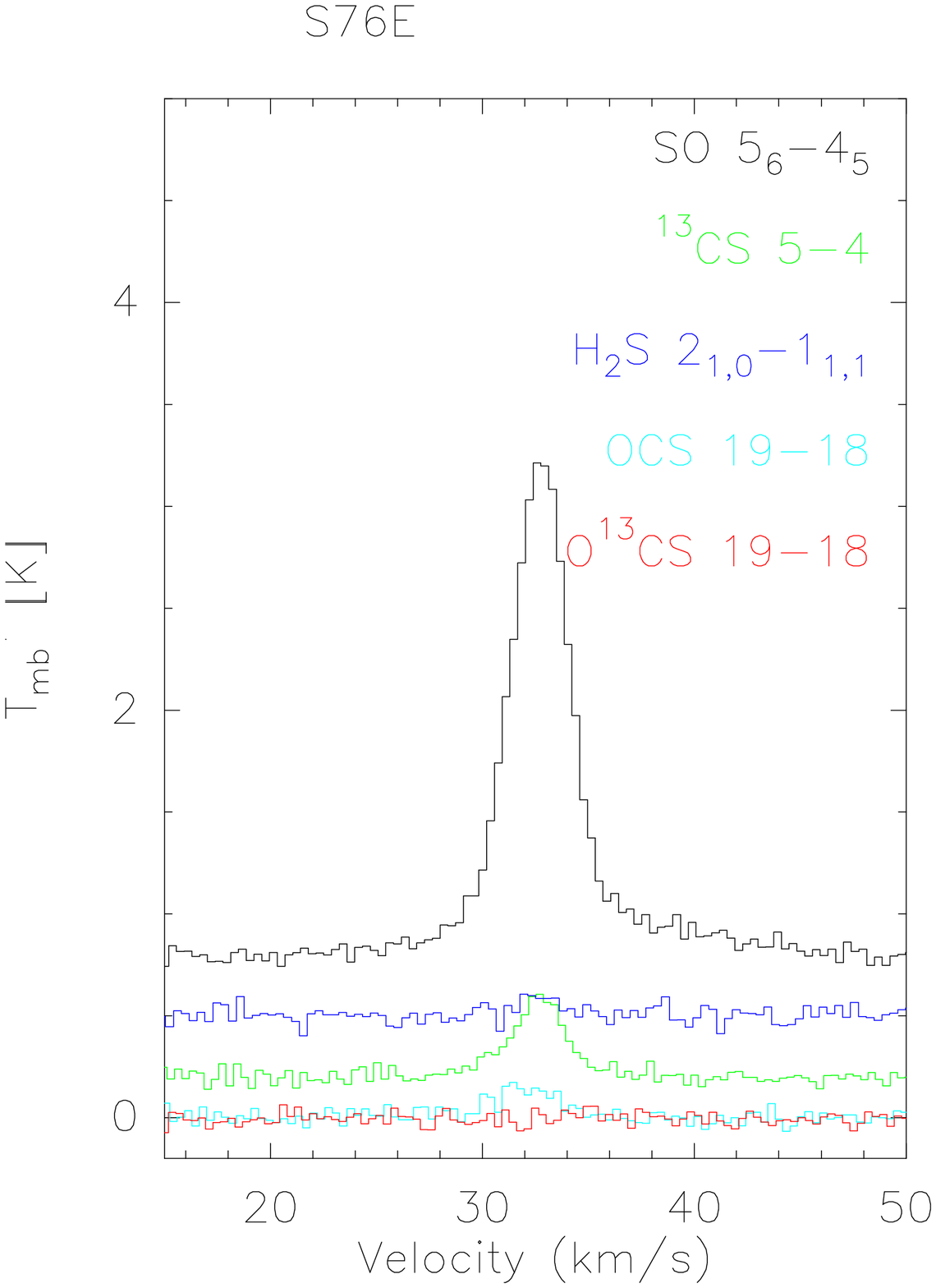}
\includegraphics[scale=0.3]{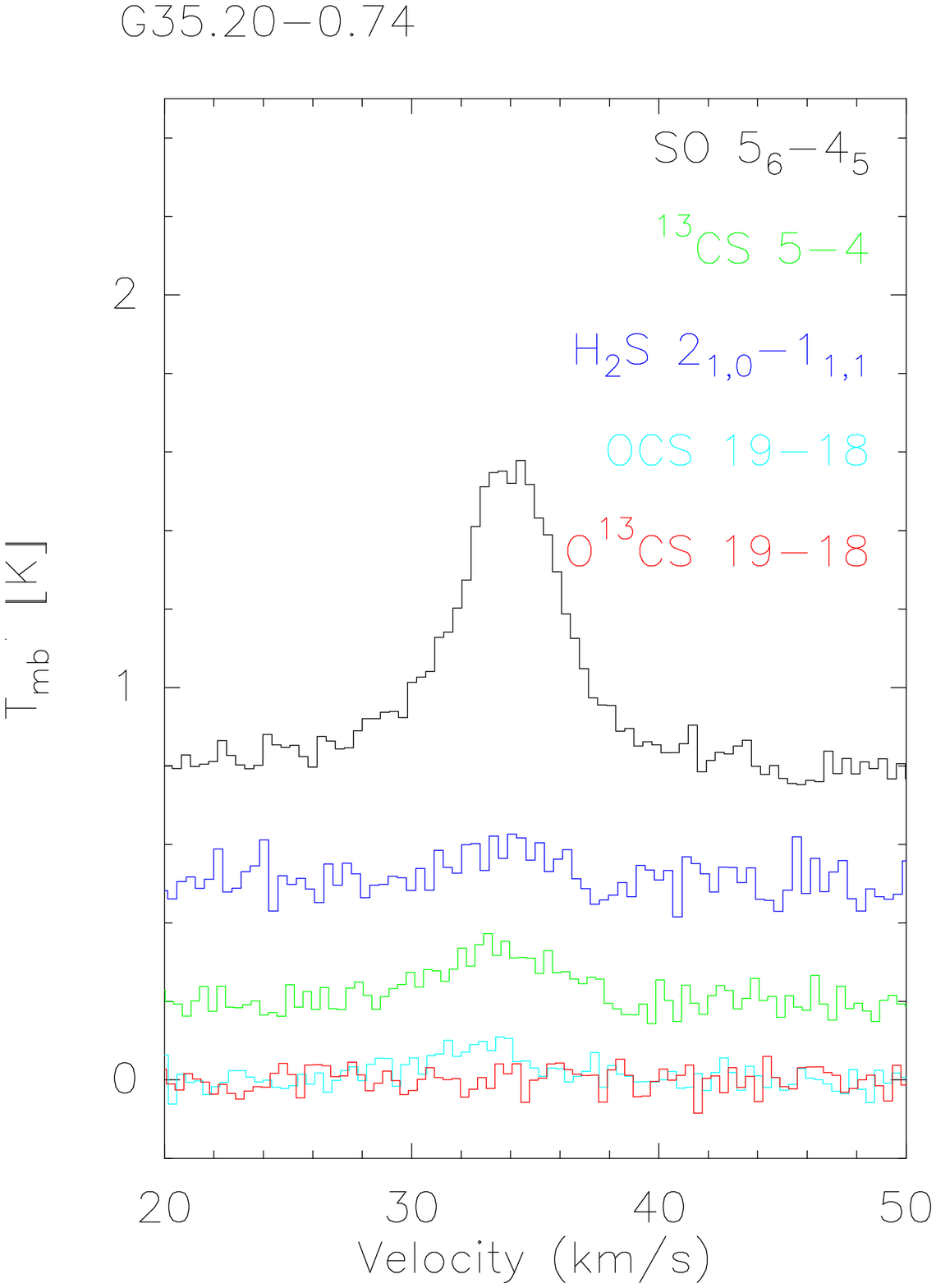}
\includegraphics[scale=0.3]{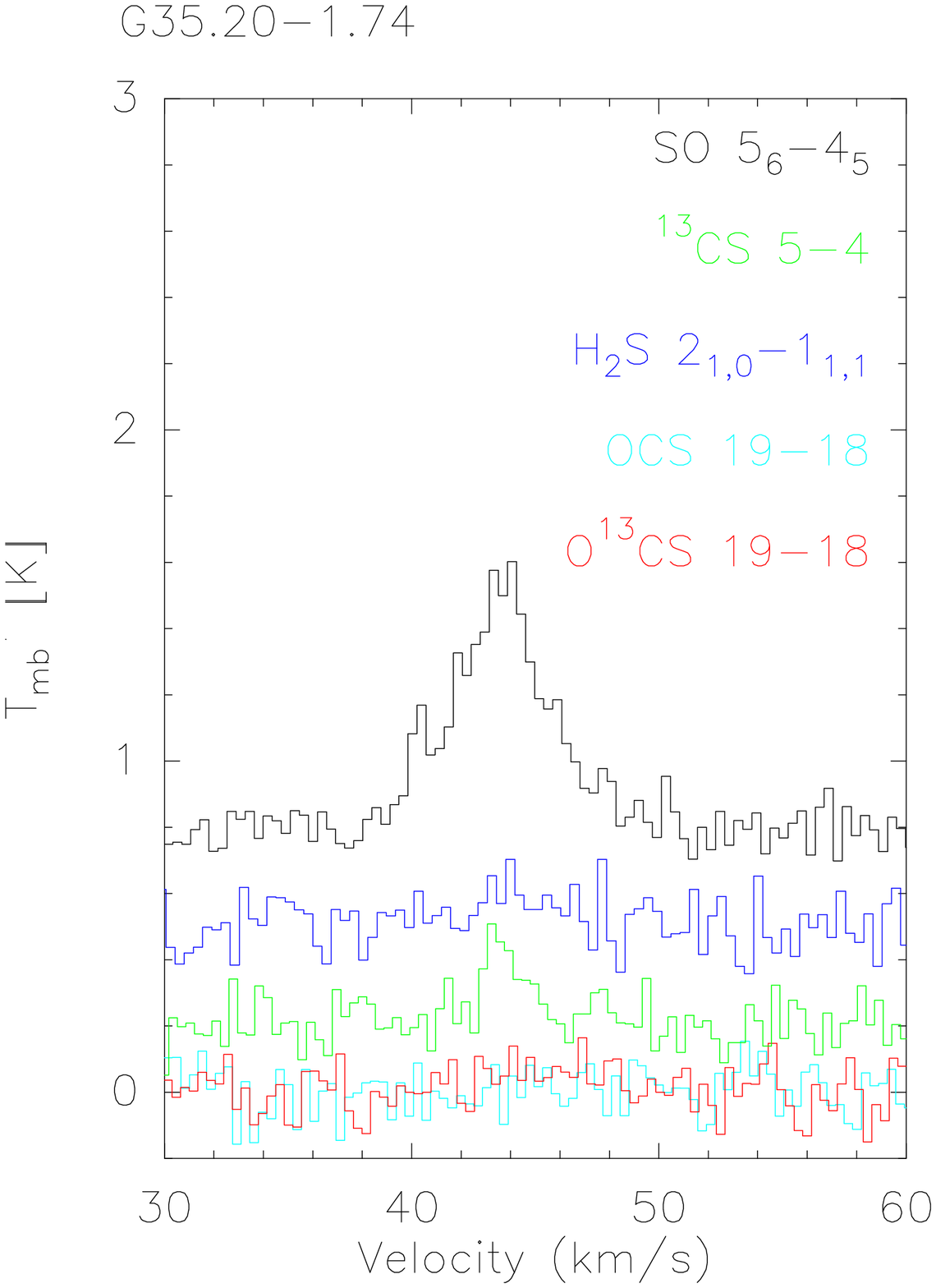}
\includegraphics[scale=0.3]{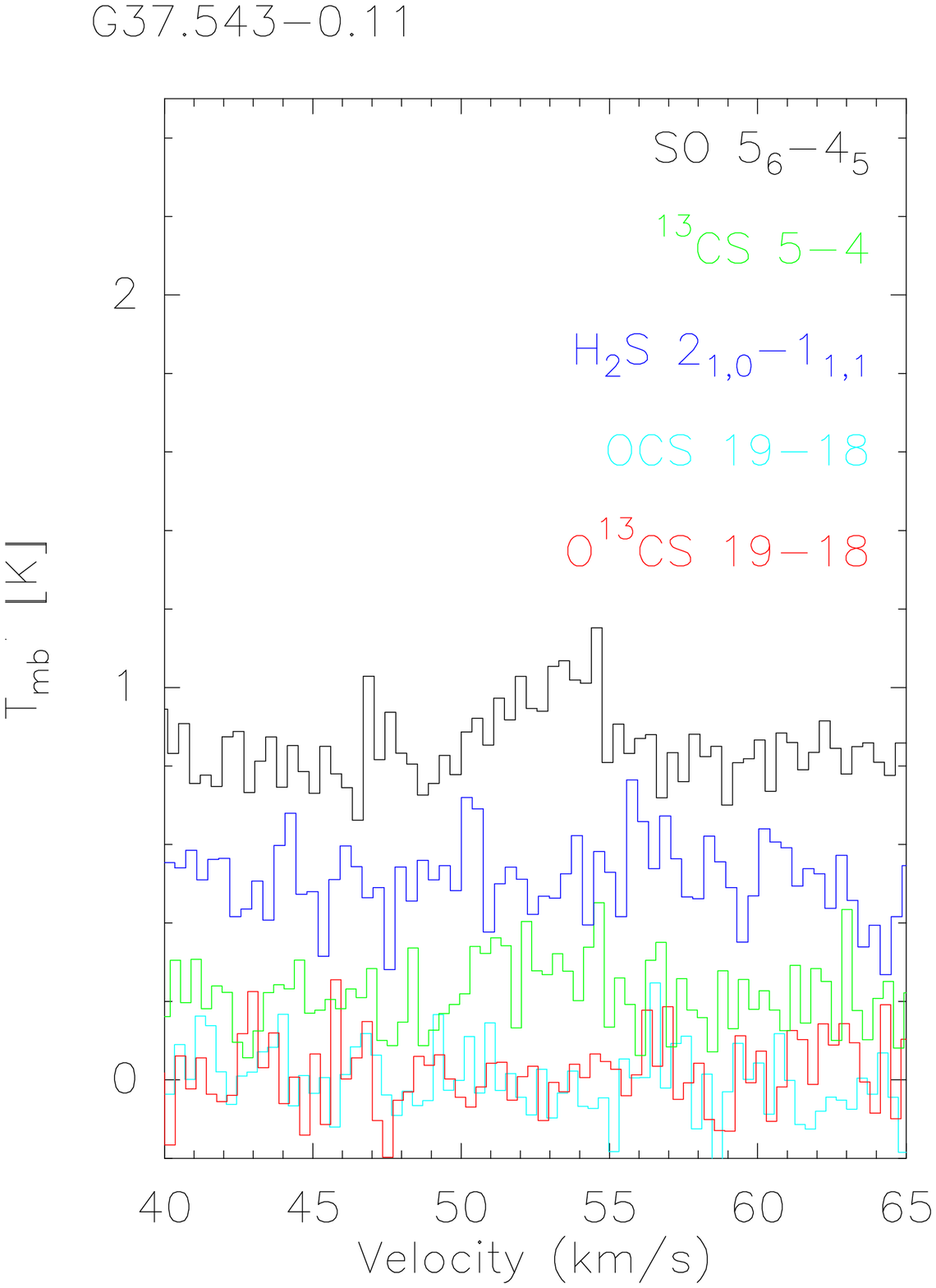}
\includegraphics[scale=0.3]{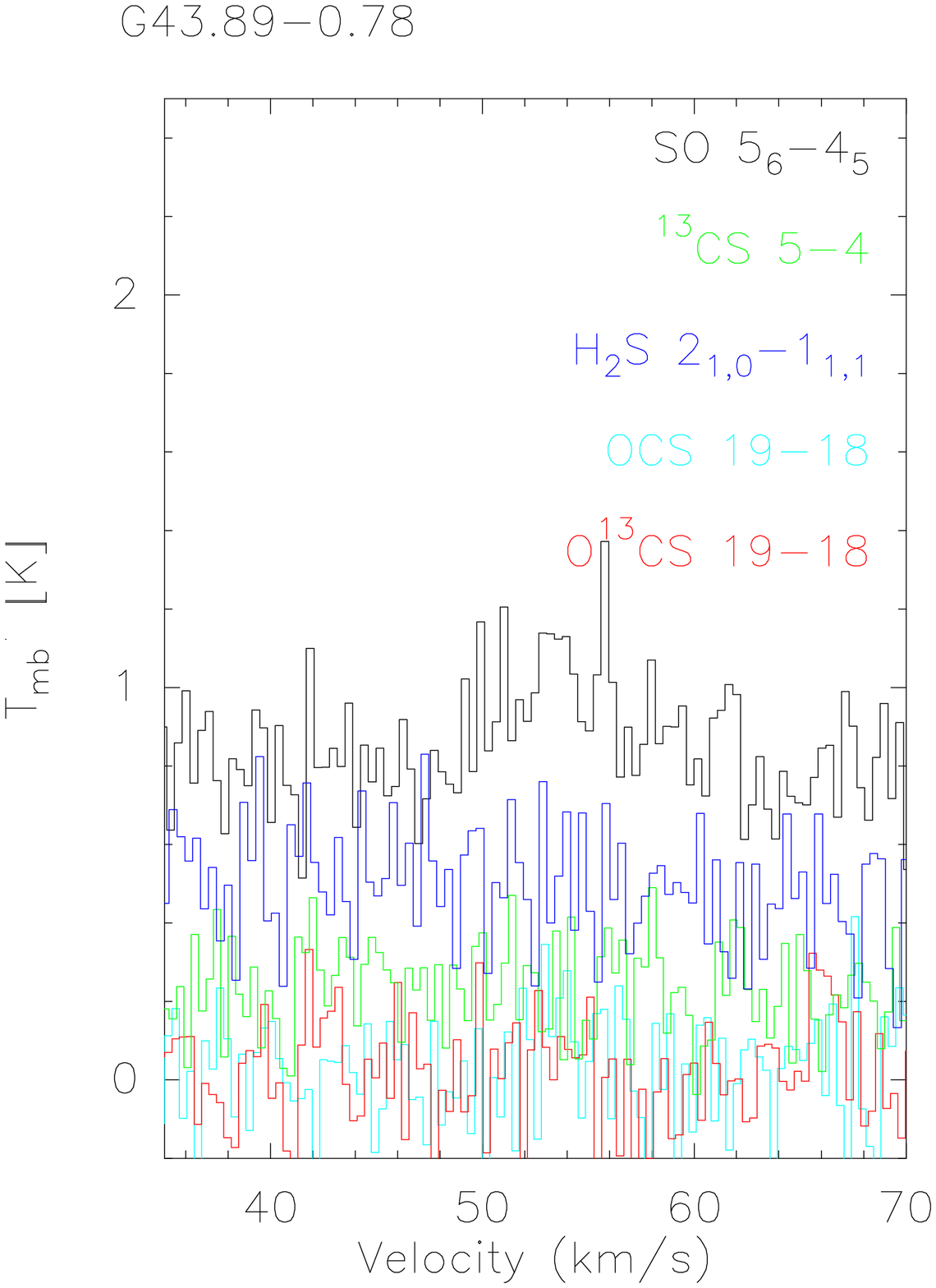}
\includegraphics[scale=0.3]{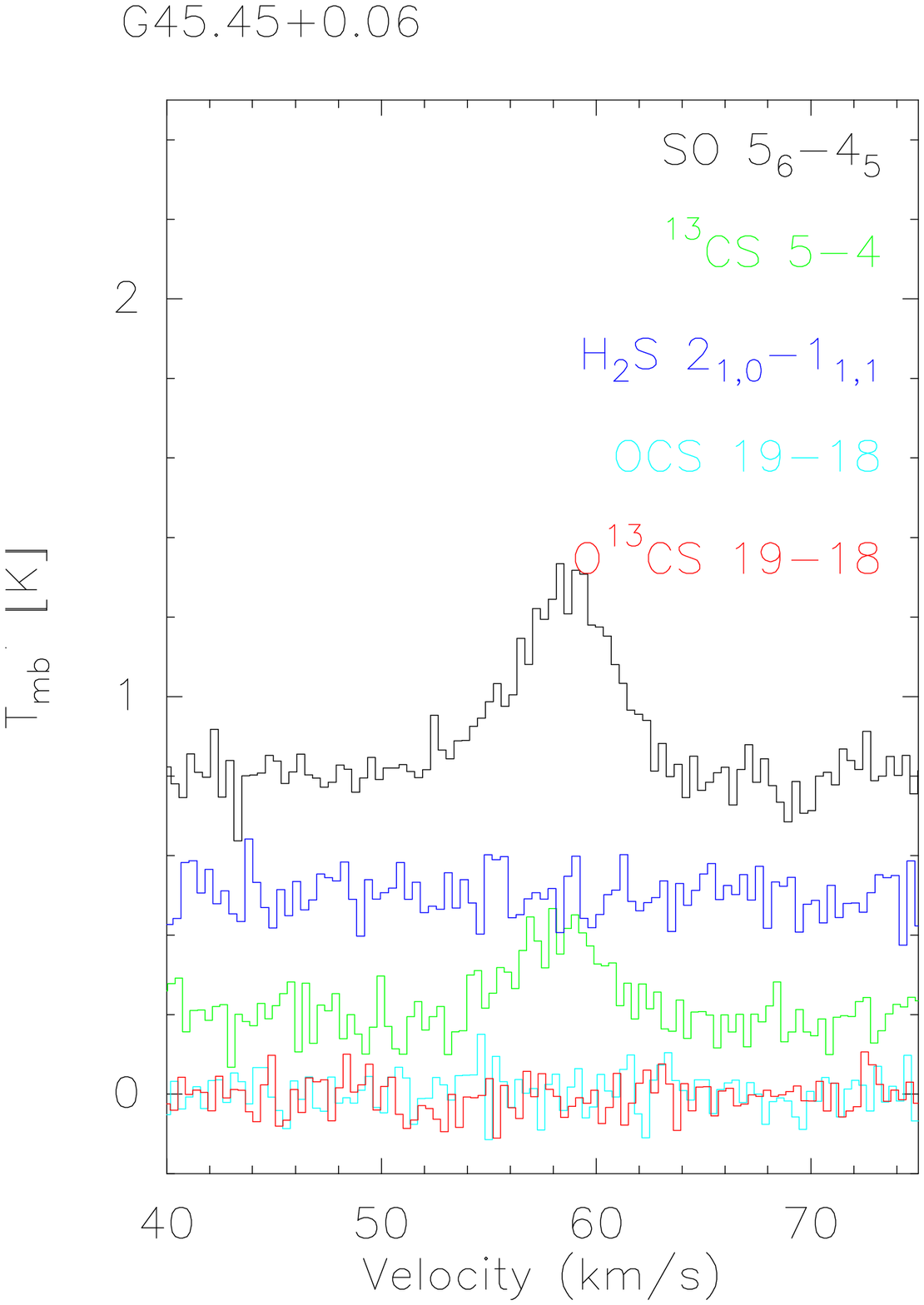}
\includegraphics[scale=0.3]{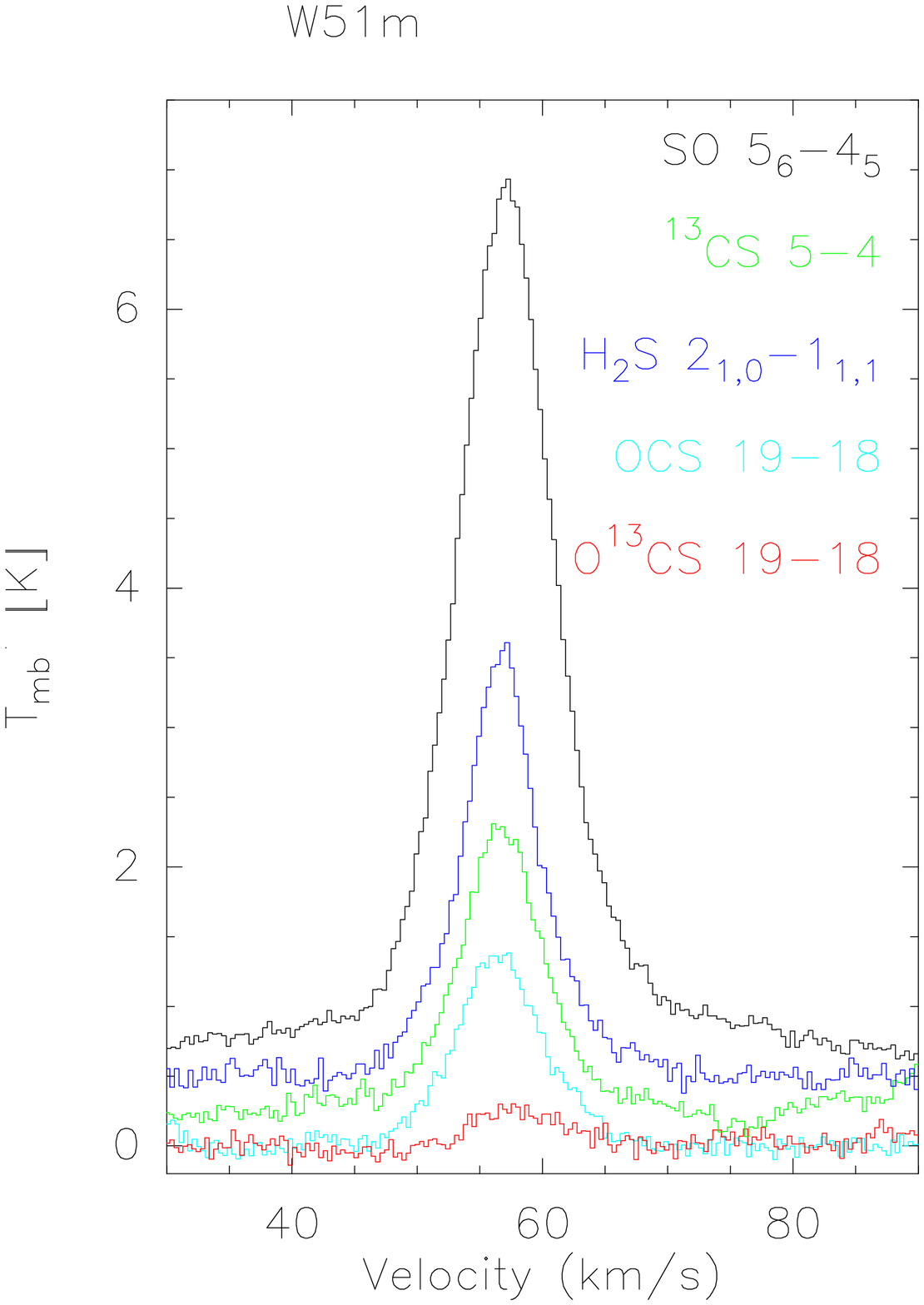}
\includegraphics[scale=0.3]{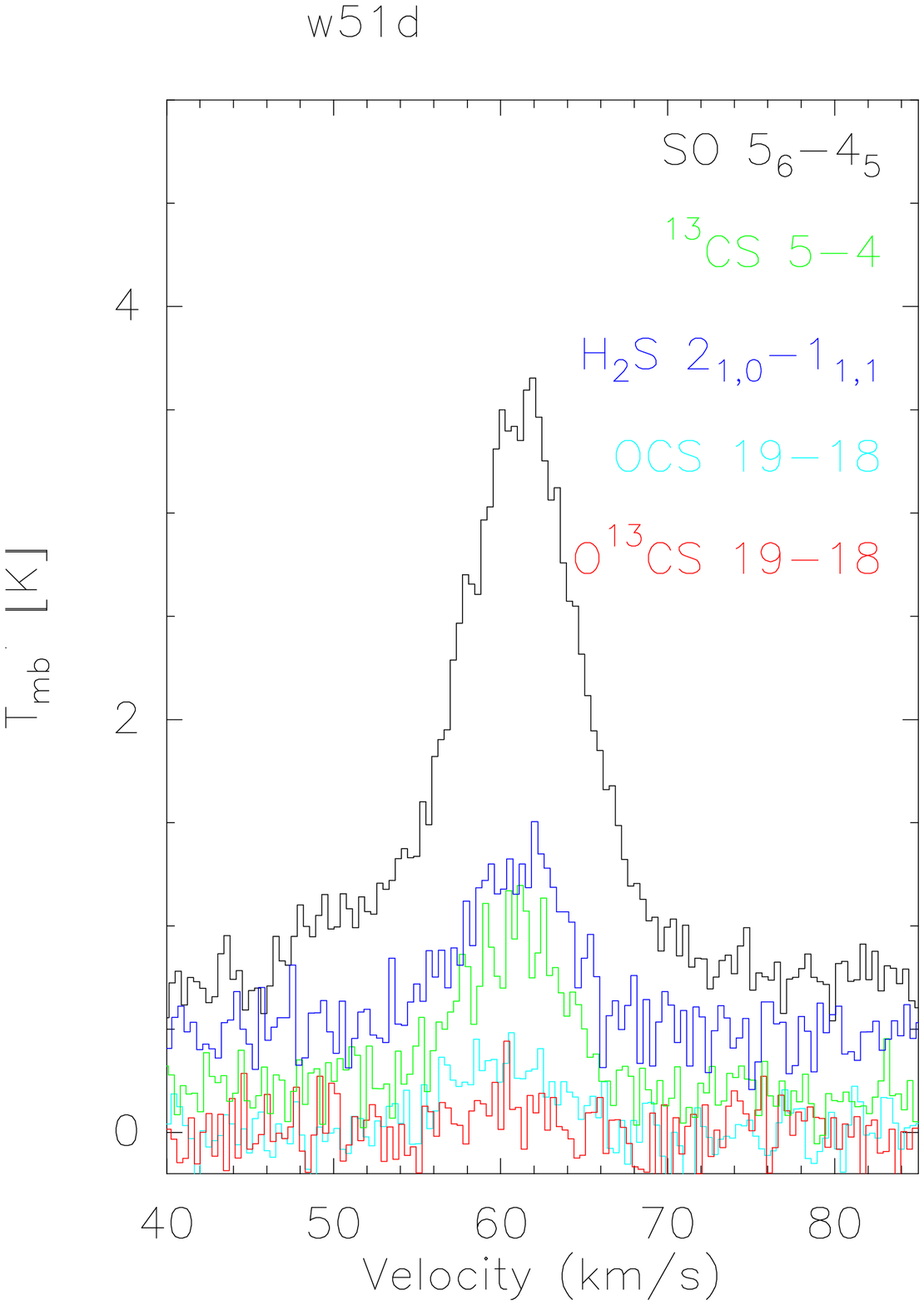}
\includegraphics[scale=0.3]{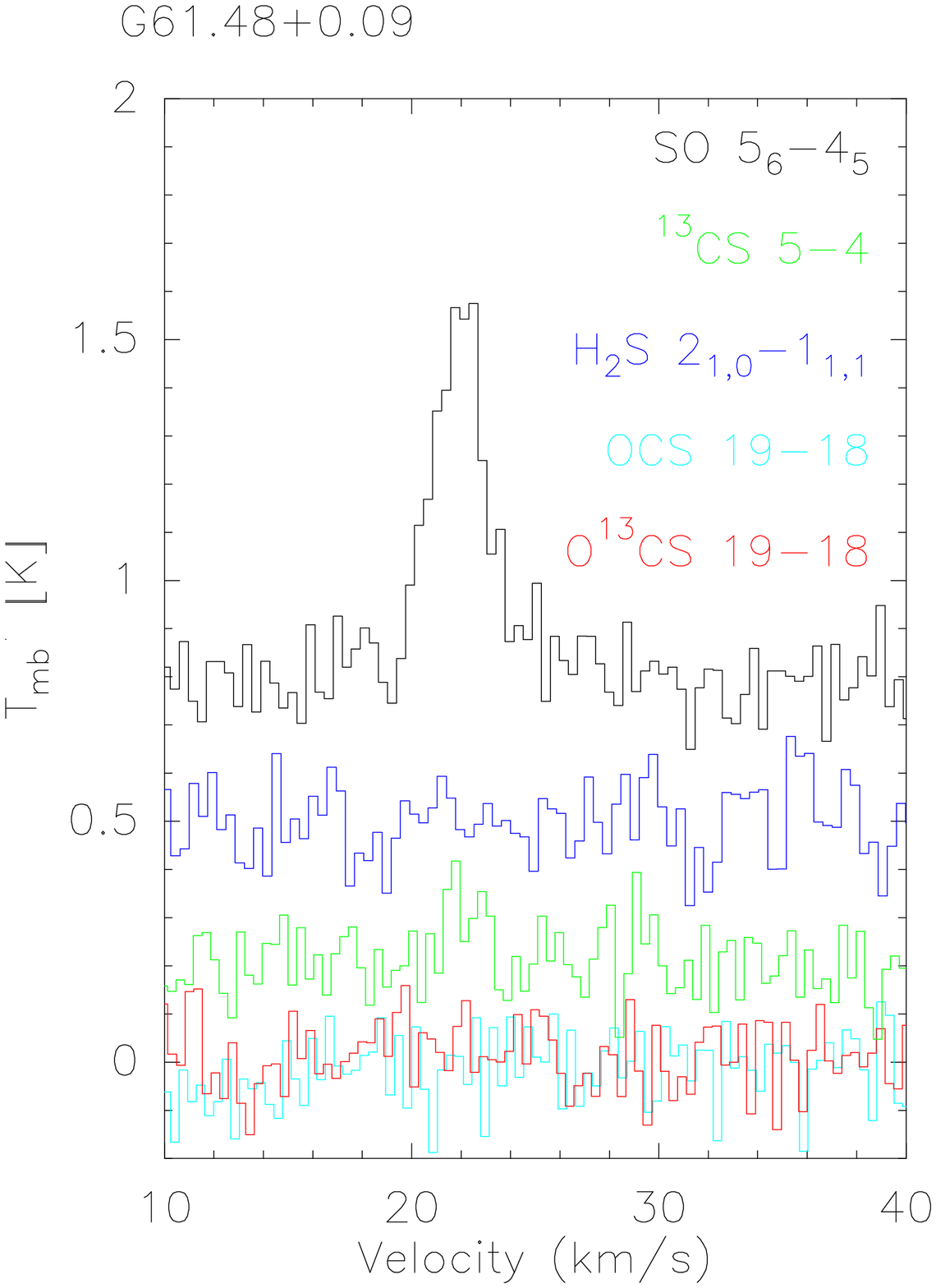}
\caption{Figure 1 continued.}
\end{figure}

\begin{figure}[htbp]
   \centering
\includegraphics[scale=0.3]{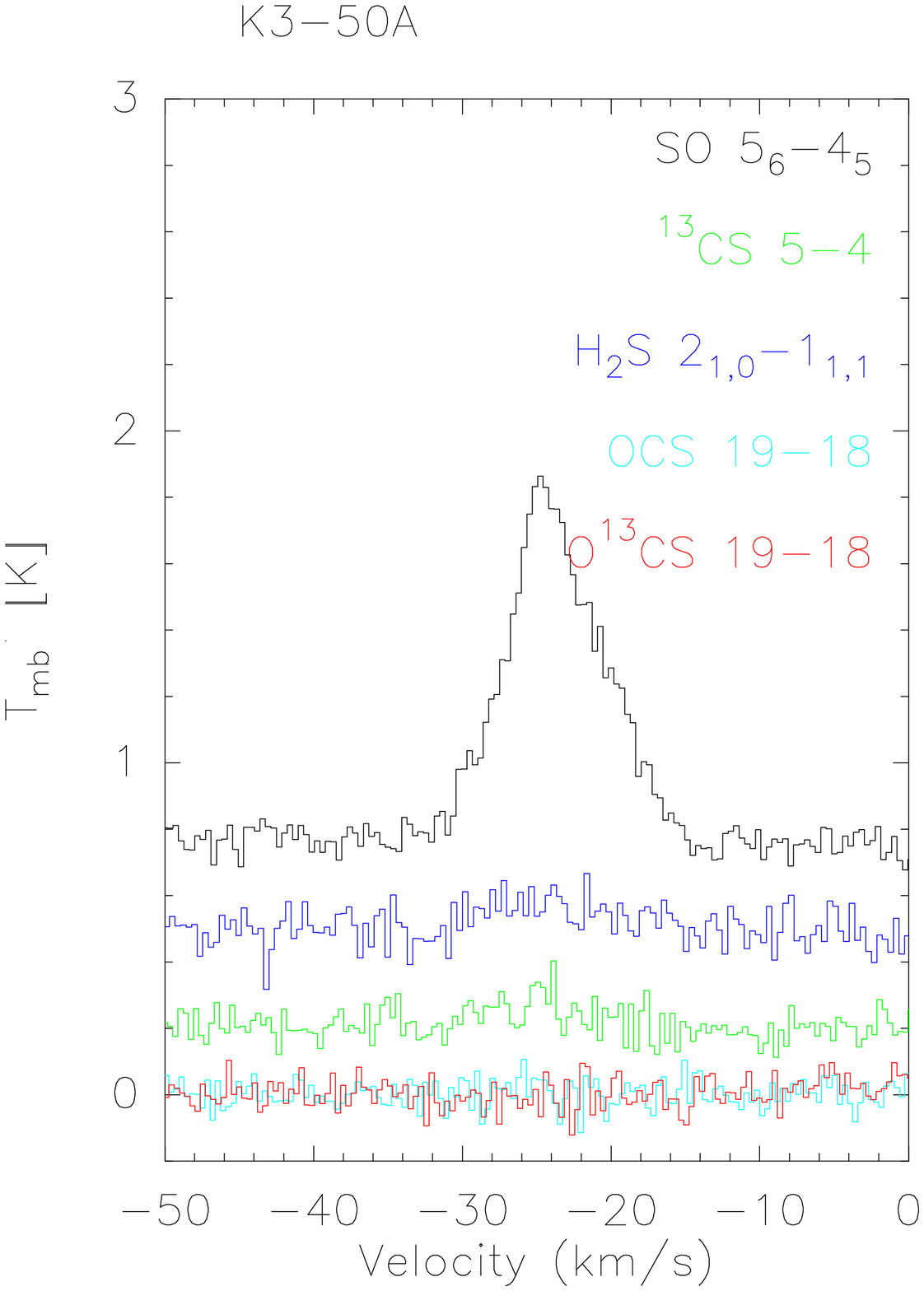}
\includegraphics[scale=0.3]{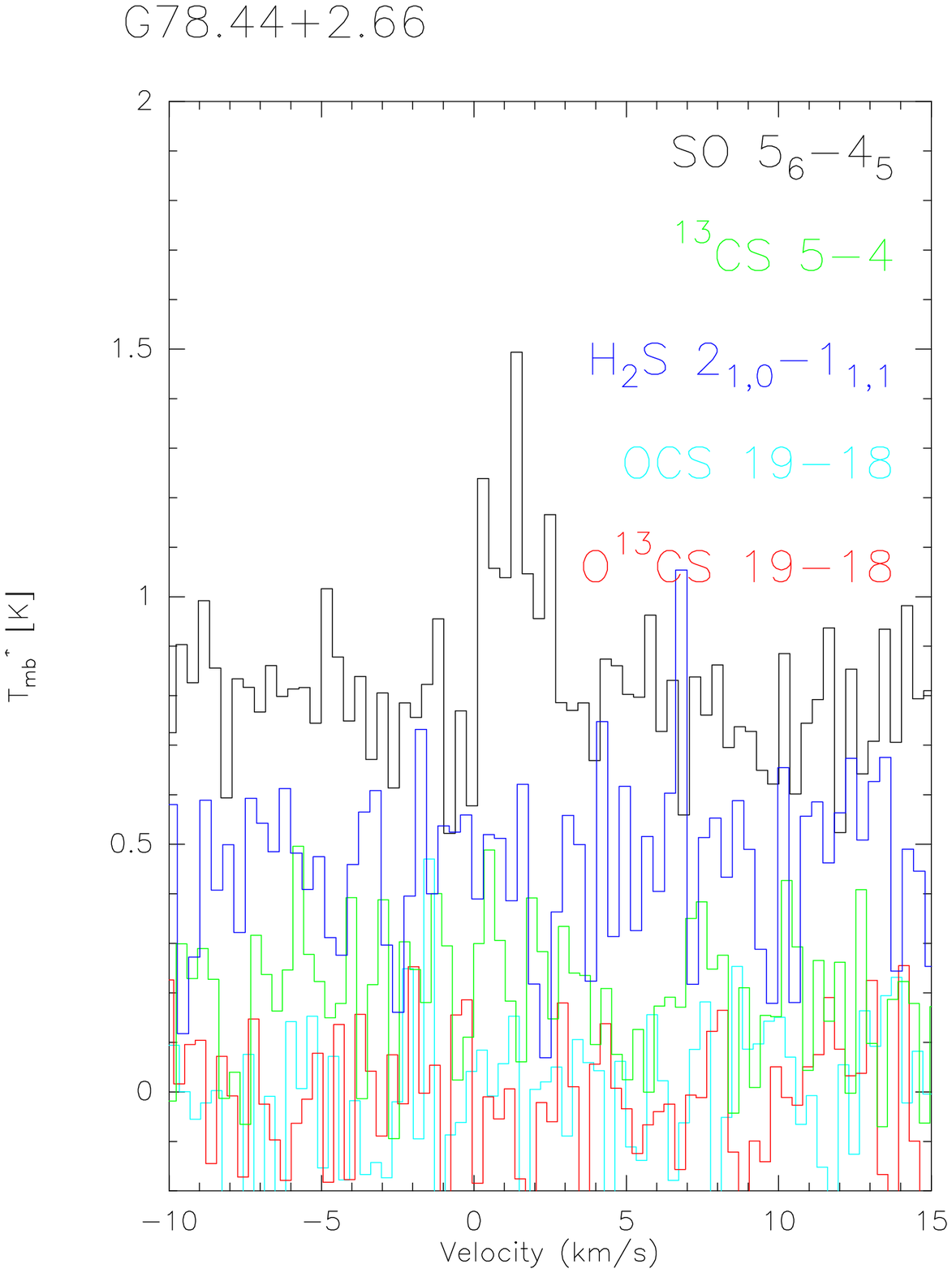}
\includegraphics[scale=0.3]{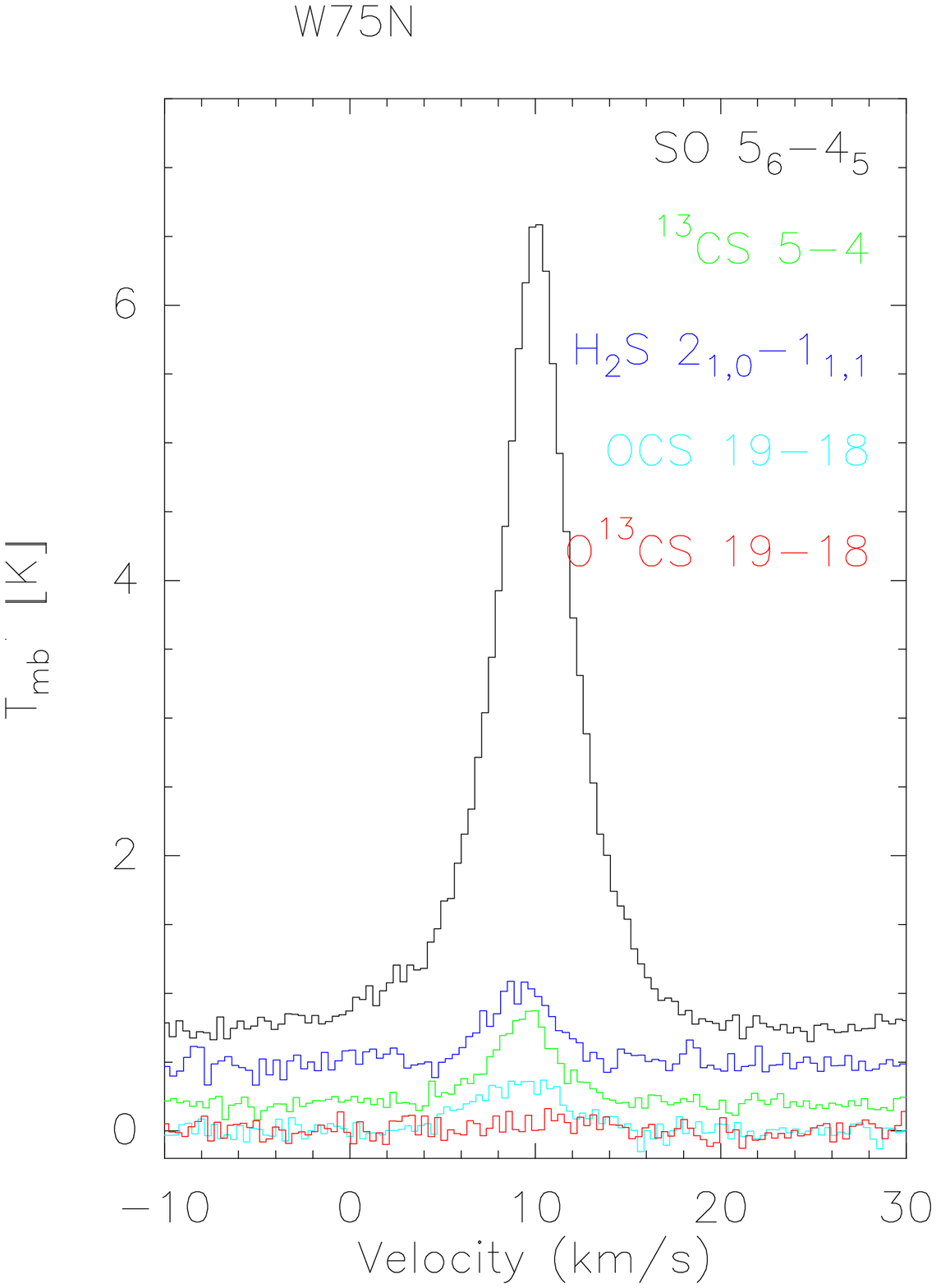}
\includegraphics[scale=0.3]{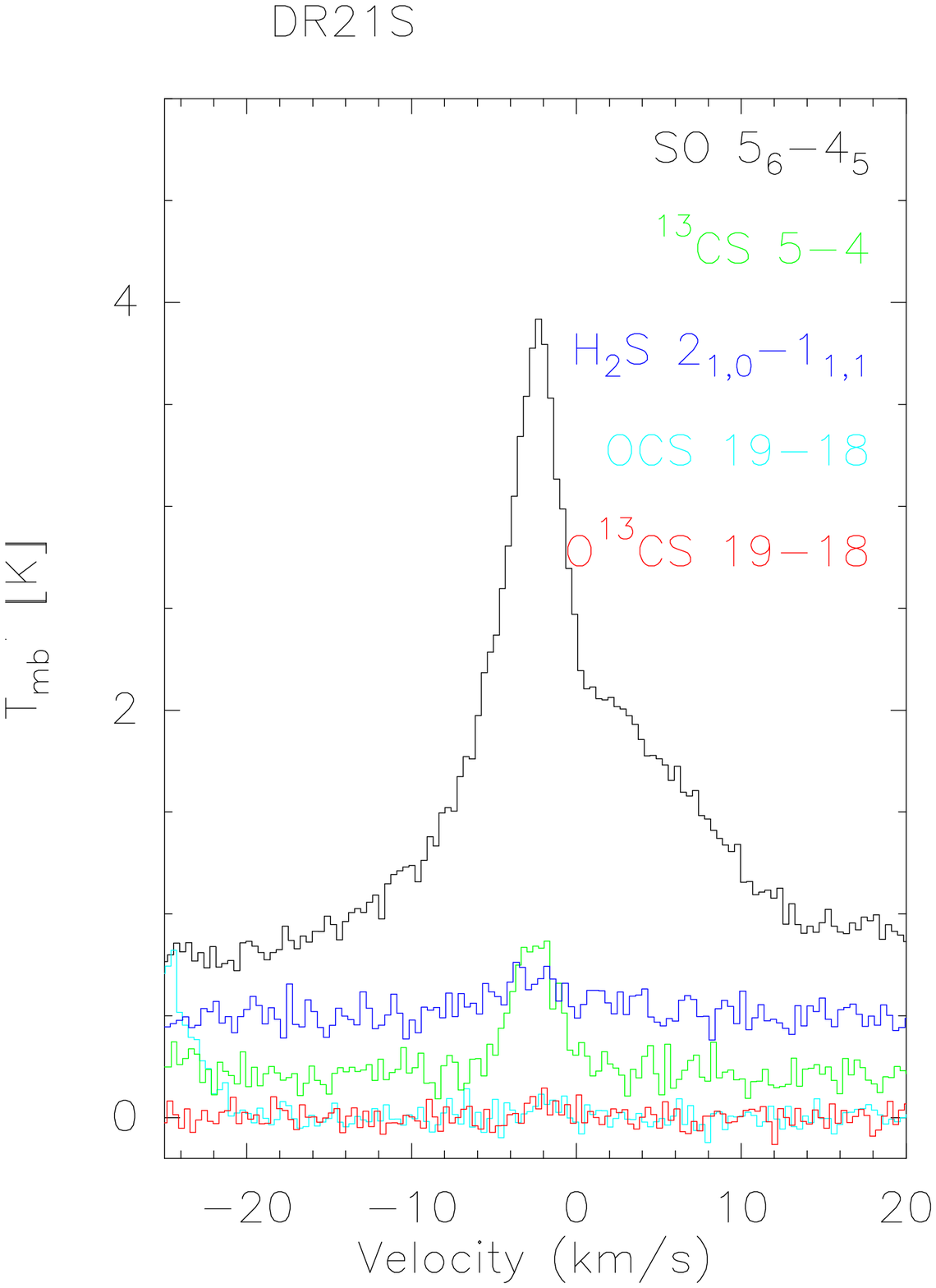}
\includegraphics[scale=0.3]{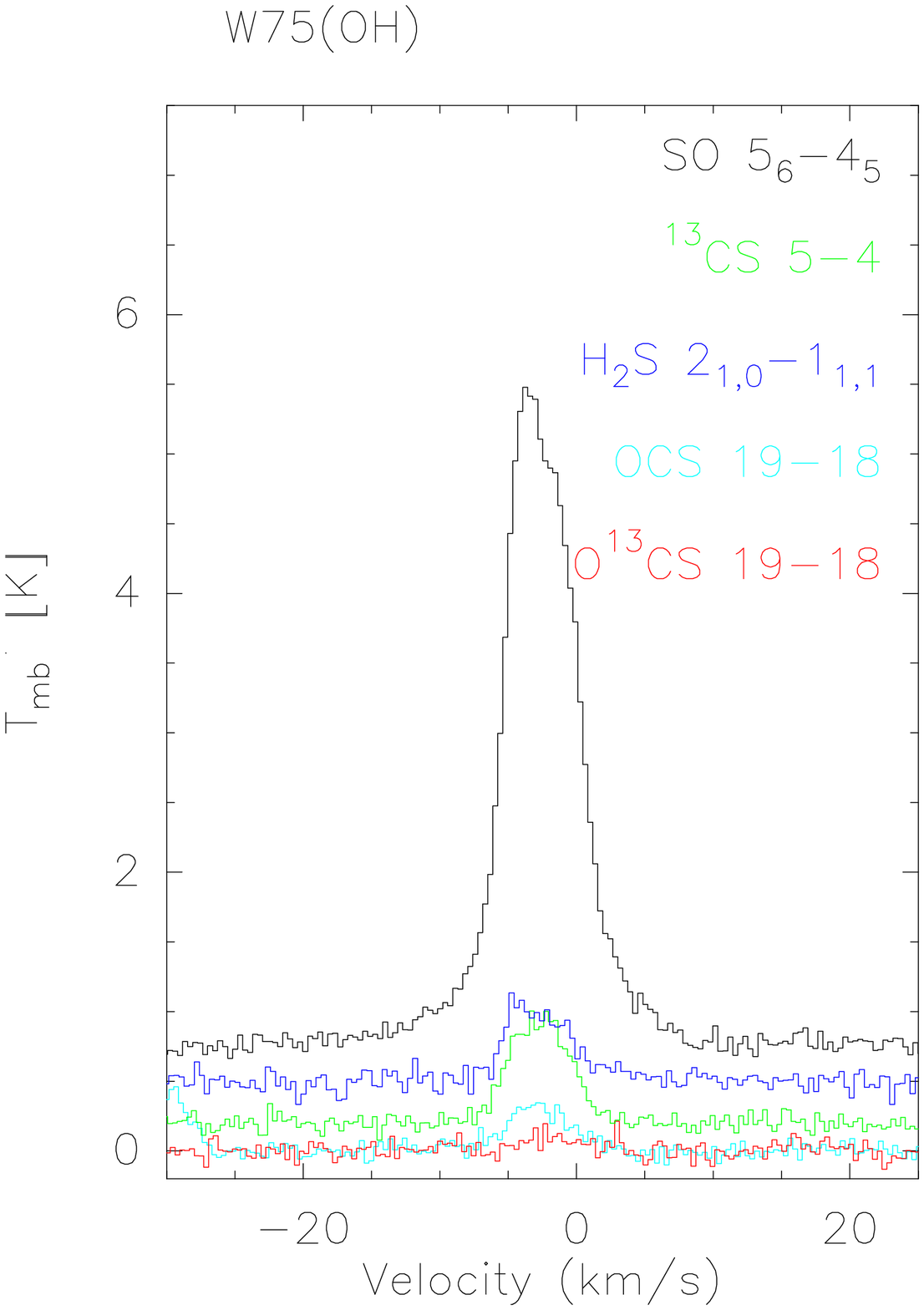}
\includegraphics[scale=0.3]{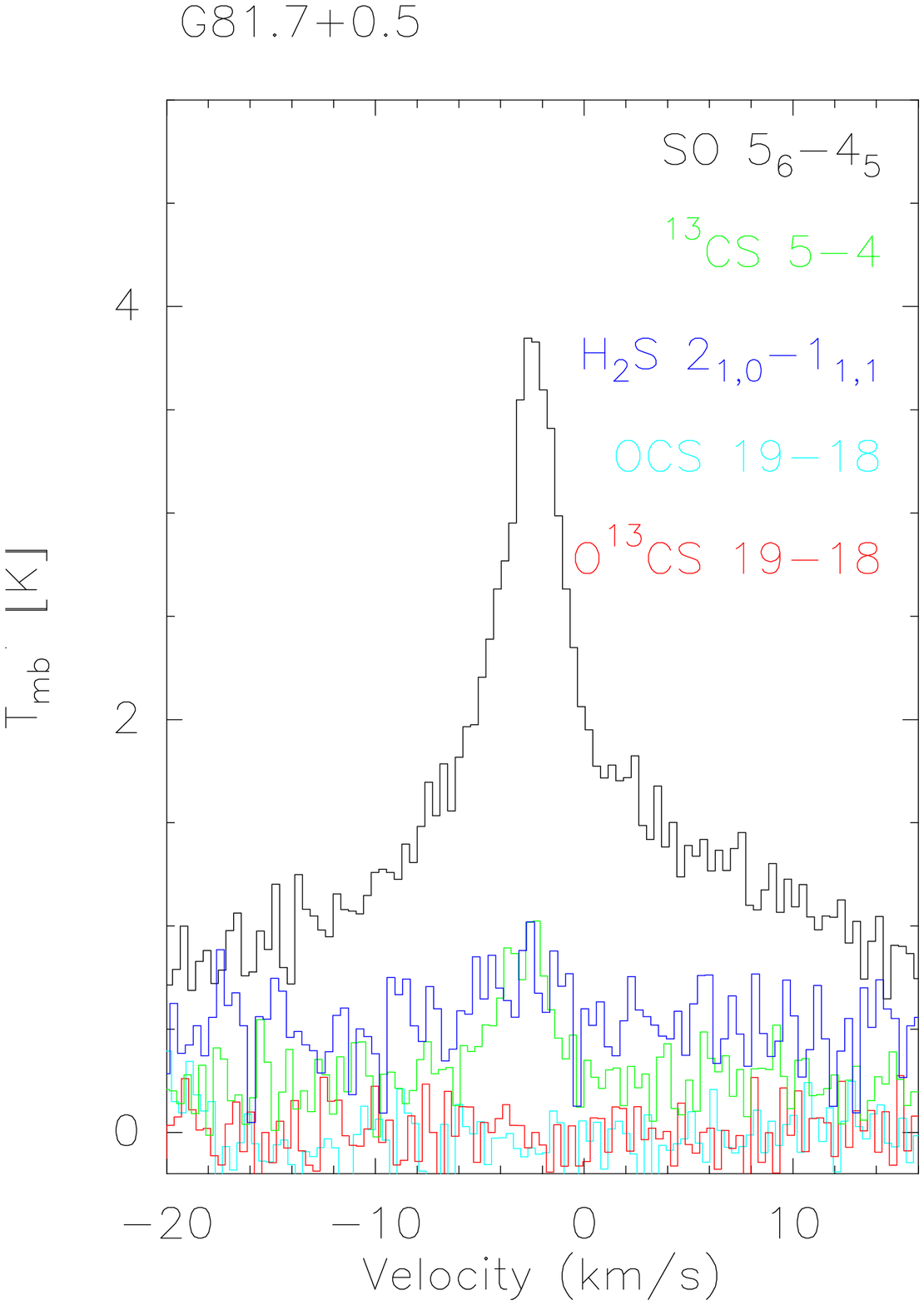}
\includegraphics[scale=0.3]{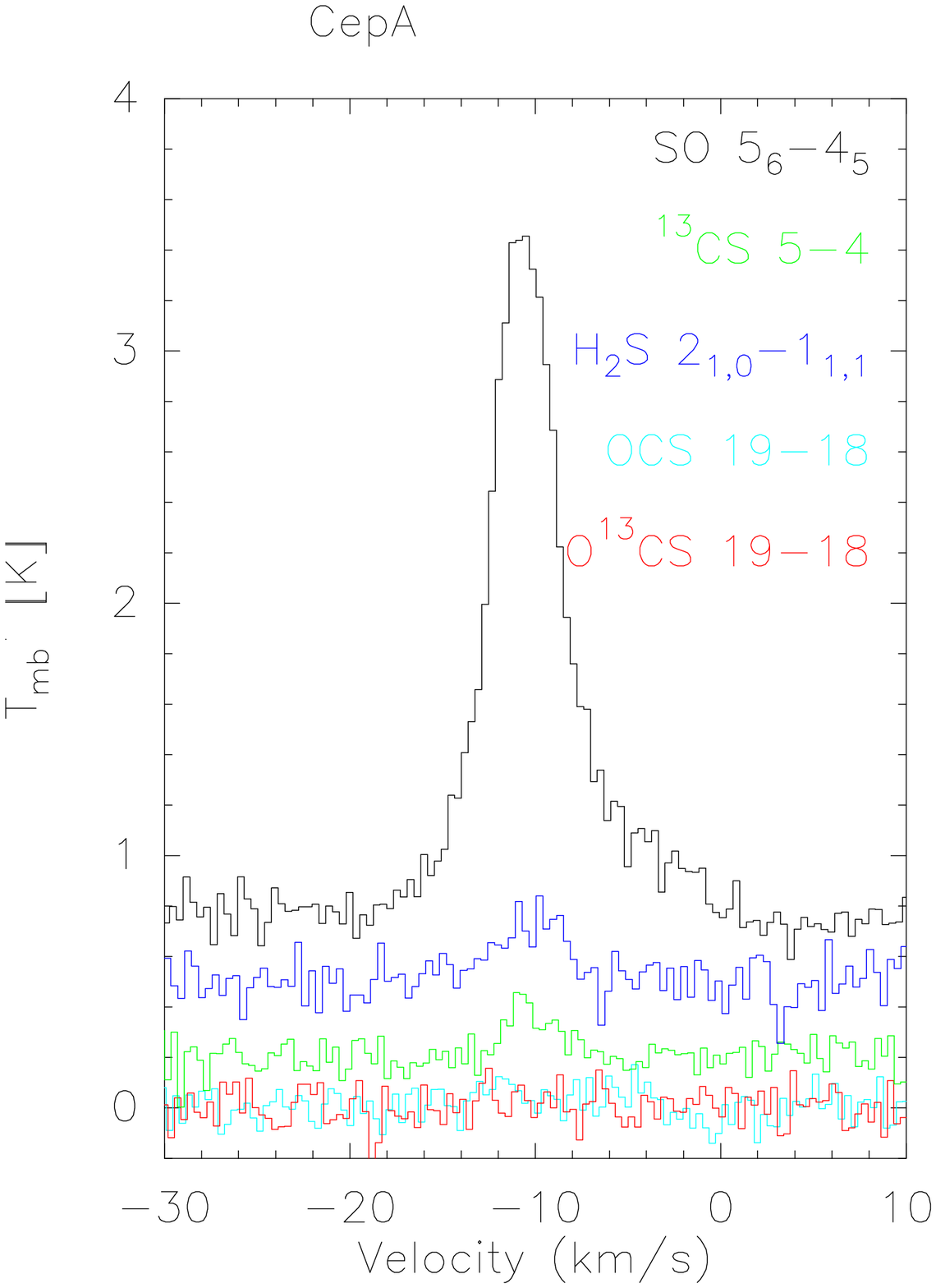}
\includegraphics[scale=0.3]{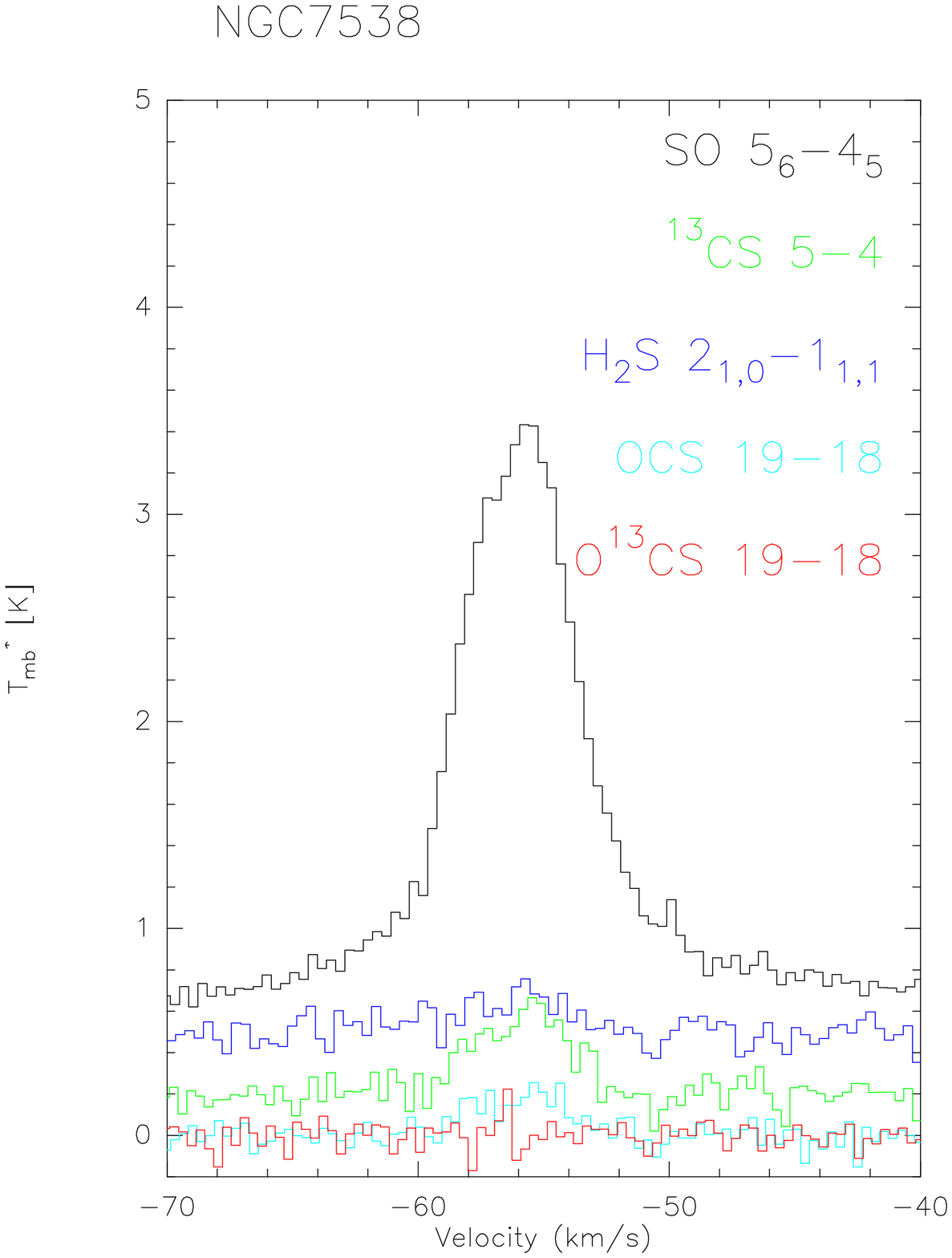}
\includegraphics[scale=0.3]{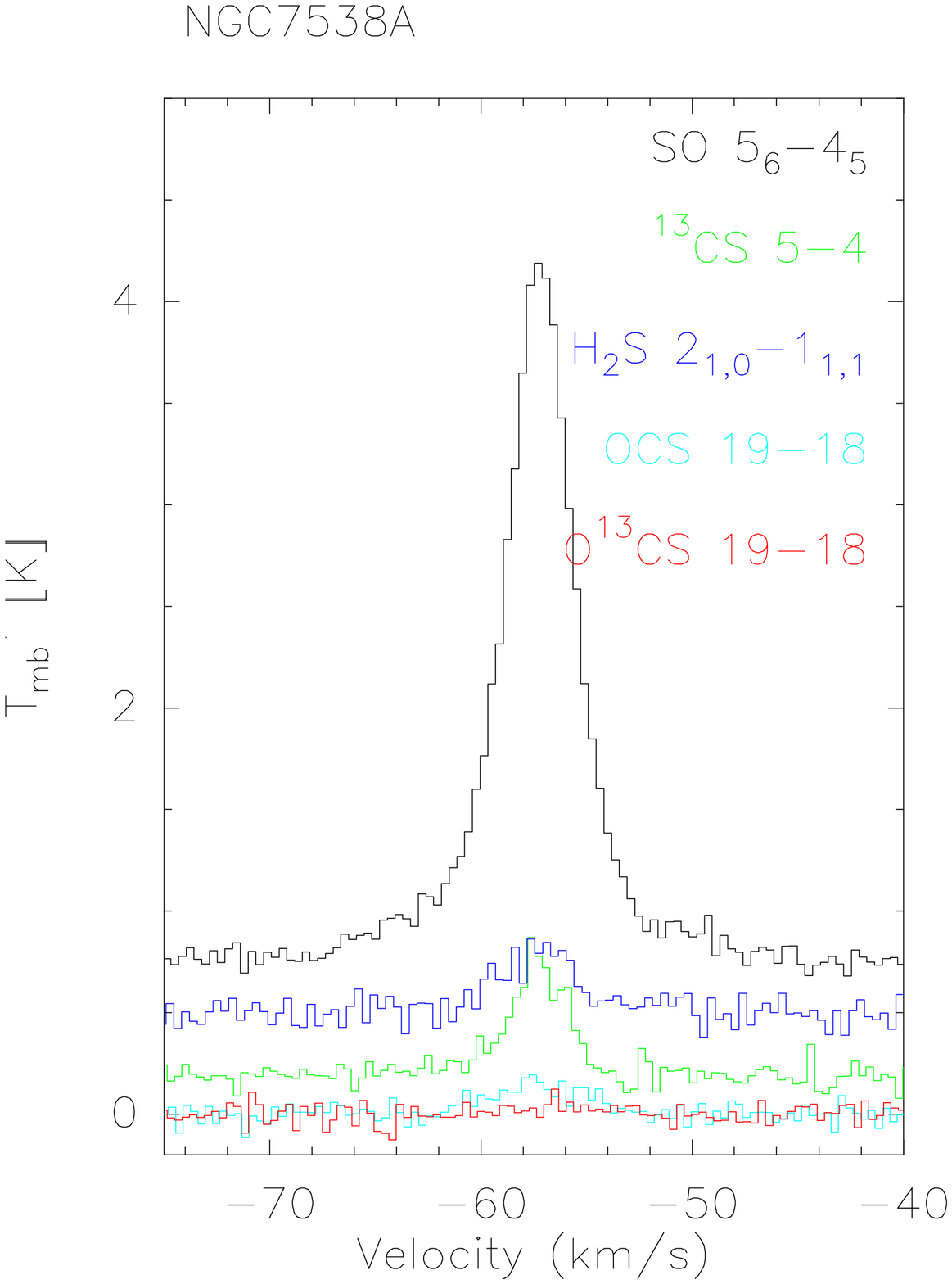}
\caption{Figure 1 continued.}
\end{figure}

\begin{figure}[htbp]
   \centering
\includegraphics[scale=0.8]{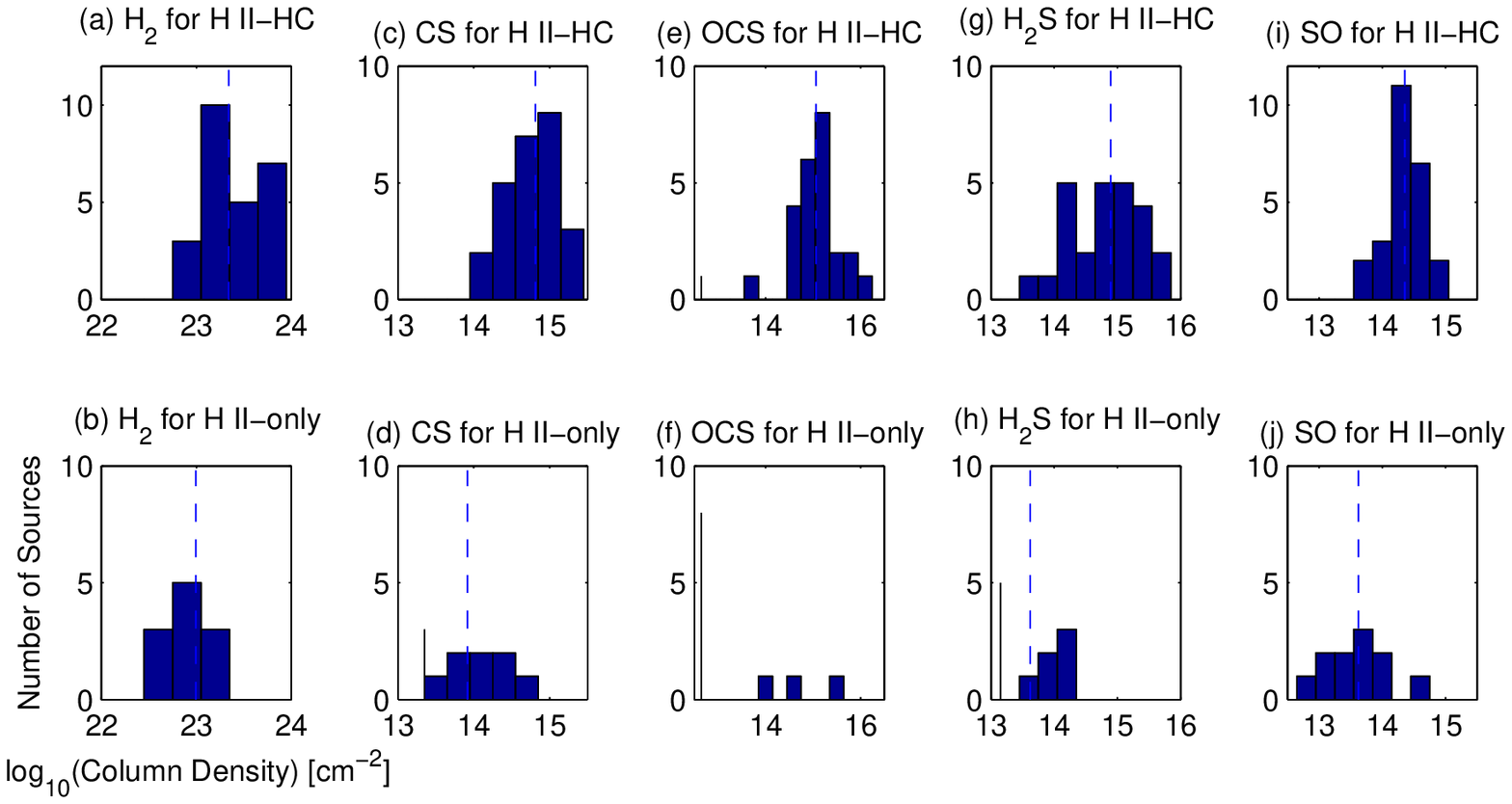}
\caption{Histograms of the number distributions of \hh, CS, OCS, \hhs\ and SO column densities for H II-HC and H II-only sources. The vertical dashed lines indicate the
median values of the column densities for each distribution. Median values of the column densities are given in Table 3. The vertical bars stand for sources undetected in particular molecular line emission.}
\end{figure}

\begin{figure}[htbp]
   \centering
\includegraphics[scale=0.8]{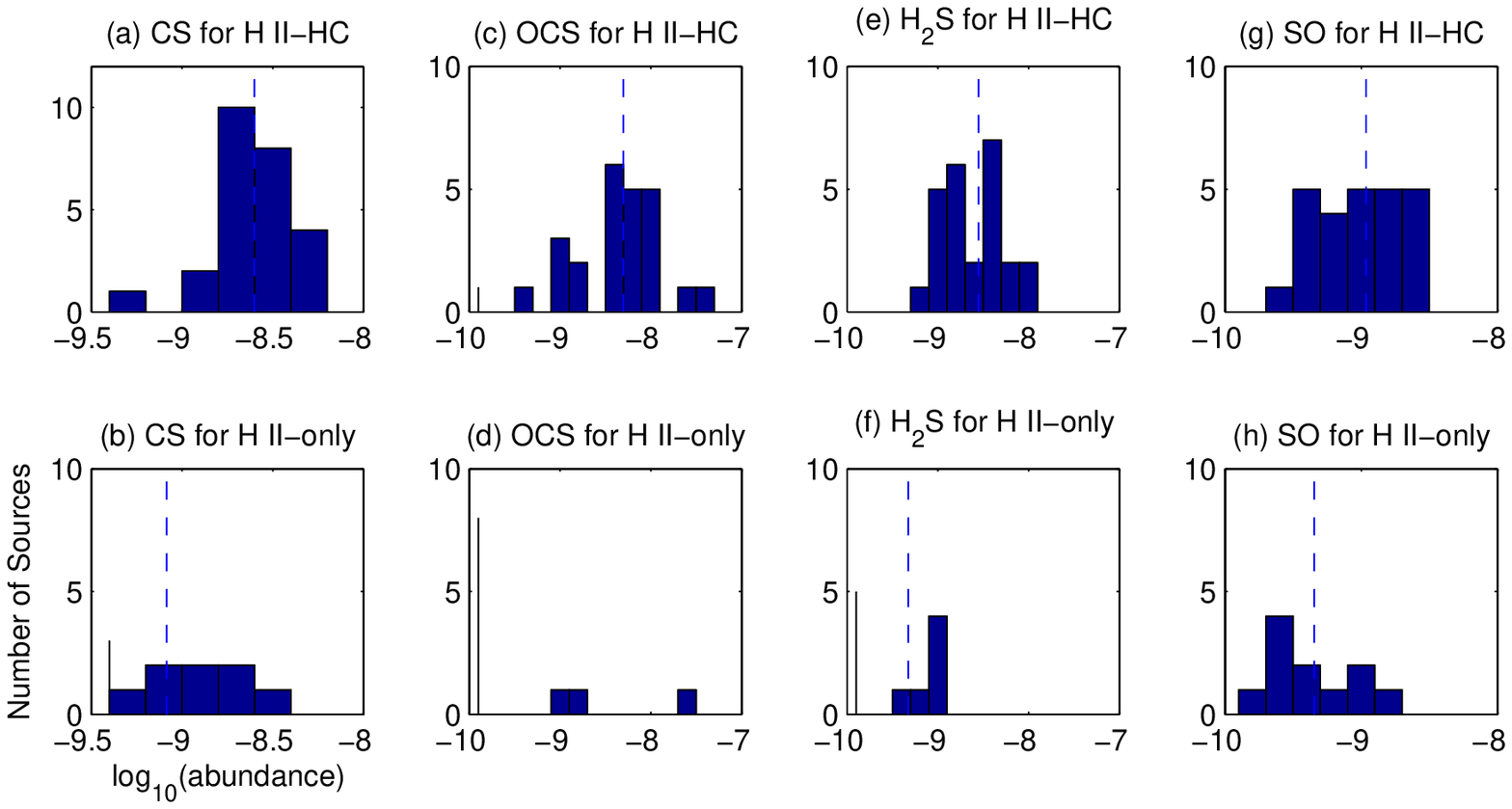}
\caption{Histograms of the number distributions of CS, OCS, \hhs\ and SO abundances for H II-HC and H II-only sources. The vertical dashed lines indicate the
median values of the column densities for each distribution. Median values of the column densities are given in Table 3. Similar to Figure 5, the vertical bars also stand for sources undetected in particular molecular line emission.}
\end{figure}


\end{document}